\documentclass[sort&compress,11pt]{elsarticle}

\usepackage[utf8]{inputenc}
\usepackage[T1]{fontenc}
\usepackage{graphicx}
\usepackage{color}
\usepackage{tikz}
\usepackage{here}
\usepackage{url}
\usepackage{amstext,amssymb}
\usepackage[pdftex,pdfpagelabels,plainpages=false,linktocpage=true]{hyperref}
\usepackage{comment}
\usepackage{xspace,commath}

\hypersetup{colorlinks}

\graphicspath{{./pictures/}} 

\definecolor{dgreen}{rgb}{0.,0.,0.}

 \definecolor{dblue}{rgb}{0.,0.,0.}

\def\checkmark{\tikz\fill[scale=0.4](0,.35) -- (.25,0) -- (1,.7) -- (.25,.15) -- cycle;}

\newcommand{\cmmnt}[1]{}

\newcommand{\Knud}{\ensuremath{\operatorname{\mathit{K\kern-.20em n}}}\xspace}

\makeatletter
\providecommand{\doi}[1]{%
  \begingroup
    \let\bibinfo\@secondoftwo
    \urlstyle{rm}%
    \href{http://dx.doi.org/#1}{%
      doi:\discretionary{}{}{}%
      \nolinkurl{#1}%
    }%
  \endgroup
}
\makeatother

\begin{document}
\begin{frontmatter}

\title{Accuracy and Performance Evaluation of Low Density Internal
       and External Flow Predictions using CFD and DSMC}

\author[add1,add2,add3]{Surya Kiran Peravali} 
\author[add2]{Vahid Jafari}
\author[add3,add5]{Amit K. Samanta}
\author[add3,add4,add5]{Jochen Küpper}
\author[add3]{Muhamed Amin}
\author[add2]{Philipp Neumann}
\author[add1]{Michael Breuer \corref{cor1}} \ead{breuer@hsu-hh.de} 

\cortext[cor1]{Corresponding author}

\address[add1]{Professur f\"ur Strömungsmechaik, 
               Helmut-Schmidt-Universit\"at / Universit\"at der Bundeswehr Hamburg,
               22043 Hamburg, Germany}

\address[add2]{Professur f\"ur High Performance Computing,
                 Helmut-Schmidt-Universit\"at / Universit\"at der Bundeswehr Hamburg,
                 22043 Hamburg, Germany}
 \address[add3]{Center for Free-Electron Laser Science CFEL, Deutsches Elektronen-Synchrotron DESY, 22603 Hamburg, Germany}

 \address[add4]{Department of Physics, Universit\"at Hamburg, 22761 Hamburg, Germany}
 \address[add5]{Center of Ultrafast Imaging, Universit\"at Hamburg, 22761 Hamburg, Germany}

\begin{abstract}
The Direct Simulation Monte Carlo (DSMC) method was
widely used to simulate low density gas flows
with large Knudsen numbers. However, DSMC encounters limitations
in the regime of lower Knudsen numbers ($\Knud<0.1$).
In such cases, approaches from classical computational
fluid dynamics (CFD) relying on the continuum assumption
are preferred, offering accurate solutions
at acceptable computational costs. 
In experiments aimed at imaging aerosolized
nanoparticles \emph{in vacuo} a wide range of Knudsen numbers
occur, which motivated the present study on the
analysis of
the advantages and drawbacks of DSMC and CFD simulations of 
rarefied flows in terms of accuracy and
computational effort. Furthermore, 
the potential of hybrid methods is evaluated.
For this purpose, DSMC and CFD
simulations of the flow inside a
convergent-divergent nozzle (internal expanding flow) and the
flow around a conical body (external shock generating flow)
were carried out. CFD
simulations utilize the software OpenFOAM and
the DSMC solution is obtained using the software
SPARTA. The results of these simulation
techniques are evaluated by comparing them 
with experimental data (1), evaluating the time-to-solution (2)
and the energy consumption (3), and assessing the feasibility of
hybrid CFD-DSMC approaches (4).
\end{abstract}

\begin{keyword}
 DSMC; SPARTA; Continuum assumption; Transition regime; Rarefied flow;
 high-performance computing
\end{keyword}

\end{frontmatter}


\section{Introduction}\label{sec:intro}

The Boltzmann equation is valid at any Knudsen number $\Knud$, i.e., from very low $\Knud$ (hydrodynamic
flows) up to very high $\Knud$ (extremely rarefied gas flows). While flows at low $\Knud$ are
traditionally simulated by methods from computational fluid dynamics (CFD) solving the Navier-Stokes
equations, they fail to provide sufficiently accurate results for rarefied gases since the continuum
mechanics assumption is violated. However, in the rarefied gas flow regime, where the mean free path
of the molecules $\lambda$ approaches or surpasses the characteristic length scale of the flow,
micro-scale effects invalidate the continuum hypothesis. Consequently, particle-based methods,
such as the direct simulation Monte Carlo (DSMC) method, often emerges as
the preferred choice. The DSMC method as introduced by~\citet{Bird1994, Bird2013}, tackles the
Boltzmann equation through Monte Carlo simulations. Unlike molecular dynamics approaches which
trace the trajectory of every particle in the fluid based on Newton's equations of motion, DSMC
stochastically generates collisions using scattering rates and post-collision velocity
distributions derived from the kinetic theory of dilute gases~\cite{francis}.  Notably,
DSMC exhibits superior accuracy in high Knudsen number regimes~\cite{Bird1994,Shen2005}.

However, none of these two methods -- CFD and DSMC -- is entirely
optimal for simulating gas flows in the intermediate-Knudsen-number regime $\Knud\approx0.01\ldots10$, especially when $\Knud$ changes drastically over the flow field. While CFD methods may suffer
from significant inaccuracies stemming from the neglect of
molecular effects, the computational demands for accurate
DSMC simulations increases with $\Knud^{-4}$~\cite{torre}
rendering them prohibitively expensive for $\Knud<0.05 $.
Consequently, there has been an increasing interest in using
hybrid methods, especially DSMC/CFD methods to simulate
rarefied gas flows with high Mach and intermediate Knudsen numbers~\cite{wang,glass,ghazan,virgile,malaikannan,espinoza,farber,giannandrea,kurtz}.

Furthermore, to decrease the time-to-solution for the
arising pure or hybrid simulations, high-performance
computing techniques were often applied~\cite{gallis,plimp,li,ivanov,roohi}.
While numerous works study the accuracy of the DSMC and CFD
methods and how different parametrizations, e.g., the choice
and parameters for the collision model or the number of
simulation particles, impact this accuracy~\cite{Fedo,francis,torre,chung,ghazan,koura,liu,moss},
they do not take into account the cost and efforts of the
methods and how much resources in terms of energy they
consume. On the other hand, in studies on
the computational performance of the DSMC method ~\cite{roohi,li,braunstein, gallis,plimp,ivanov}
the influence of the different parametrizations on
the simulation results was not investigated.

There is a long list of publications assessing the DSMC and CFD conundrum,
c.f.~\cite{braunstein, khanlarov, roohi,li,gallis,plimp,ivanov,
   celoria,chung,ghazan, koura,liu,moss,Fedo,francis,torre}. However, a detailed explanation of the
impact of the great variety of parametrization possibilities of the solvers, in particular for DSMC,
is often incomplete or tailored to very specific setups that are not easily generalizable.
Furthermore, the precision of the results and the performance of the DSMC method are not considered
simultaneously to judging if the accuracy is worth the amount of time and energy that the simulation consumes.

Here, we briefly revise the CFD and DSMC methods and compare them in terms of accuracy and
computational effort in different flow regimes. Considering both an external flow around a
conical body and an internal nozzle flow (2D and 3D), parametrizations are detailed and
their impact on the solution is analyzed. The software packages OpenFOAM (CFD)~\cite{openfoam} and SPARTA (DSMC)~\cite{sparta2022} were used.
Besides accuracy, the computational costs (run-time and energy consumption) are analyzed.

After summarizing related work in \autoref{sec:related}, the 
details of the test cases and the
parameters for both DSMC and CFD simulations are described in \autoref{sec:testcases}. Sections~\ref{sec:cfd}
and~\ref{sec:dsmc} explain the CFD and DSMC solvers employed in this work, respectively, as well as
the fundamental physical modeling of these approaches. The simulation approaches described in
Sections~\ref{sec:numerical} to \ref{sec:hybrid} are underpinned by validation with experimental
results, c.f.\ \autoref{sec:result}. A conclusion and an
outlook to future work is provided in \autoref{sec:conclusion}.
\\

\noindent The main goals of this work are:
\begin{itemize}
   \itemsep=0pt
\item Presenting a comprehensive study of rarefied gas flows for internal and external configurations

\item Using a combination of 2D/3D configurations of DSMC, CFD and hybrid DSMC-CFD simulation

\item Highlighting the effect of different parameters having major impact on the simulation such as
   the mesh size, number of particles, time step, collision model, boundary conditions, and
   computational speed 

\item Validating and evaluating the simulations against experiments

\item Reporting the performances and computational efforts of the employed approaches in terms of
   scalability and energy consumption
\end{itemize}

\section{State of the art}
\label{sec:related}

Numerous studies were performed to analyze gas flow fields, including
rarefied and continuum flows, most of them using DSMC or CFD
methods. We provide a brief summary of the literature,
  which cannot be comprehensive but primarily presents studies that
  depict specific aspects relevant to our study yet collectively
  demonstrate remaining questions in the field.

The majority focused on the accuracy and parameter studies
of DSMC simulations such as the choice of the mesh or the collision
model~\cite{gopal,ze,VSSparam2,stefanov,L-B1,L-B2} and validated the results partially against
experimental results. Some studies were related to more practical cases for engineering
purposes~\cite{ghazan,xiao,wang,torre,glass,liu,moss}. Since the DSMC method can be very time
consuming, many studies targeted the performance of the approach: While the computational
performance of the DSMC method on different computer architectures in terms of runtime is discussed in~\cite{plimp,ivanov, gallis,khanlarov,kloth}, algorithms and
optimizations to speed up the DSMC method can be found in~\cite{roohi,li,braunstein,celoria}. To the
authors' knowledge, energy consumption has not been addressed so far. Discussions on the impact of
the mesh dependency, collision model and boundary conditions are not included in the aforementioned
works.

Flows through micro-nozzles
were studied based on DSMC and a compressible Navier-Stokes solver
applying slip and no-slip conditions~\cite{torre}. The computational results are compared
and the most important outcome is that a better agreement between DSMC and CFD is
observed when a slip-wall boundary condition is implemented. It is also shown that the CFD and DSMC
results differ in the divergent part of the nozzle, especially close to the outflow, where the
breakdown parameter $\Knud$ is relatively high or, in other words,
strong rarefied effects start to
appear. No evaluation of the results with regard to a comparison with experimental findings or
computational expenses are included. The effect of different parameters such as inlet
and wall boundary conditions and the Reynolds number on DSMC and Navier–Stokes approaches for a
micro-nozzle flow with a relatively small Knudsen numbers was studied~\cite{liu}. Furthermore, 
it was investigated in which part of the micro-nozzle DSMC and CFD provide the best results. It is
shown that the CFD results exhibit obvious deviations from the DSMC results as $\Knud $ exceeds
$ 0.045 $. The computational performance of large-scale parallel DSMC on
homogeneous (CPU) and heterogeneous (CPU+GPU) systems was studied and
different programming approaches (MPI, hybrid MPI+OpenMP and OpenACC)
were discussed~\cite{li}.

Extensive research on the development of hybrid DSMC-CFD methods
was carried out,
c.f.~\cite{wang,glass,ghazan,virgile,malaikannan,espinoza,farber,giannandrea,kurtz}.
With regard to their application,
simulation cases were mostly very specific, and the comparison was focused typically either on
numerical accuracy~\cite{wang,glass,ghazan,farber,giannandrea,kurtz,wu} or on
performance~\cite{virgile}. For example, the advantages of a hybrid
DSMC-CFD approach over pure DSMC
is indicated in~\cite{wang}. For this purpose, the authors simulate a hypersonic flow over a
two-dimensional wedge. A comparison of the flow field predicted by a 3D CFD-DSMC simulation
considering a space capsule geometry is given in~\cite{glass}, including a validation against
experimental results from wind tunnel tests. The study also
investigates the effect of the mesh
dependency on CFD and DSMC methods. A comparison study \cite{ghazan} explores
the performance difference between coupled CFD-DSMC and pure CFD methodologies
in simulating a gas centrifuge handling $ ^{235}UF_6 $ gas.  
Pure DSMC results around the Mars
pathfinder and Mars micro-probe capsules are studied in~\cite{moss}, however, 
without a validation against experimental results. Computational efficiency
of massively parallel (stand-alone) DSMC for
different cases is studied, amongst others, in~\cite{plimp} and~\cite{kloth}.
Another comparison study between DSMC and CFD results of a low-density nozzle
flow and the experimental evaluation is discussed in \cite{chung}. 

The accuracy of a DSMC simulation depends on a number of numerical
parameters such as the time step size,
the cell size, and the number of samples. Furthermore, 
the choice of the collision model plays an important role.
An analysis of (statistical and deterministic) numerical errors
corresponding to numerical parameters in the DSMC method 
is provided in~\cite{Fedo} based
on a simple heat transfer problem between two parallel plates in
1D and 2D. The statistical error analysis of the DSMC method applied to hypersonic and nozzle flows is provided in~\cite{CHEN_boyd}. A further error analysis
considering various numerical parameters (sampling cells,
sampling time step and sample sizes) is provided in~\cite{plotnikov}. 

\autoref{tab:relworks} summarizes which related
studies covered the different areas of DSMC/CFD simulations
in terms of accuracy and computational
performance and classifies the current paper accordingly.
It should provide recommendations and suggestions on DSMC,
CFD and hybrid methods as well as a calibration of those DSMC
parameters having a major impact on the simulations to achieve
sufficiently accurate results in an acceptable amount of time.
 
\begin{table}[ht!]
	\color{black}
 {\scriptsize
	\begin{tabular}{lcccccc}
		\hline
		\textbf{Paper}&\textbf{CM}$^a$&\textbf{$\Knud $ }&\textbf{GSI}$^b$&\textbf{Run-}&\textbf{EC}$^c$\cmmnt{&\textbf{MU}$^d$}&\textbf{Validation} \\
&  & &  &\textbf{Time} &  &\textbf{against Exp.} \\
		\hline
		Boyd~\cite{Boyd1}& &\checkmark  & & & &\checkmark\\

        \citet{wang}&  & \checkmark & & &  &\\

         \citet{glass}& &\checkmark  & & & & \checkmark \\

		\citet{ghazan}& &\checkmark &  &  & &\\

      	\citet{gallis}& \checkmark & \checkmark & &\checkmark &  &\\

        \citet{plimp}&  &  & & \checkmark& \cmmnt{&\checkmark }&\checkmark\\

        \citet{li}& &  & &\checkmark & \cmmnt{&\checkmark } &\\

        \citet{roohi}&  & & &\checkmark &  &\\

        \citet{torre}&  &\checkmark & & & &\\

		\citet{chung}& \checkmark&\checkmark  &\checkmark & &  &\checkmark\\
        
        \citet{MaxCLL_nasa}& & &\checkmark & &  & \\

		\citet{koura}& \checkmark&  & & & & \\

		\citet{liu}& &\checkmark  & & & & \\

		\citet{moss}& &\checkmark  & & &  &\\

        \citet{khanlarov}&  &  & & \checkmark& \cmmnt{&\checkmark } &\\

        \citet{gopal}& \checkmark &  & & &  &\\

        \citet{VSSparam2}& \checkmark &  & & & \cmmnt{&\checkmark }&\checkmark\\

		\citet{stefanov}&  &\checkmark & &  & &\\

		\citet{L-B1}& \checkmark &  & &  & &\\

		\citet{L-B2}& \checkmark & \checkmark & & & & \\

     	\citet{xiao}& \checkmark &\checkmark & &  & &\checkmark\\

        \citet{kloth}& &  & &\checkmark & \cmmnt{&\checkmark } &\checkmark\\

        \citet{CHEN_boyd}&\checkmark  &\checkmark  & &\checkmark & & \\

        \citet{plotnikov}&  &\checkmark  & & & & \\

        \citet{Pfeiffer}&  &\checkmark  & &\checkmark & &\checkmark \\

		\citet{falchi}&  & \checkmark& & & & \\
		\textbf{\textit{Current study}}& \checkmark & \checkmark& \checkmark& \checkmark& \checkmark&\checkmark \\
		\hline
	\end{tabular}}
	\caption{\label{tab:relworks} Overview of related work on DSMC/CFD simulations and classification of the present
     study. Here $\Knud$ refers to studies
    related to the continuum breakdown using global and/or local Knudsen numbers \newline
	($^a$ Collision Model,
	$^b$ Gas-Surface-Interaction,
    $^c$ Energy Consumption).
}
\end{table}

\section{Test Cases}\label{sec:testcases}

\subsection{Internal flow -- Low density nozzle flow}

\begin{figure}[ht]
\centering
\includegraphics[width=0.7\textwidth]{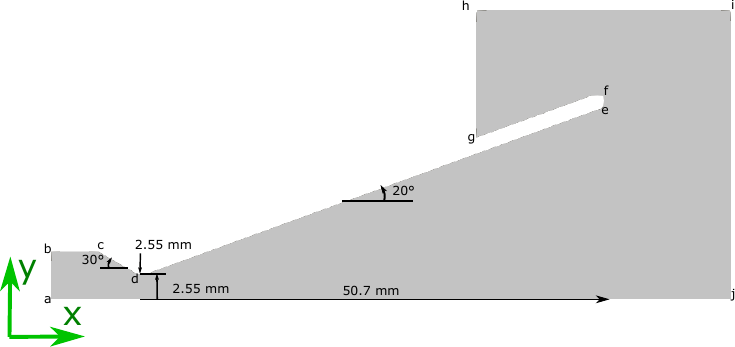}
   \caption{Computational domain of the low density nozzle flow (cases i.I to i.III).}
\label{fig:temp474}
\end{figure}

\autoref{fig:temp474} indicates the 2D simulation domain of the low density nozzle.
The test case is based on the experiment by Rothe~\cite{rothe} where the low
density flow properties are measured inside the nozzle using the electron beam
fluorescence technique. The inflow boundary is located on the extreme
left (line ab), i.e., at the inlet of the pressure chamber. An outflow boundary
condition is assumed along the extreme right  of the geometry (line fghij).
The segment (aj) in \autoref{fig:temp474} is the axis of symmetry. 
For 2D simulations, an axisymmetric boundary condition was imposed.
3D simulations were carried using the complete geometry.

The test gas used in the simulation is nitrogen with a stagnation temperature of 300~K. Three different stagnation pressure configurations are tested according to the setup described in \autoref{tab:internalcases}. That leads to the three internal flow cases i.I to i.III. 
The nozzle Reynolds number $Re_N$ is calculated using $Re_N = {\rho_o \, \hat{u} \, r_*}/{\mu_o}$ where $\rho_o$ is the stagnation density, $\hat{u}$ is the adiabatic speed, $r_*$ is the radius of the throat and $\mu_o$ is the  viscosity based on the stagnation condition.
 The Knudsen number is calculated based on the stagnation condition and the throat diameter.

\begin{table}[h!]
	\centering
	\color{black}
	\begin{tabular}{lccc}
		\hline
		\textbf{Parameters}&\textbf{i.I}&\textbf{i.II} &\textbf{i.III} \\
		\hline
		Test gas& $N_2$& $N_2$ & $N_2$ \\
		
		Stagnation temperature, $T_o$ [\text{K}] & 300  & 300   & 300 \\
		
		Stagnation pressure, $P_o$ [\text{Pa}]& 474  & 209  & 141  \\
		
		Wall temperature, $T_w$ [\text{K}]& 300  & 300  & 300  \\
		
		Nozzle Reynolds number, $Re_N$& 590  & 260 & 175  \\
		
		Throat Knudsen number, $\Knud_t$& $2.3 \times 10^{-3}$ & $6.17 \times 10^{-3}$  & $9.15 \times 10^{-3}$\\
		
		Back pressure, $P_a$ [\text{Pa}]& 1.8  & 0.64   & 0.45 \\
		\hline
	\end{tabular}
	\caption{\label{tab:internalcases}Flow-condition parameters
          of the low-density-nozzle cases i.I to i.III.}
\end{table}

\subsection{External flow -- Blunt cone \& sharp cone}

\begin{figure}[H]
	\centering
	\begin{tabular}{cc}
	\includegraphics[width=0.49\textwidth]{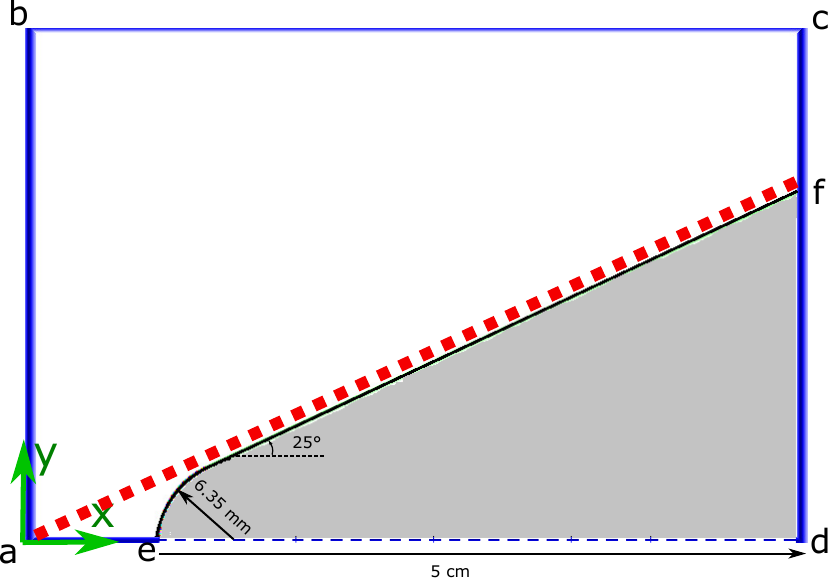}&\includegraphics[width=0.49\textwidth]{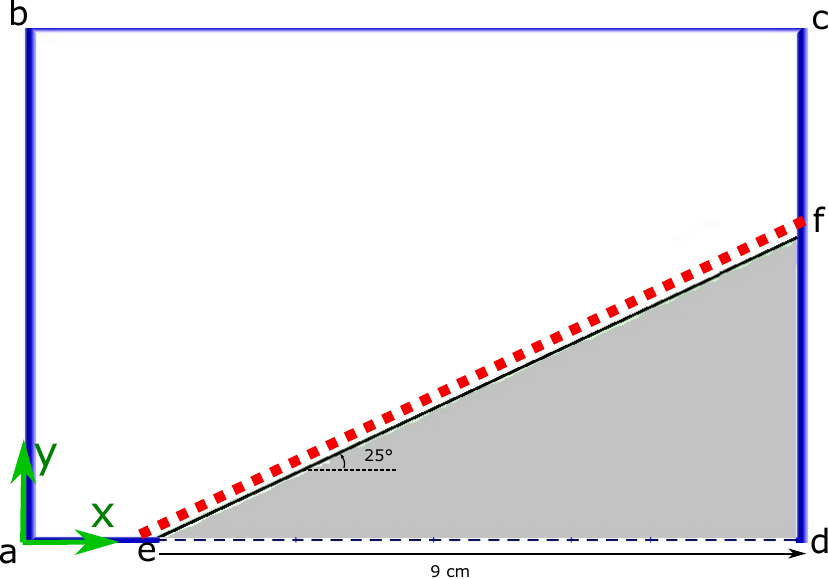}\\
	(a)&(b)\\\\
	\end{tabular}
	\caption{Computational domains for the external flow test cases. (a)~Half body cone with blunted nose, 2D setup. (b)~cone with sharp nose, 2D setup. }
	\label{fig:domainext}
\end{figure}

 Figure~\ref{fig:domainext} demonstrates the 2D computational domains
 of the external flow simulations considering two geometries.
 The first geometry is a cone-shaped half body with a blunted nose and the second one
 possesses a sharp nose. Both geometries are assumed to be infinitely
 long in positive $x$-direction. The symmetry line of the half body is
 aligned with the free-stream. For the first half body, only the
 first 0.05~m of the blunted nose with a radius of
 $6.35 \times 10^{-3}$~m (0.25 in) is considered
 For the second half body, the first 0.09~m of the sharp nose is taken 
 into account. The computational domains around these bodies are depicted in 
 Fig.~\ref{fig:domainext}.

The fluid considered in both cases was pure nitrogen and
the ambient conditions at different altitudes are listed in \autoref{tab:externalcases}.
Other parameters needed for CFD and DSMC simulations such as the
pressure and number density were calculated correspondingly.
The flow direction was in \mbox{$x$-direction} such that the
left and right surfaces of the domain represent the inlet and
outlet, respectively (see Fig.~\ref{fig:domainext}).
For the 2D external flow, only half of the computational domain
in $y$-direction was considered, i.e., a cut through the
computational domain ($x$-$y$ plane) as visible in \autoref{fig:domainext}.
The axisymmetric boundary condition was imposed on the
line a-d.

\begin{table}[h!]
	\centering
	\color{black}
	\begin{tabular}{l c c c}
		\hline
		\textbf{Parameters}&\textbf{e.I}&\textbf{e.II}\\
		\hline
        
		Geometry& Blunt cone&Sharp Cone \\
		
		Gas mixture& N2&N2 \\
		
		Ambient velocity, $ V_\infty~[\frac{\text{m}}{\text{s}}] $ & 2764.5 &2072.6\\
		
		Ambient temperature, $ T_\infty~[\text{K}] $ & 144.4 &42.61 \\
		
		Ambient pressure, $ p_\infty~[\text{Pa}] $ & 21.91 & 2.23 \\
		
		Ambient density, $ \rho_\infty~[\frac{\text{kg}}{\text{m}^3}] $ & $ 5.113\times10^{-4}$ & $ 1.757\times10^{-4}$\\
		
		Mach number, $M_\infty $& 11.3 & 15.6\\
		
		Wall temperature, $T_w~[\text{K}] $& 297.2& 297.2\\
		
		Knudsen number, $\Knud_\infty$& $4.168 \times 10^{-3}$  &$ 1.236 \times 10^{-2}$\\
		
  Reynolds number, $Re_\infty$& 4020 & 1862\\
		\hline
	\end{tabular}
	\caption{\label{tab:externalcases} Flow conditions for external test cases.}
\end{table}

\section{Numerical Approach} \label{sec:numerical}
\subsection{Continuum Flow Solver:
Navier-Stokes}\label{sec:cfd}

For this study, CFD simulations were performed using OpenFOAM, version v2112.
OpenFOAM is a finite-volume based code which has a collection of
libraries dedicated to the solution of partial differential
equations (Navier-Stokes equations). Among the standard solvers available
in the OpenFoam library, the \verb|rhoCentralFoam| solver is used for
investigating the present test cases. \verb|rhoCentralFoam| is a
density-based transient solver used for transonic and supersonic
flow regimes of a compressible gas \cite{Greenshields}.

First/second-order schemes are used for the discretization of the governing equations. For gradient and divergence terms, a second-order Gauss linear scheme is used. 
For the diffusive terms 
in the governing equations, the Gauss scheme is the only choice of discretization and requires an interpolation scheme for the diffusion coefficient (linear). For the transient cases, the temporal derivatives
are discretized using either the first-order bounded implicit Euler scheme or the backward scheme. The Kurganov scheme \cite{Kurganov, Kraposhin} is used to compute fluxes at the cell interfaces which prevents spurious oscillations around shocks.
 
 In OpenFOAM, the computational domain is generally split up
 into a set of patches and the boundary conditions are then
 assigned as attributes to the patches and to the field variables
 on a patch. Various kinds of boundary
 conditions are available in the OpenFOAM library. 
 The boundary conditions assigned to the test cases 
 are listed in \autoref{tab:BCI} and \autoref{tab:BCE}
 for the internal and external flow cases, respectively. 
 For both internal- and external-flow cases, the CFD calculations
 rely on structured grids specified and depicted
 in~\ref{app:grids}.
 
\begin{table}[h!]
	\centering
	\color{black}
	\begin{tabular}{l c c c}
		\hline
		\textbf{Boundary}&\textbf{U}&\textbf{P}&\textbf{T} \\
		\hline
		Inlet& zero gradient & fixed value & fixed value \\
		
		Outlet& zero gradient  & wave transmissive & zero gradient \\
		
		Wall& no-slip  & zero gradient   &  fixed value\\
		\hline
		
	\end{tabular}
	\caption{\label{tab:BCI} Boundary conditions for the 
                             low-density-nozzle flow (internal flow).}
\end{table}

\begin{table}[h!]
	\centering
	\color{black}
	\begin{tabular}{l c c c}
		\hline
		\textbf{Boundary}&\textbf{U}&\textbf{P}&\textbf{T} \\
		\hline
		Inlet& fixed value & zero gradient & fixed value \\
		
		Outlet& inlet outlet  & wave transmissive & zero gradient \\
		
		Boundaries & supersonic free-stream  & zero gradient   &  fixed value\\
		
		Solid walls& no-slip  & zero gradient   &  fixed value\\
		\hline
	\end{tabular}
	\caption{\label{tab:BCE}  Boundary conditions for the flow around
                a conical body (external flow).}
\end{table}

\subsection{Kinetic Approach: DSMC Method}\label{sec:dsmc}

The Direct Simulation Monte Carlo method is a discrete particle simulation
technique that provides a numerical approximation of the solution
of the Boltzmann equation. In this method each particle represents
a large number of real molecules \cite{Bird2013,Shen2005} while
maintaining the overall phase space distribution.
The inter-molecular and the molecule-surface collisions
are calculated using probabilistic models.
The momentum term and  the collision term 
in the Boltzmann  equation are solved in a decoupled
manner \cite{francis}. The DSMC method is a widely used methods
for simulating rarefied gas flows. Its accuracy depends on
the number of particles per grid cell, the size of the grid
cells, the choice of the time steps and the collision models,
i.e., for particle-particle and particle-surface collisions.
It is general practice to limit 
the maximum cell size to one third
of the mean free path and the time step 
to below one fourth of the
mean collision time~\cite{Shen2005}. In the subsequent
analysis process, the microscopic-flow properties are
sampled by averaging the particle properties per grid cell
to obtain the macroscopic flow quantities. These sampled
values are stored for the geometric center of the DSMC grid
cells.

\noindent We used the SPARTA 
(Stochastic PArallel Rarefied-gas Time-accurate 
Analyzer~\cite{SPARTA}) DSMC code to simulate
transitional and rarefied flows. SPARTA is a highly
benchmarked tool \cite{spartaBench} that can simulate
systems with a few to millions or billions of particles.
It exhibits a good scalability and memory usage~\cite{SPARTA,kloth}. 
Presently, the no-time-counter (NTC) method is used as
collision-sampling technique \cite{Bird2013} along with the
variable-hard-sphere (VHS) and variable-soft-spheres
(VSS) molecular models~\cite{Shen2005, Bird1994} as well
as the Larsen and Borgnakke (L-B) model~\cite{L-B1,L-B2} to handle
internal energy exchange. In the L-B model, only a fraction of the
collisions are assumed to be inelastic which is defined by an
average probability of the internal energy exchange~$\phi$.
This parameter is used to determine the rate of the relaxation
process of the energy which can also be given as the inverse
of the relaxation collision number $Z$ ($\phi = 1/Z$)~\cite{chung,Shen2005}.
SPARTA uses the aforementioned Larsen and Borgnakke model with
constant and variable relaxation~\cite{Bird1994}. In the current study,
the vibrational mode and chemical reactions are assumed to be
frozen. The collision parameters used are given in 
\autoref{tab:collisionparam}. 

\begin{table}[h!] 
	\centering
	\color{black}
	\begin{tabular}{c c c c c c c}
		\hline
		\textbf{Gas}&\textbf{Molecular diameter}&\textbf{$\omega$}&\textbf{$T_\text{ref}$}&\textbf{$\alpha$}&\textbf{$\xi_{rot}$} &$\phi_{rot}$ \\
		\hline
		$N_2$& $4.07 \times 10^{-10}$~m & 0.74 & 273.15~K & 1.36 & 2 & 0.2 \\
		\hline	
	\end{tabular}
	\caption{\label{tab:collisionparam} Collision model parameters
          of simulated gas molecules
          \cite{VSSparam2,Bird2013}. $\omega$ is the
            power exponent of the temperature in the viscosity law,
            $T_\text{ref}$ is the reference temperature, $\alpha$ is
            the scattering coefficient and $\xi_{rot}$ is the
            rotational degree of freedom.}

\end{table}

For the variable relaxation model, the parameter $\phi_{rot}$ in SPARTA
is calculated using equation~(A.5) in~\cite{Bird1994}.
The unknowns in the equation were inferred from experiments~\citep{parker,lordi}.
Both models can be used for temperature ranges from 0 to 1200~K. 


The boundary conditions in DSMC can be classified in two types:
Free-stream boundary conditions (inlet and outlet) and wall
boundary conditions, where the gas-surface interactions are
predominant. At the inlet, the flux and the thermal state of
the molecules are defined according to the flow conditions
and the particles are introduced into the simulation
domain~\cite{Bird1994}. The outlet removes the simulation
particles leaving the simulation domain (see~\ref{app:machno}).

\subsection{Gas-surface interactions}\label{sec:dsmc_bc}

The gas-surface interaction model plays a dominant role
for the accuracy of the DSMC simulation and choosing an
accurate model is subjected to many factors. The most widely used
gas-surface interaction models in DSMC are
the Maxwell model and the Cercignani-Lampi-Lord
(CLL) model which include several unknown parameters.
In the following, these models are briefly discussed to
establish some guidelines in choosing the appropriate model along with
the corresponding parameters.

\begin{figure}[H]
	\centering
	\begin{tabular}{cc}
		\includegraphics[width=0.45\textwidth]{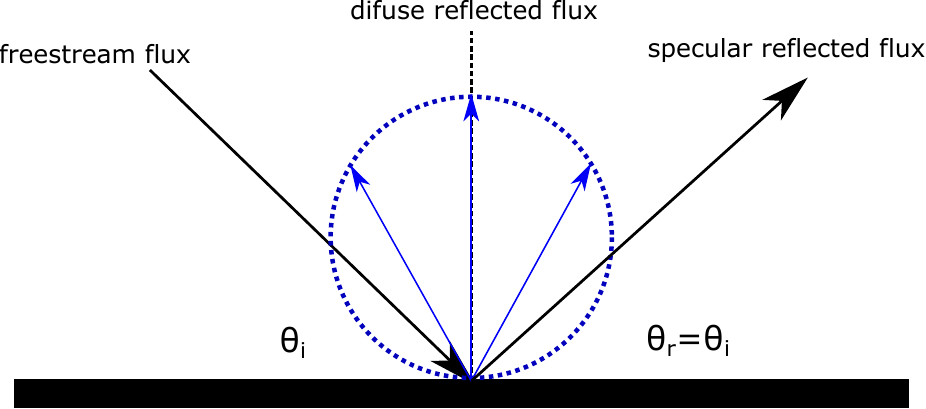}&\includegraphics[width=0.45\textwidth]{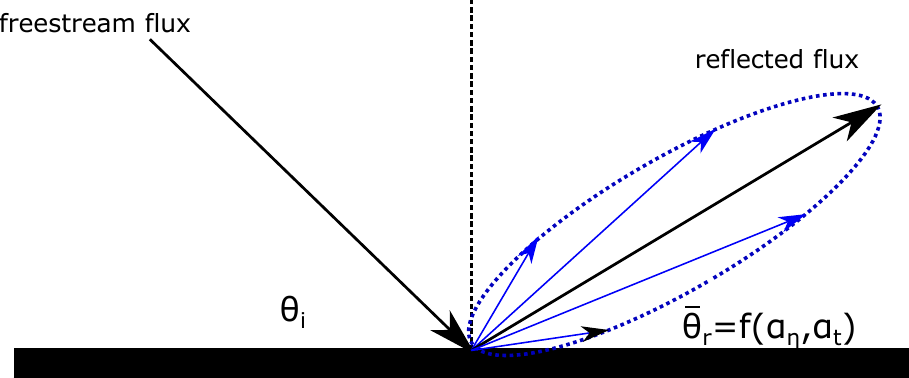}\\
		(a)&(b)\\\\
		
	\end{tabular}
	\caption{Schematic diagrams of the (a)~Maxwell model and (b)~CLL  models.}
	\label{fig:maxacc}
\end{figure}

In the Maxwell model (\autoref{fig:maxacc}~(a)), it is assumed that
the molecules are either reflected diffusively with complete energy
accommodation ($\alpha_E =100~\%$) or reflected specularly with no
energy exchange. The Maxwell model takes a fraction $\epsilon$ of
incident molecules to be scattered diffusely and the remaining 
fraction of molecules ($1-\epsilon$) are scattered
specularly. This model is useful for describing the thermodynamic
behavior of the gas, however, it does not describe the molecular
behavior of the gas which is frequently observed in fundamental
gas-surface scattering experiments~\cite{lobularExp}. 

The Cercignani-Lampis-Lord model (CLL)~\cite{Lord_ext1, Lord_ext2} is based on the assumption that there is no coupling of the normal and tangential velocity components
during gas-molecule reflections from a surface. Therefore, this model uses two coefficients $\alpha_n$ and $\alpha_t$, which represent energy-accommodation coefficients associated with normal and tangential components of the velocity, respectively. The scattering distribution of the molecules is centered around an average scattering angle $\overline{\theta}_r$ which is a function of the two accommodation coefficients (\autoref{fig:maxacc}~(b)). This scattering distribution has a lobular shape similar to the one observed in experiments~\cite{lobularExp}. This model also accounts for the internal energy exchange by introducing accommodation coefficients for rotational and vibrational modes. Furthermore, CLL has the capability to produce diffuse scattering with incomplete energy accommodation ($\alpha_E = \alpha_n = \alpha_t < 100~\%$) by changing the scattering distribution. The details of the implementation of this model are described in \cite{Shen2005}.


\section{Hybrid CFD/DSMC}\label{sec:hybrid}

The one-way coupling of CFD and DSMC is implemented as follows:
The fluid flow in the entire computational domain is simulated based on
the continuum solver (OpenFoam). An interface position (plane) is chosen
in the computational domain and the fluid flow data are extracted at
this position. This position is determined according to a
continuum-breakdown parameter calculated from the flow field.
The computational domain of the DSMC method is generated by
splitting the former computational domain from the interface
position toward the outflow exit. The extracted fluid flow
data are introduced as an inflow boundary condition to the
DSMC simulation, which is then carried out to resolve the
transitional rarefied region.  

In order to quantify the continuum-breakdown parameter,
different definitions of the Knudsen number are used such as 
(a) a global Knudsen number $\Knud = \lambda / L$ and  
(b) a local Knudsen number or Boyd's Gradient-Length-Local
Knudsen number $\Knud_{GLL,Q} = \lambda |\nabla Q | / Q$~\cite{Boyd1}; $\lambda$
is the mean free path, $L$ is the characteristic length, $Q$ represents
a macroscopic flow property such as the density $\rho$, the
velocity $\textbf{v}$ or the temperature $T$. A breakdown
Knudsen number $\Knud_B$ is calculated based on
the maximum of the local Knudsen numbers and the global Knudsen number
in the computational domain~\cite{Pfeiffer, torre}, i.e, 
    \begin{equation}\label{eq:Kn_break}
        Kn_B = \max(Kn, Kn_{GLL,\rho}, Kn_{GLL, T}, Kn_{GLL,|\textbf{v}|}).
    \end{equation}

When $\Knud_B>0.05$, the continuum breakdown is assumed~\cite{torre, espinoza}.


\section{Results and Discussion} \label{sec:result}
\subsection{Internal Flow: DSMC, CFD, Hybrid}\label{sec:internal_1}

Simulations were performed for the test cases presented in
\autoref{tab:internalcases} using the continuum approach
(OpenFoam), the DSMC method and the hybrid CFD/DSMC.
The full DSMC simulations are inefficient in the low Knudsen
regime ($\Knud < 0.05$) and demand higher computational effort. 
Therefore, the hybrid DSMC method was used in order to
accelerate the simulations. The results obtained by the continuum approach
were used to estimate the breakdown Knudsen number $\Knud_B$
(eq.~(\ref{eq:Kn_break})). By determining $\Knud_B$ throughout
the simulation domain, it was observed that the continuum breakdown
occurred in the simulation domain right after the throat.
Hence, the interface between the continuum and DSMC domain was
positioned at the throat using the
approach described in Section~\ref{sec:hybrid}. The continuum simulations
were performed for a 3-D simulation domain 
and the hybrid DSMC using both the 2D axisymmetric and 3D
configuration. The simulation results were validated against
experiments conducted by Rothe~\cite{rothe} and the
deviations are presented in the following figures.
The sensitivity studies concerning different simulation parameters
are described in Section~\ref{sec:internal_accuracy}. Due to the limited
availability of experimental data, the cases i.I and i.III
were studied in more detail than case i.II, where only the
centerline temperature data is available. 

\begin{figure}[H]
	\centering
	\begin{tabular}{cc}
	\includegraphics[width=0.5\textwidth]{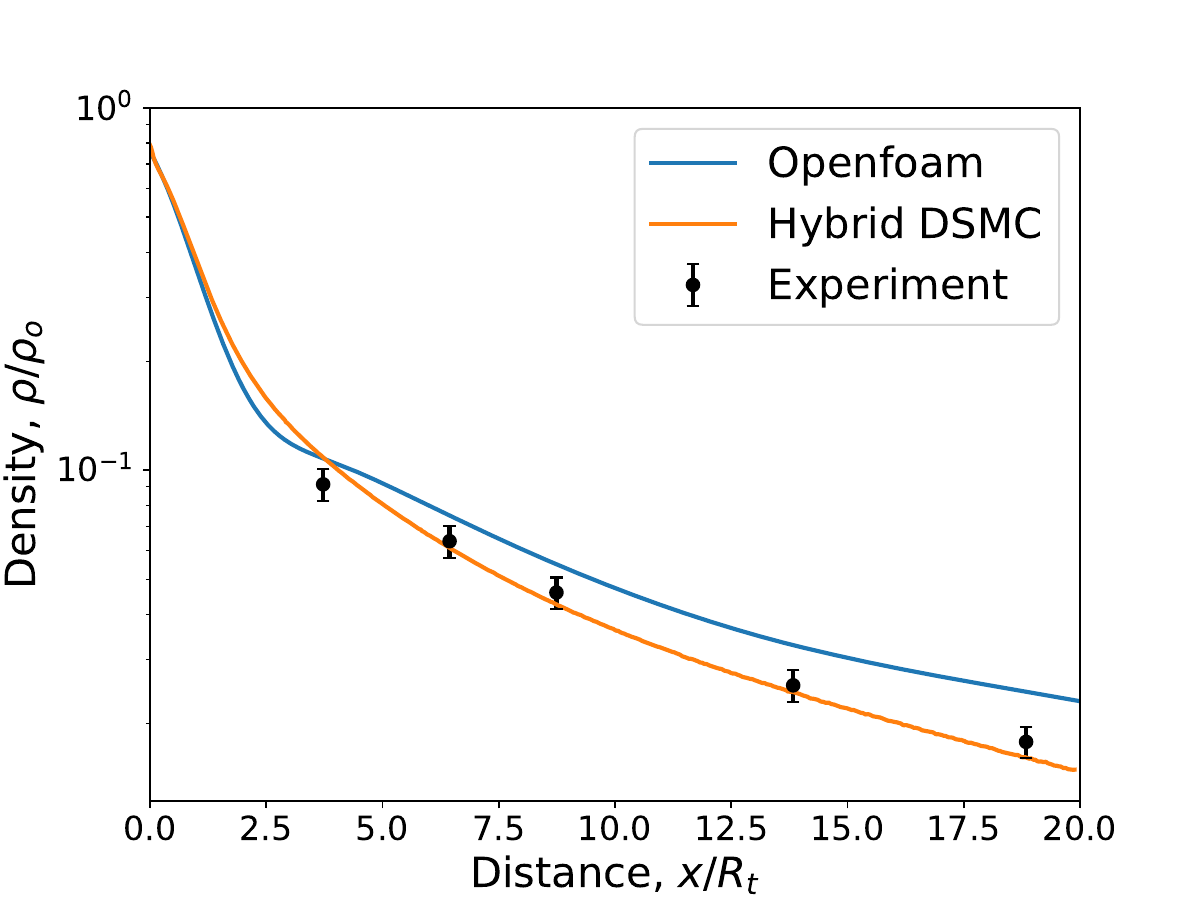}&\includegraphics[width=0.5\textwidth]{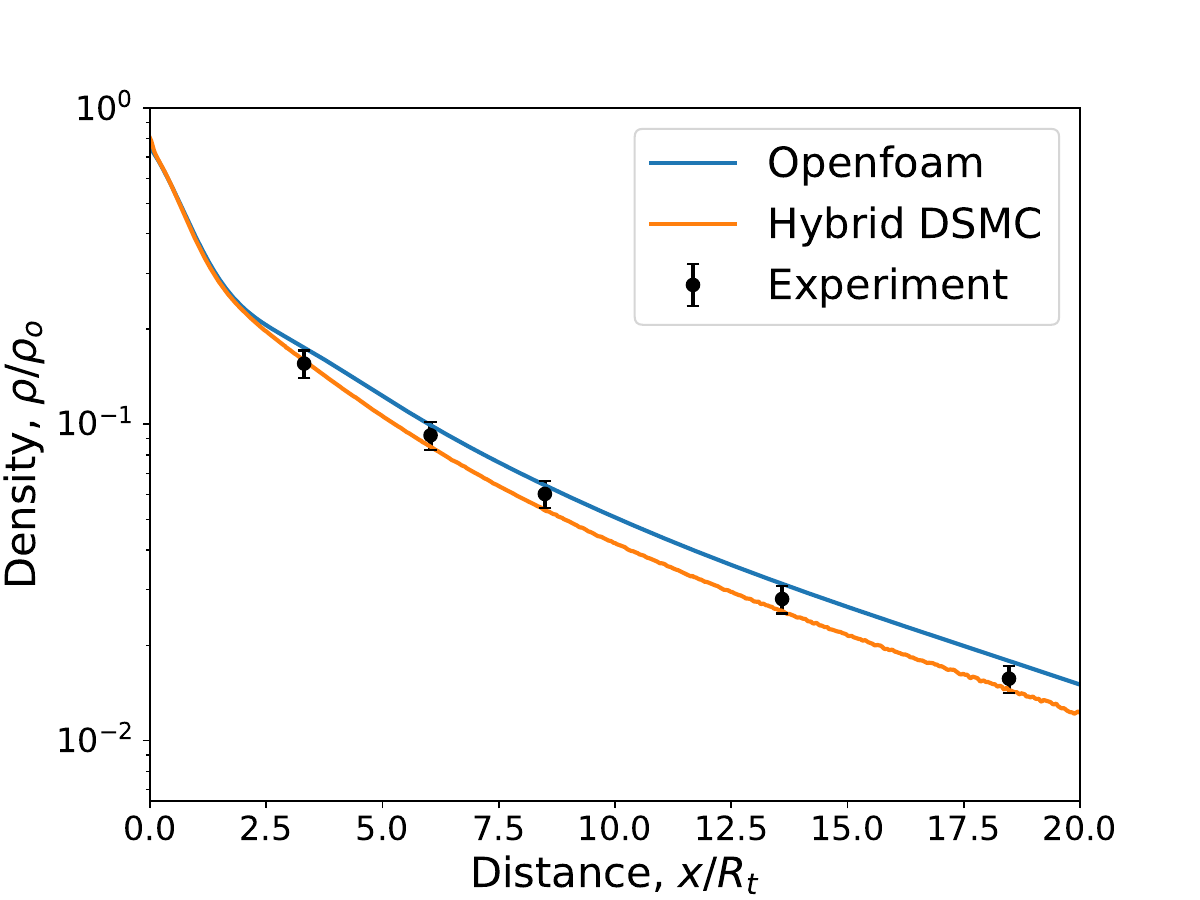}\\
	(a)&(b)\\
	
	\end{tabular}
	\caption{Comparison of density variation along the nozzle
              centerline: (a) Case i.I, (b) Case i.III.}
	\label{fig:474&141validation_clineDensity}
\end{figure}

\begin{figure}[H]
	\centering
	\begin{tabular}{cc}
	\includegraphics[width=0.5\textwidth]{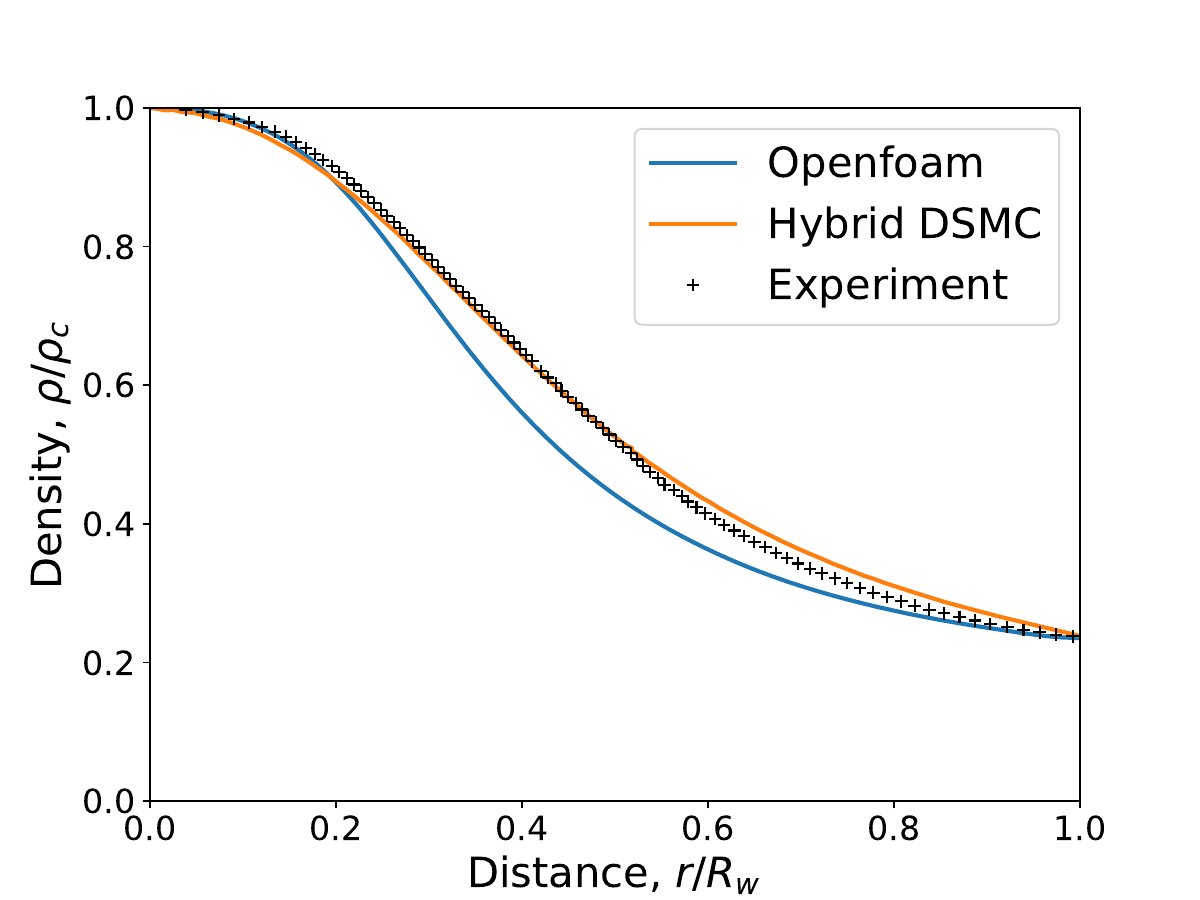}&\includegraphics[width=0.5\textwidth]{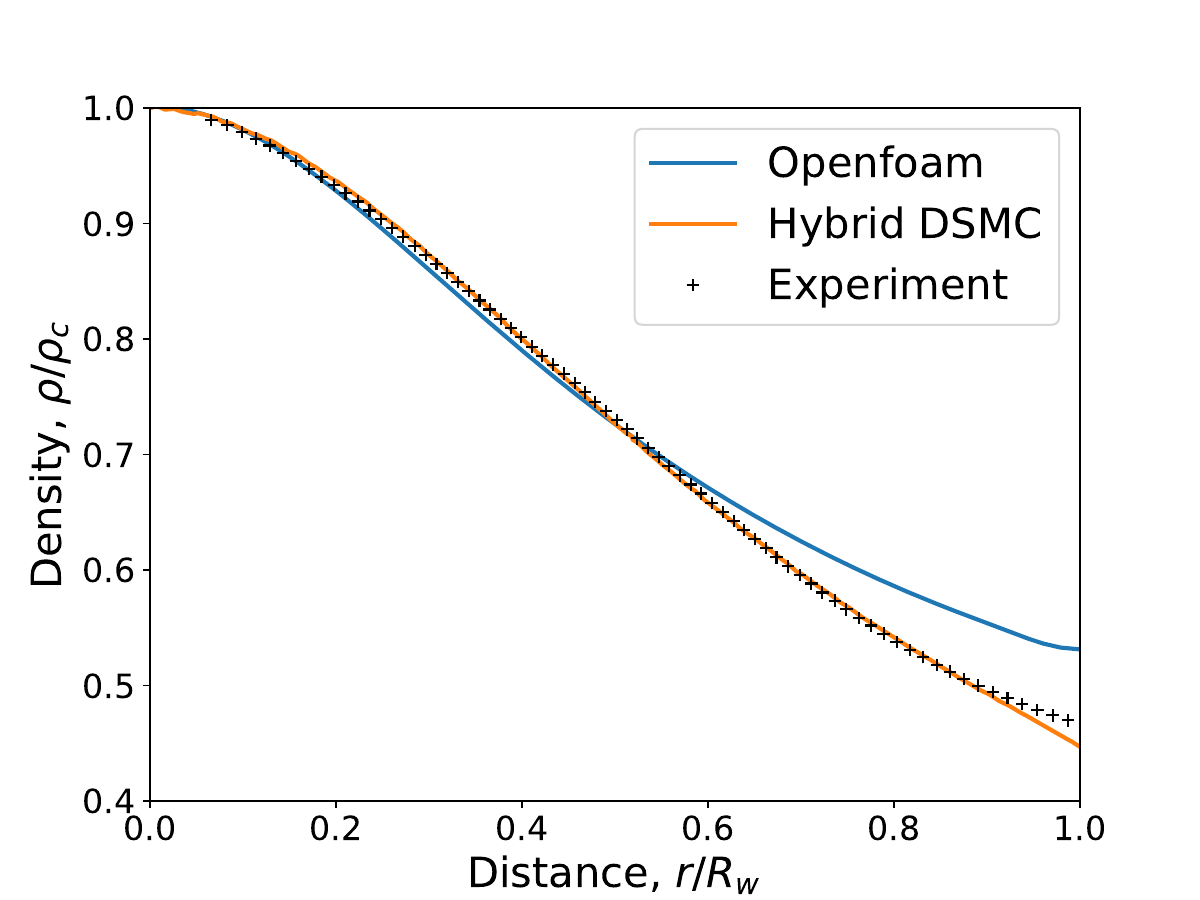}\\
	(a)&(b)\\
	
	\end{tabular}
	\caption{Comparison of density variation along the radial direction at the cross-section $x/R_t=13.7$ from the throat for the cases (a) i.I and (b) i.III.}
	\label{fig:474&141validation_13Density}
\end{figure}

\begin{figure}[H]
	\centering
	\begin{tabular}{cc}
	\includegraphics[width=0.5\textwidth]{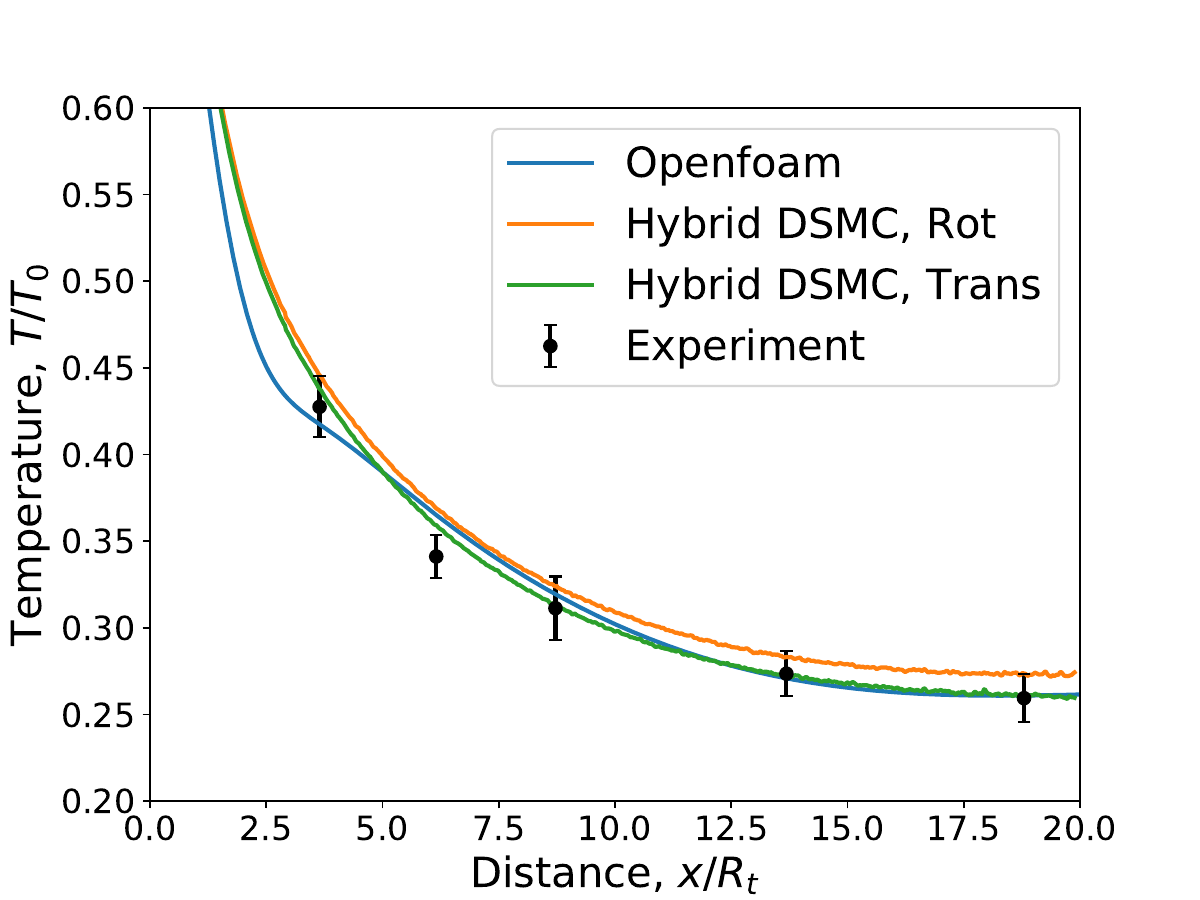}&\includegraphics[width=0.5\textwidth]{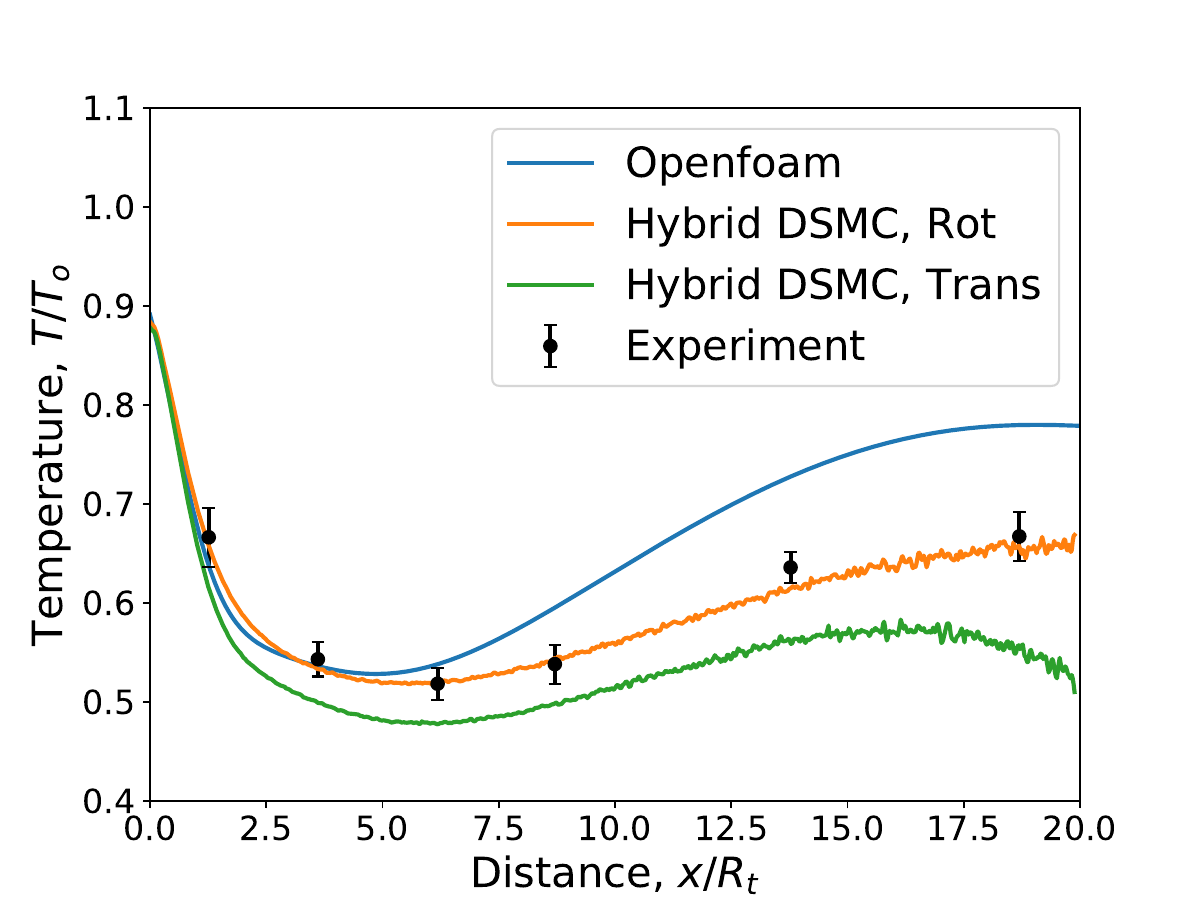}\\
	(a)&(b)\\
	
	\end{tabular}
	\caption{Comparison of temperature variation along the nozzle centerline
             for the cases: (a) i.I and (b) i.III.}
	\label{fig:474pa&141clineTempvalidation}
\end{figure}

\begin{figure}[H] 
	\centering
	\begin{tabular}{cc}
	\includegraphics[width=0.5\textwidth]{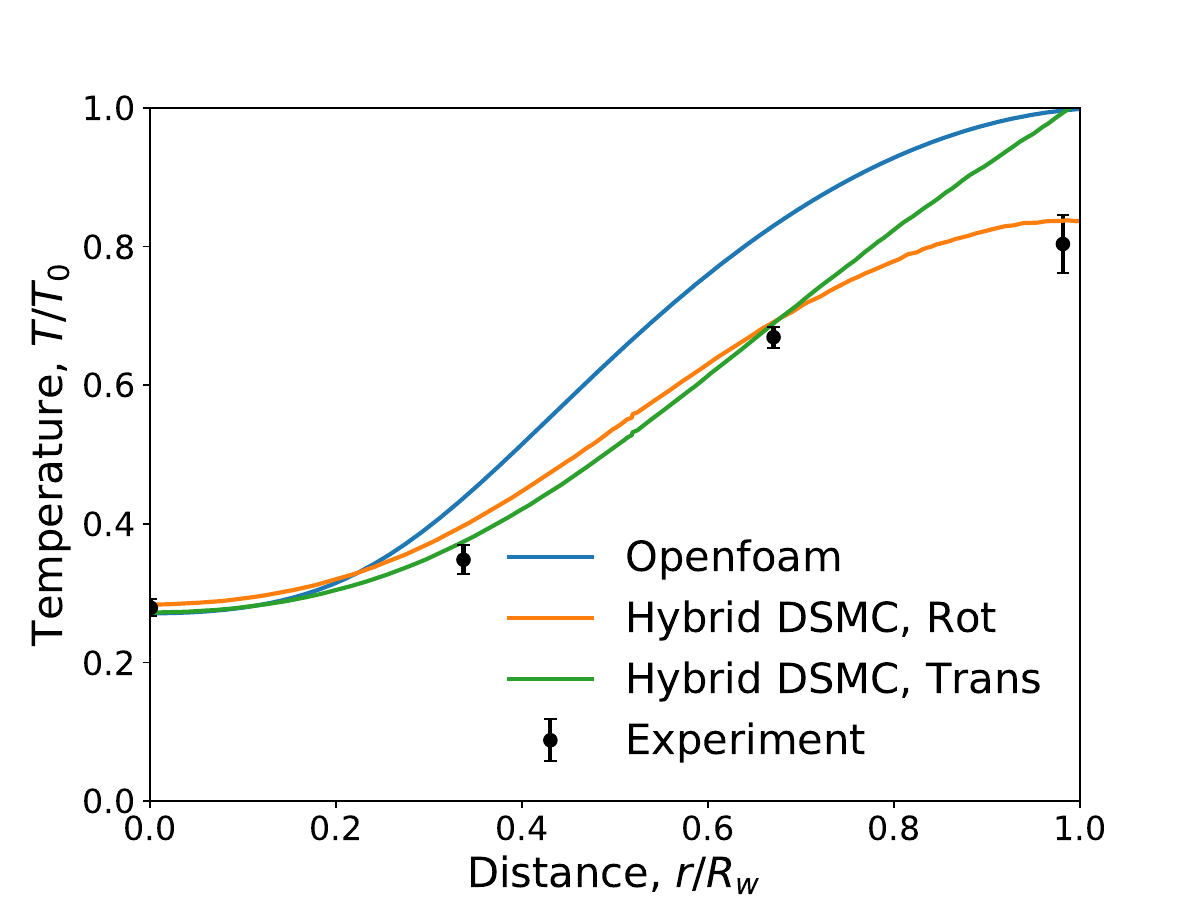}&\includegraphics[width=0.5\textwidth]{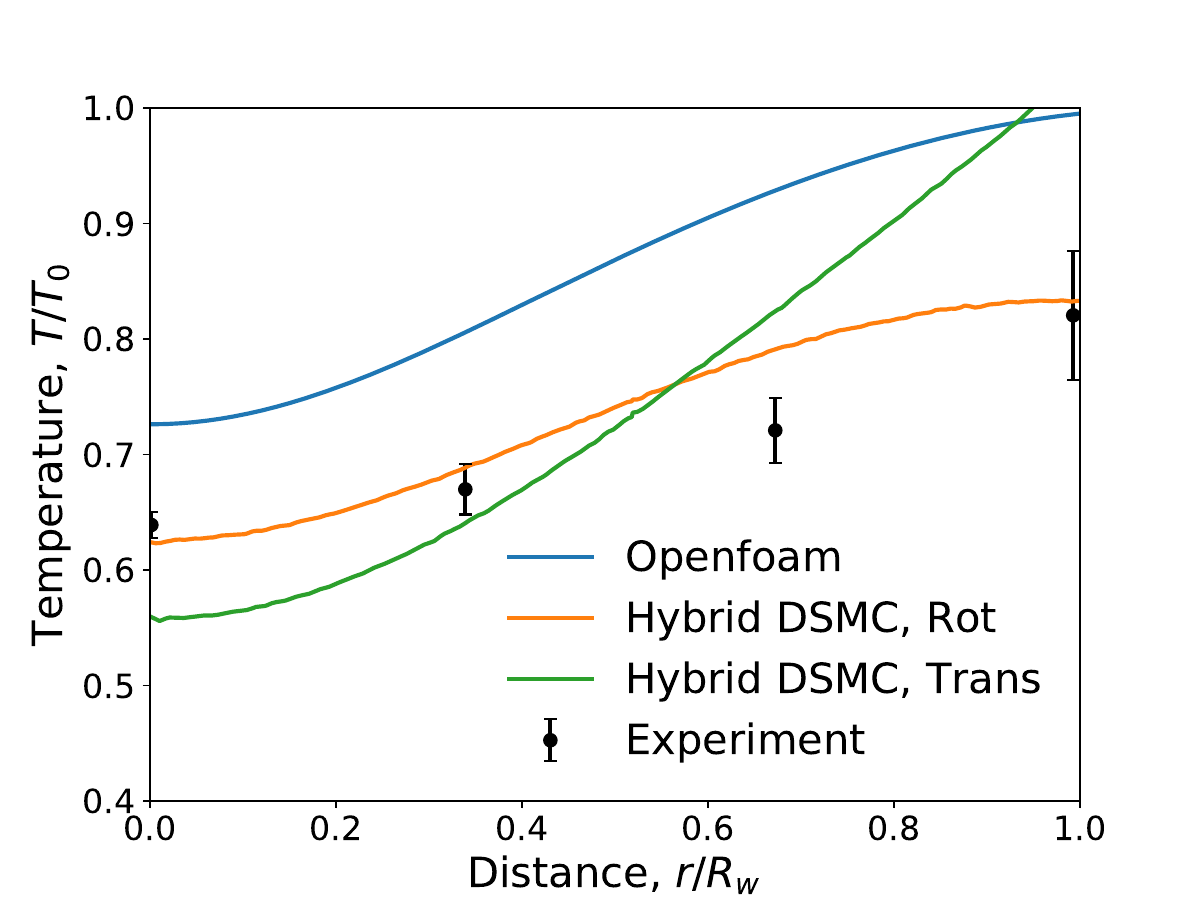}\\
	(a)&(b)\\
	
	\end{tabular}
	\caption{Comparison of temperature profiles in the radial direction at a non-dimensional distance $x/R_t = 13.7$ from the throat for the cases (a) i.I 
     and (b) i.III.}
	\label{fig:474&141radialTempvalidation}
\end{figure}

\autoref{fig:474&141validation_clineDensity}
to \ref{fig:474&141radialTempvalidation} show various flow
parameters obtained using the continuum approach and the 
hybrid DSMC approach. For the purpose of 
validation, the corresponding experimental data~\cite{rothe} are included.
\autoref{fig:474&141validation_clineDensity}~(a) and (b) show
the centerline density profiles in the nozzle for cases i.I and
i.III, respectively. Here, the
densities $\rho$ were normalized by the stagnation density $\rho_o$
and this ratio was plotted against the non-dimensional axial distance, i.e.,
the ratio of the axial position from the throat $x$ and the radius of
the throat $R_t$. The radial variation of densities normalized by
the maximum cross-sectional density $\rho_c$ found at the axis is
studied at the cross-sectional position $x/R_t=13.7$ from
the throat for case~i.I in \autoref{fig:474&141validation_13Density}~(a)
and for case~i.III in \autoref{fig:474&141validation_13Density}~(b).
The radial density profiles are also studied for other cross-sectional
positions depicted in \autoref{fig:radialDensity_bc} in Section~\ref{sec:internal_accuracy}. The density profiles obtained by the hybrid DSMC method show good agreement with the experimental data compared to the continuum method. It should be noted that the error limits in the measured experimental densities reported by Rothe~\cite{rothe} are $\pm 10~\%$ along the centerline and $\pm 5~\%$ for the relative densities along the cross-sections.

\noindent \autoref{fig:474pa&141clineTempvalidation} shows the
centerline temperature profiles of cases i.I and i.III,
respectively. The equilibrium temperature was calculated by the
continuum method and the translational and rotational temperatures
were predicted by the hybrid DSMC method. Similar to the densities,
all temperatures were normalized by the stagnation temperature $T_o$
and plotted against $x/R_t$. For case i.I, where $Re_N > 500$, the
computed temperatures from all three models agree well with the
experimental data, while for $Re_N < 300$ there are significant
differences: Unlike the case i.I ($Re_N > 500$) where the temperatures
decrease monotonically from the throat to the exit of the nozzle
(\autoref{fig:474pa&141clineTempvalidation}~(a)), in the cases i.II
and i.III ($Re_N < 300$) the temperatures reduce to a minimum at $
x/R_t \approx 6$ and increase toward the exit of the nozzle (see
\autoref{fig:474pa&141clineTempvalidation}~(b) and
\autoref{fig:fnum_int_ext}. This is a result of stronger rarefaction
effects where the flow gets thermalized due to viscous dissipation,
i.e., due to more molecule-surface collisions than molecule-molecule
collisions. This effect also has an influence on the Mach
  number profiles shown in~\autoref{fig:mach_nozzle}. However, it can
be observed that the rotational temperatures predicted by DSMC match
best with the measurements as the experimental data \cite{rothe} are
rotational temperatures.

\noindent \autoref{fig:474&141radialTempvalidation} shows the normalized
temperature variation against the radial distance normalized by
the cross-sectional radius at 
$x/R_t=13.7$ from the throat. The rotational temperatures are again 
in good agreement with the experimental data. The rotational
temperature is always greater than the translational temperature
due to flow expansion~\citep{rothe}. The equilibrium temperature obtained by the
continuum method can be compared with the translational temperatures from the
hybrid DSMC method. The differences between these two temperatures, 
particularly for $Re_N < 300$, and the differences in densities observed
above are due to the inaccuracy of the continuum method in the rarefied flow
regimes. 

\subsection{External Flow: DSMC, CFD}
\label{sec:external_1}

External flow simulations were carried out for the test
cases listed in \autoref{tab:externalcases} using pure CFD
and DSMC methods. The flow is assumed to be laminar, which is
in line with the fact that the Reynolds number is relatively
small. As described earlier, the results for the continuum
approach were used to estimate the breakdown Knudsen number $\Knud_B$
(Eq.~(\ref{eq:Kn_break})). 
 By determining $\Knud_B$ throughout the simulation domain, it is
 found that near the surface of the external bodies, $\Knud_B$ is
 relatively small ($\Knud_B < 0.004$). This is due to a shock wave
 formed around the bodies which eventually increases the pressure
 and the density inside the shock layer. 
The pressure coefficient $C_p$ is used as a quantity for validating
the external flow cases. In \autoref{fig:domainext} the dotted red
lines indicate the measurement line, along which the results are evaluated.

\begin{figure}[H]
	\centering
	\begin{tabular}{cc}
		\includegraphics[width=0.5\textwidth]{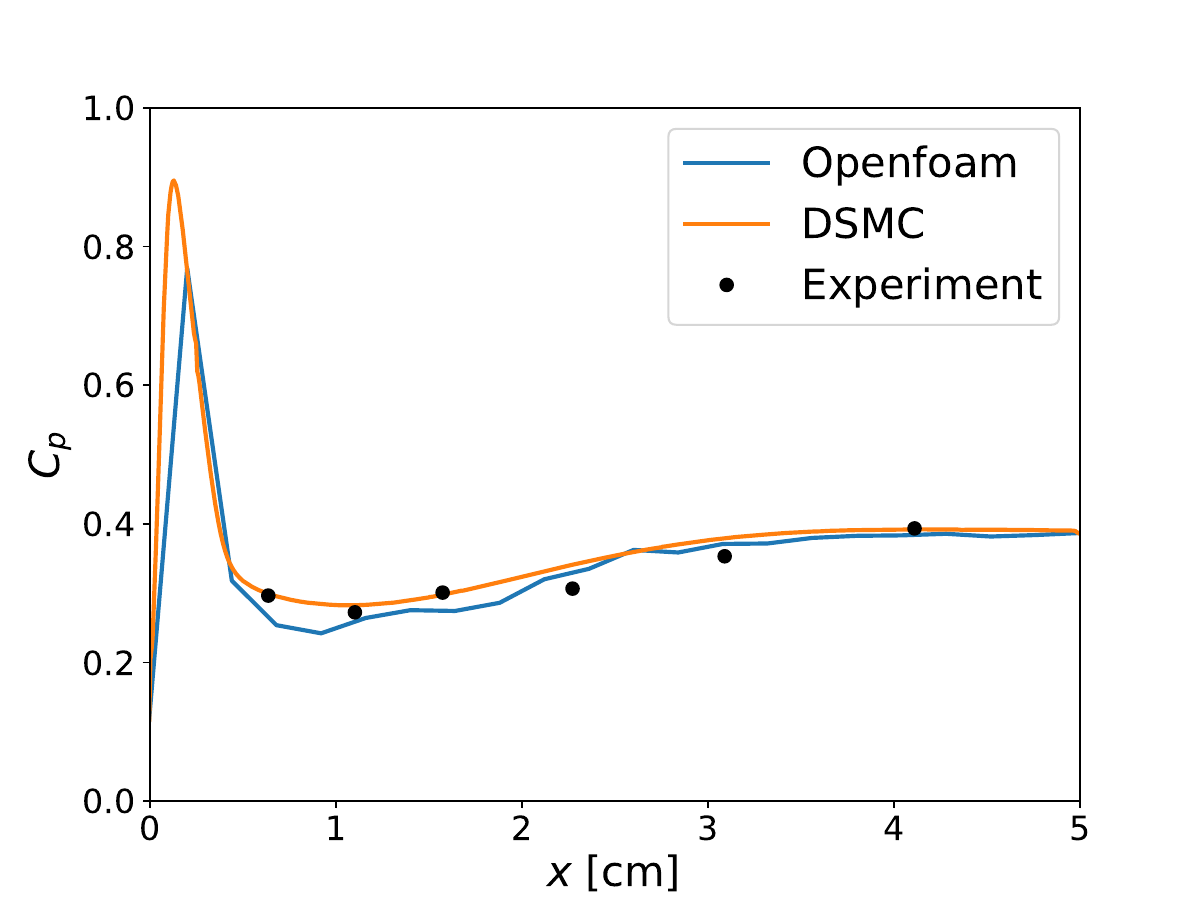}&\includegraphics[width=0.5\textwidth]{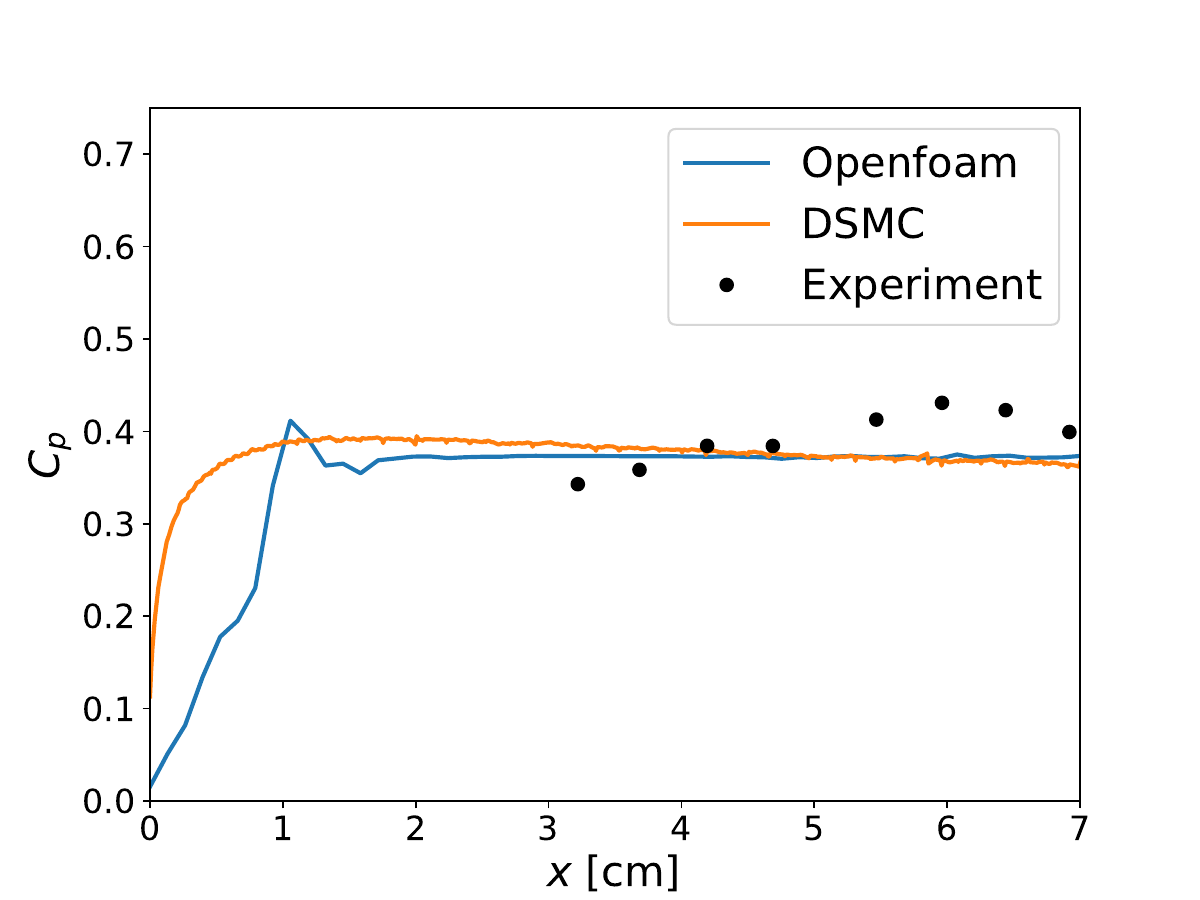}\\
		(a)&(b)\\
	\end{tabular}
	\caption{Comparison of the pressure coefficient $ C_p $ over the surface of
            the body for the cases (a) e.I and (b) e.II.}
	\label{fig:ext01}
\end{figure}

\autoref{fig:ext01} shows the variation of $C_p$ on the surface plotted
against the axial distance $x$ for cases e.I and e.II, respectively.
Due to the near-continuum flow in the vicinity of the surface, both continuum
and DSMC methods show similar trends in the results and reasonably 
match the experimental data~\cite{wang, wang_old,moss2}. However, the DSMC results show
a smoother variation in $C_p$. Furthermore, there is a significant
difference in the results of the methods near the stagnation
region ($x \approx 0$~cm to $0.5$~cm). This is because the DSMC
method predicts a much thicker shock layer around the body compared
to the continuum method particularly near the stagnation
region. This phenomenon is subject to further investigations.  
Sensitivity studies of different simulation parameters are described in Section~\ref{sec:internal_accuracy}.

\subsection{Impact of DSMC simulation parameters}\label{sec:internal_accuracy}

The results presented in the last two subsections show the level of
accuracy of the DSMC method in resolving rarefied gas flows.
However, as mentioned in Section~\ref{sec:dsmc}, several simulation
parameters are responsible for attaining these accurate results.
In this section, the sensitivity of the results with respect to
different DSMC simulation parameters is presented. 
Although the simulations for the internal flow cases were
carried out using the hybrid method, only the DSMC domain, 
i.e., the region where $\Knud_B > 0.05$, is considered for this
parameter study.

\subsubsection{Effect of simulation particles and DSMC grid size}

In order to accurately describe the rarefied gas flow,
it is important to have a sufficient number of simulation
particles in the simulation domain and per grid cell.
These represent the distribution of the actual gas
molecules and are necessary to preserve the statistical
accuracy and resolution of molecular collisions in the
simulation. In SPARTA this property is controlled by the keyword \verb|fnum|.
The parameter \verb|fnum| sets the ratio of real molecules to the
simulation particles. Therefore, the smaller the value of \verb|fnum|, the
greater the number of simulation particles and the simulation accuracy.
Once the number of simulation particles crosses a certain threshold,
the accuracy of the simulation reaches convergence in results.
Using much smaller values of \verb|fnum| compared to this threshold
value increases the computational cost as in DSMC the computational
cost scales linearly  with the number of simulation particles \cite{spartaBench}.
Therefore, it is very important to choose a trade-off value of \verb|fnum|
in order to optimize the computational costs while ensuring the
accuracy of the simulation.

The simulation domain was discretized with regular grids and the grid cell
size, e.g., $\Delta x$, of the simulation is chosen according to the
criterion mentioned in Section~\ref{sec:dsmc}, i.e.,
$\Delta x \leq \frac{1}{3} \lambda_{min}$ with the minimum
value of the mean free path in the  simulation domain $\lambda_{min}$.

\begin{figure}[H] 
	\centering
	\begin{tabular}{cc}
	\includegraphics[width=0.5\textwidth]{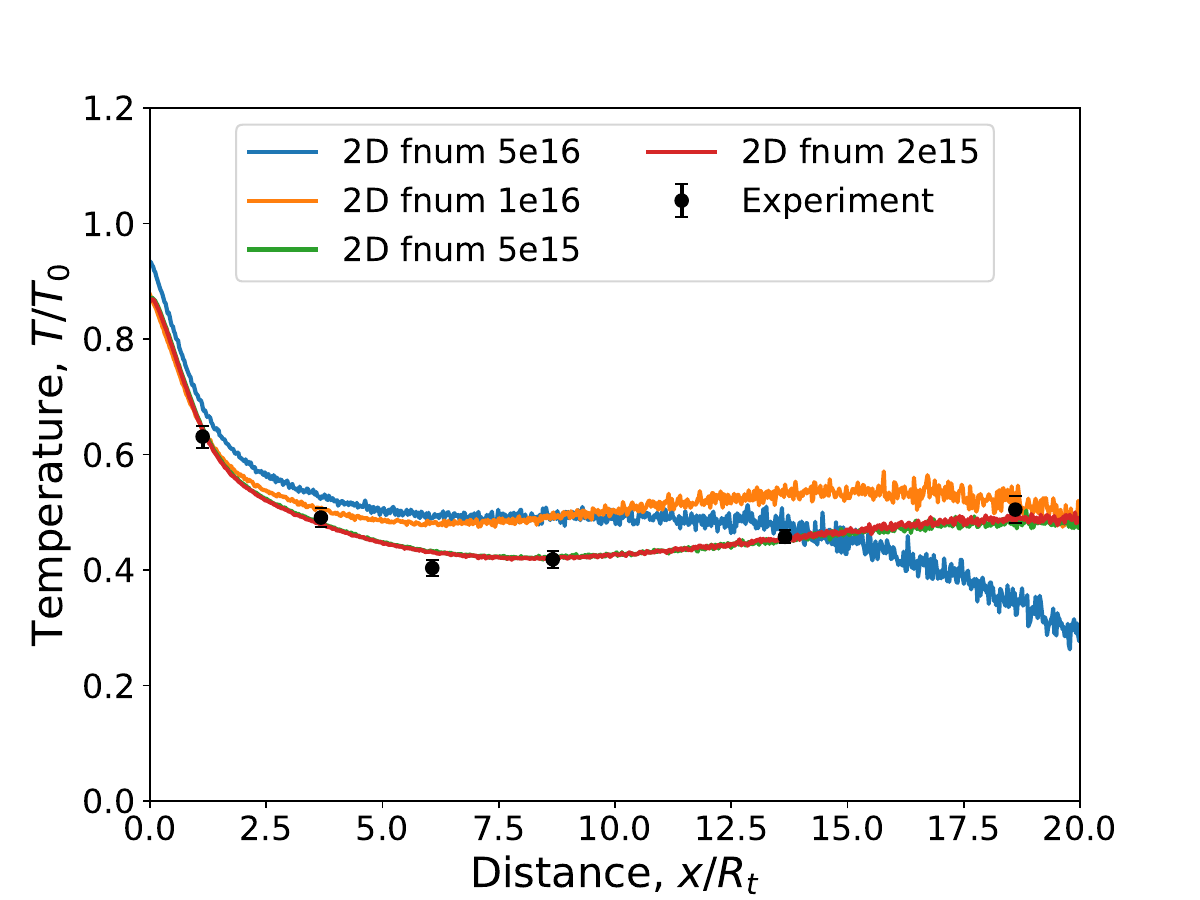}&
     \includegraphics[width=0.5\textwidth]{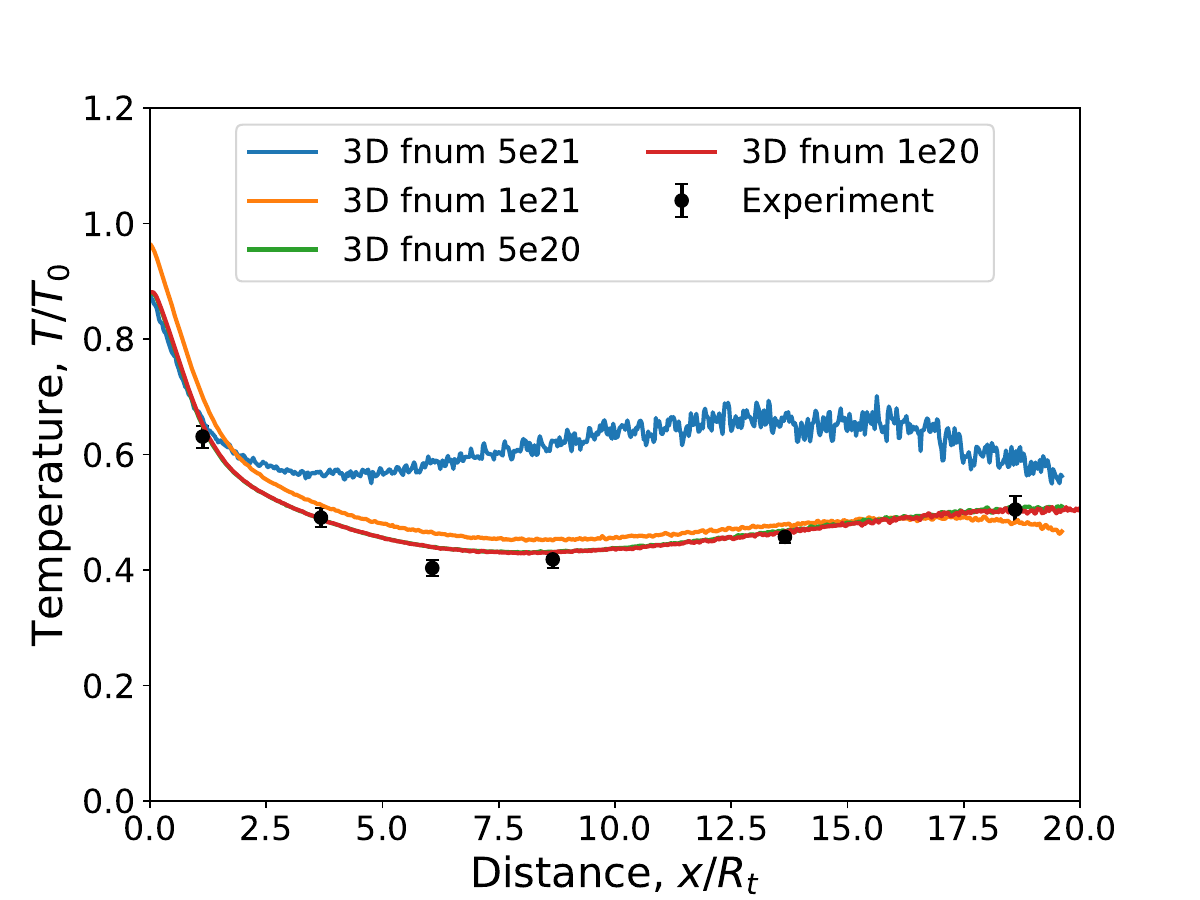}\\
     (a)&(b)\\
	\end{tabular}
    \caption{\textcolor{dgreen}{Effect of \texttt{fnum} on centerline rotational temperatures for case i.II. (a) 2D, (b) 3D.}}
	\label{fig:fnum_int_ext}
\end{figure}

\begin{figure}[H] 
	\centering
	\begin{tabular}{cc}
	\includegraphics[width=0.5\textwidth]{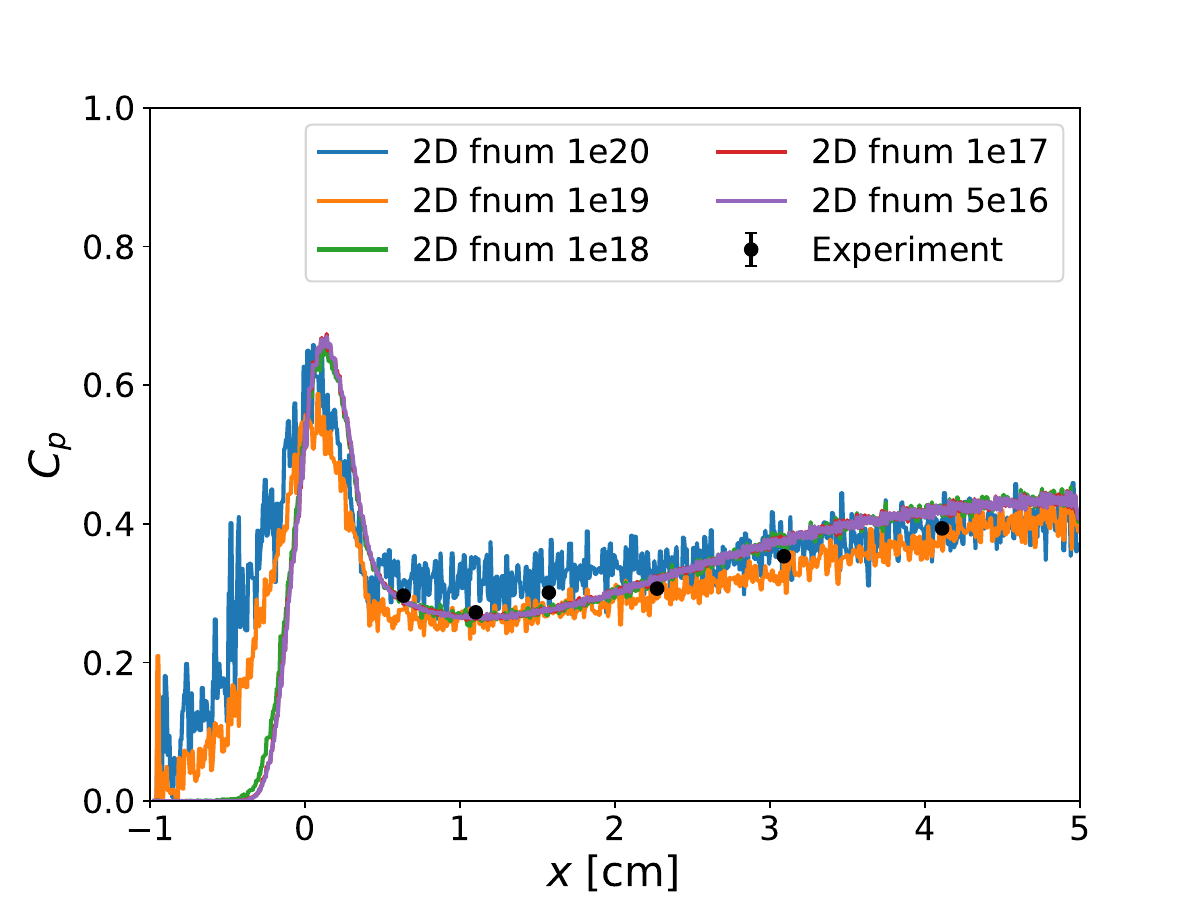}&
     \includegraphics[width=0.5\textwidth]{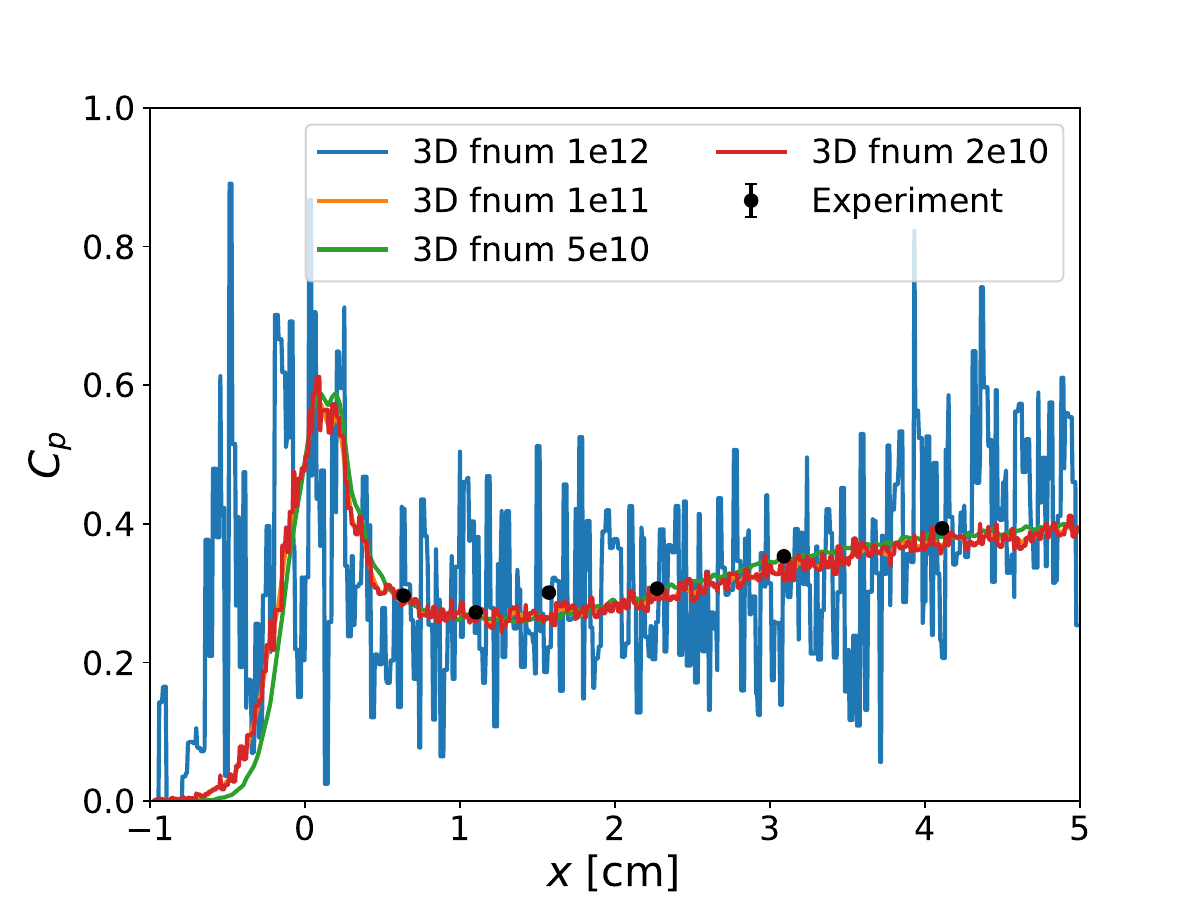}\\
     (a)&(b)\\
	\end{tabular}
\caption{\textcolor{dgreen}{Effect of \texttt{fnum} on the pressure coefficient $C_p$ for case e.I. (a) 2D, (b) 3D.}}
	\label{fig:fnum_int_ext2}
\end{figure}

The present convergence study varying \verb|fnum| has been performed
for a uniform grid of size $\Delta x$. The simulation particles were
created using the \verb|fnum| parameter and were distributed such that
each cell has roughly the same number of particles.
\autoref{fig:fnum_int_ext}~(a) shows the effect of the
parameter \verb|fnum| on the predicted rotational temperature for the
2D internal flow case i.II.
A value of \verb|fnum| $\leq 5 \times 10^{15}$
is required to reach convergence and agreement with the 
experimental data for the 2D axisymmetric
configuration. Likewise, \autoref{fig:fnum_int_ext2}~(a) shows
the effect of \verb|fnum| on the predicted 
pressure coefficient $C_p$ for the 2D external flow case e.I.
Here a value of \verb|fnum| $\leq 1 \times 10^{17}$ is required
for convergence and a reasonable agreement with the 
experimental data. Using smaller values of \verb|fnum|
increases the number of simulation particles and thereby the
computational cost. Hence, the above mentioned values of
\verb|fnum| are chosen as a trade-off for the simulation
of the flow cases.

\begin{figure}[H] 
\centering
	\begin{tabular}{cc}
	\includegraphics[width=0.5\textwidth]{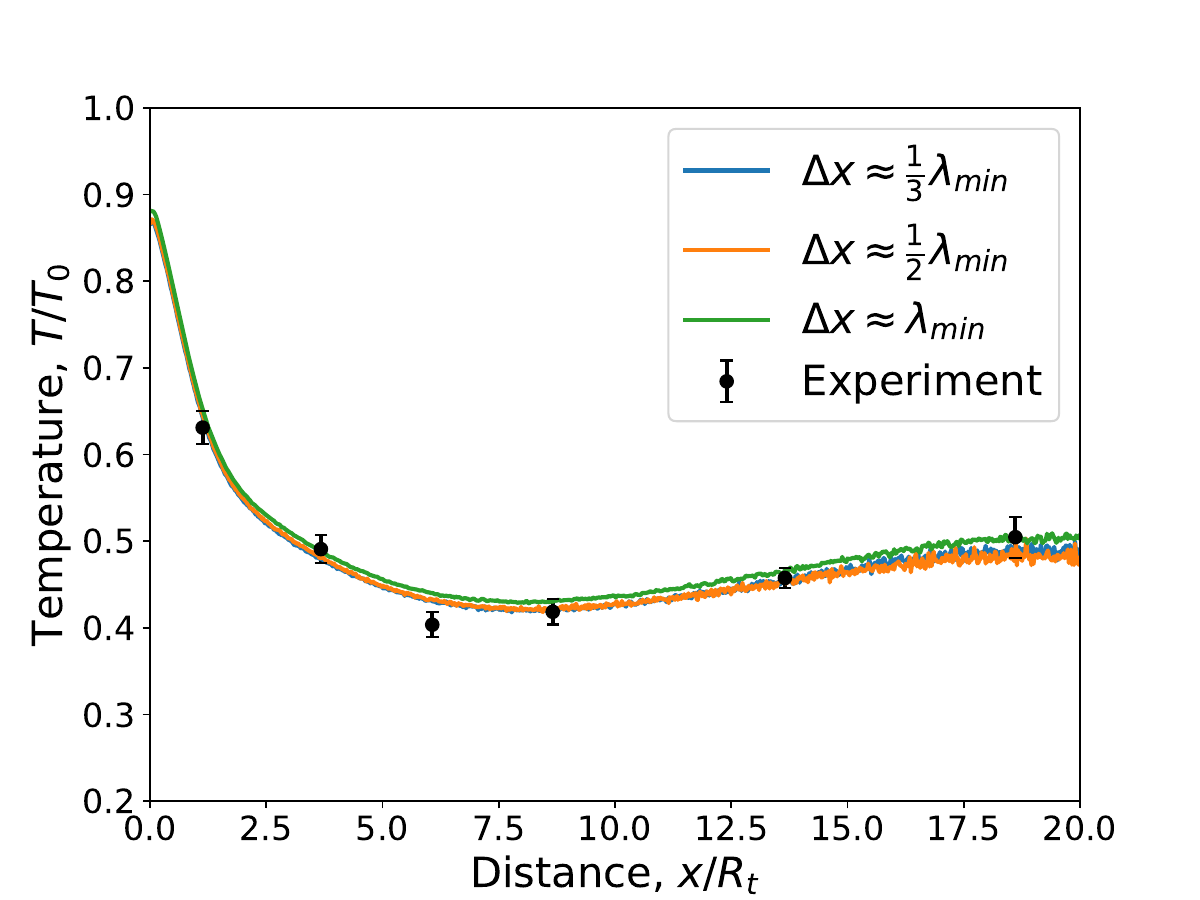}&
     \includegraphics[width=0.5\textwidth]{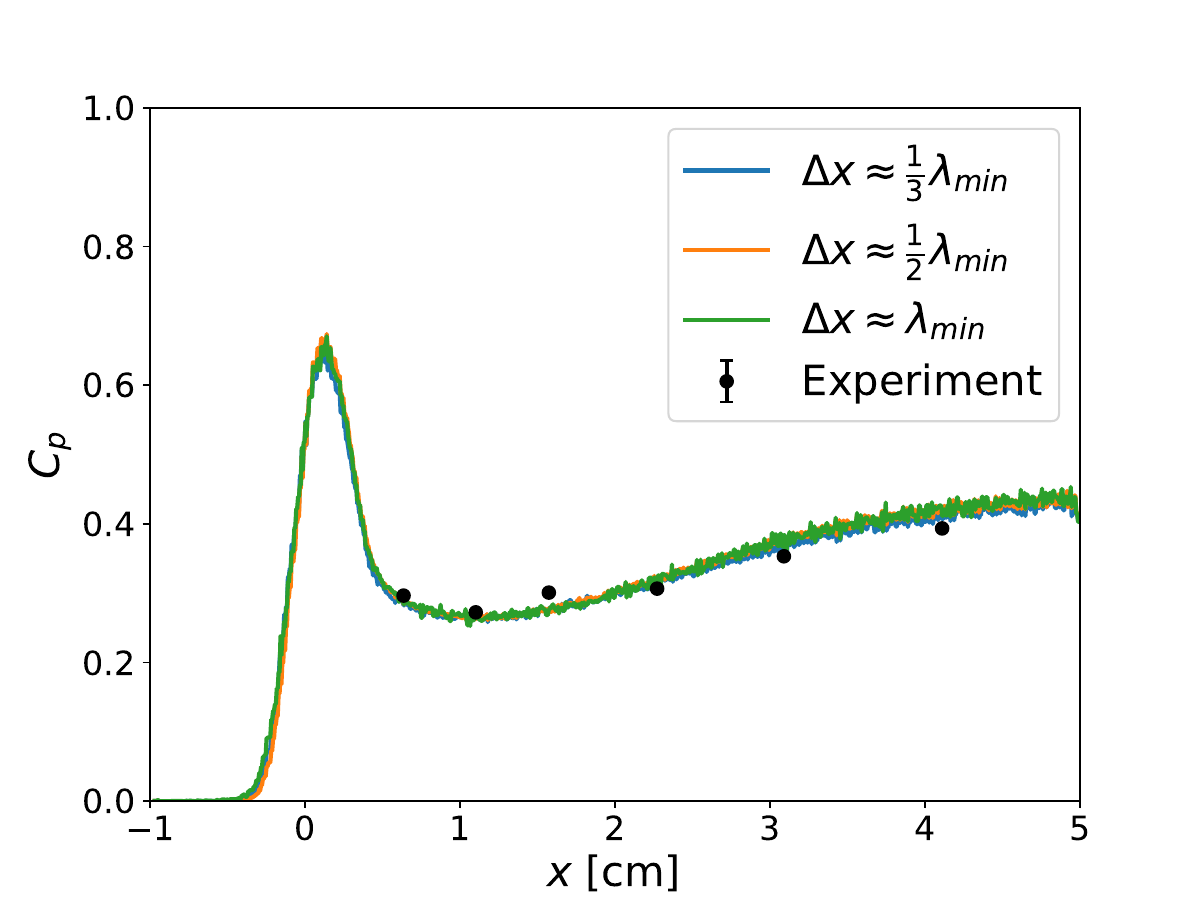}\\
	(a)&(b)\\
	\end{tabular}
\caption{Effect of grid resolution on (a) centerline rotational temperatures for case i.II and 
(b) pressure coefficient $C_p$ for case e.I.}
	\label{fig:grid_int_ext}
\end{figure}

A convergence study has also been performed with respect to the
grid resolution of the simulation. The uniform regular grid is
coarsened until the maximum grid size $\Delta x \leq \lambda_{min}$ is reached 
keeping the \verb|fnum| value constant (trade-off \verb|fnum|). 
\autoref{fig:grid_int_ext} shows the 
results achieved by the different grid sizes for
case~i.II and e.I, respectively. In the limit $\Delta x \leq \lambda_{min}$
there are no significant deviations between the calculations.
Based on these cases the optimal value of the grid size is 
$\Delta x \approx \lambda_{min}$ with \verb|fnum| $\leq 5 \times 10^{15}$
for the internal and \verb|fnum| $\leq 1 \times 10^{17}$ for
the external 2D axisymmetric configurations. 
Compared to the previous configuration $\Delta x \approx \frac{1}{3}\lambda_{min}$
these parameters increased the simulation efficiency roughly by a
factor of $3.5$ for the 
internal case~i.II and by a factor of $4.25$ for the external case~e.I.
The 3D simulations are also carried out with $\Delta x \approx \lambda_{min}$ and the corresponding trade-off value of \verb|fnum| is $\leq 5 \times 10^{20}$ for the
case~i.II and $\leq 1 \times 10^{11}$ for the case~e.I, 
see \autoref{fig:fnum_int_ext}~(b)~and~\autoref{fig:fnum_int_ext2}~(b).

The number of simulation particles and the computational effort
could be further reduced without compromising the computational
accuracy by using an adaptive mesh refinement technique. However,
this technique is not considered in the current paper.

\subsubsection{Effect of time step}

The choice of the time step is another parameter
which can significantly affect the solution of the DSMC method.
In a DSMC simulation the time step $\Delta t$ must be chosen
in relation to the mean collision time $t_\text{mct}$~\cite{Shen2005}.
The time step is estimated according to the following equation:
\begin{equation}\label{eq:mct}
    \Delta t  \; \leq \; t_\text{mct} \; = \; \frac{\lambda}{\Bar{v}} \; . 
\end{equation}

\begin{figure}[H] 
\centering
	\begin{tabular}{cc}
	\includegraphics[width=0.5\textwidth]{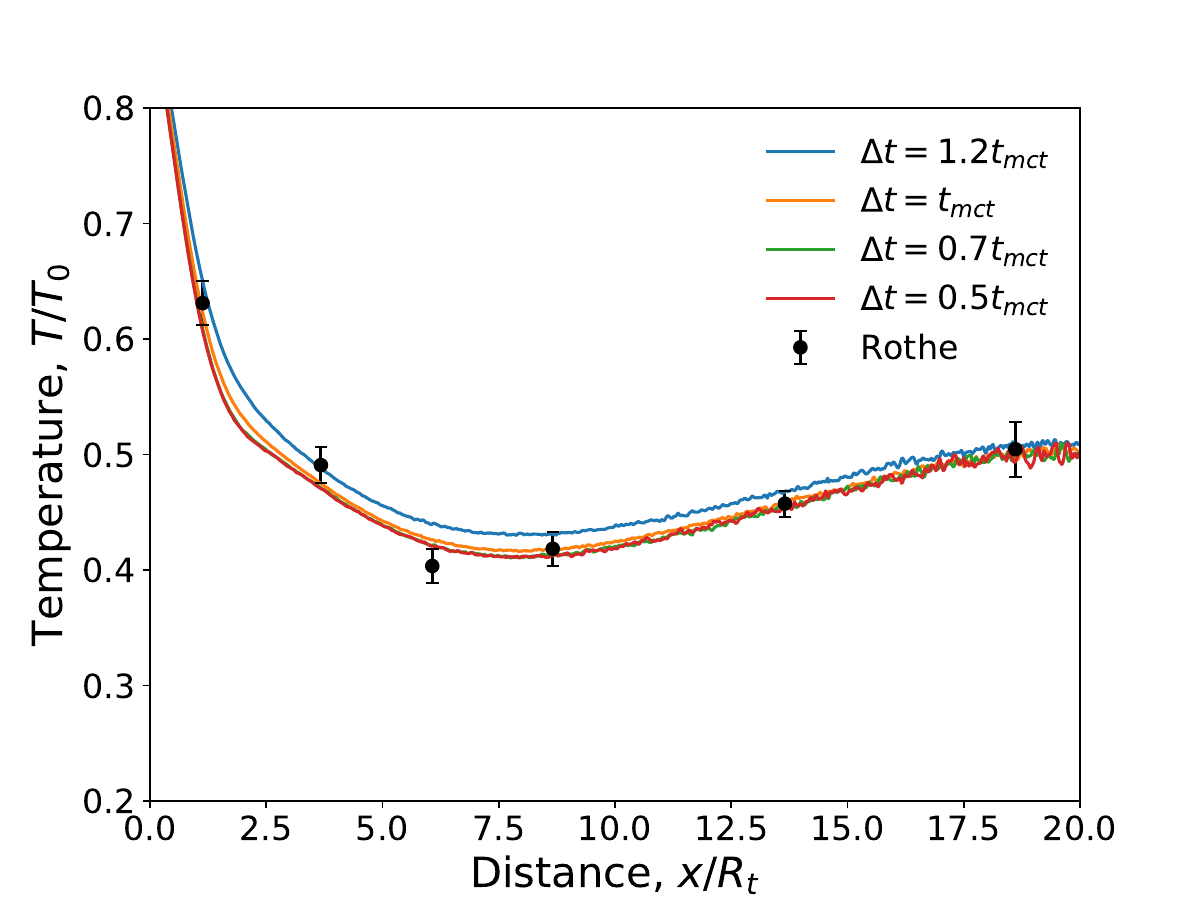}&
     \includegraphics[width=0.5\textwidth]{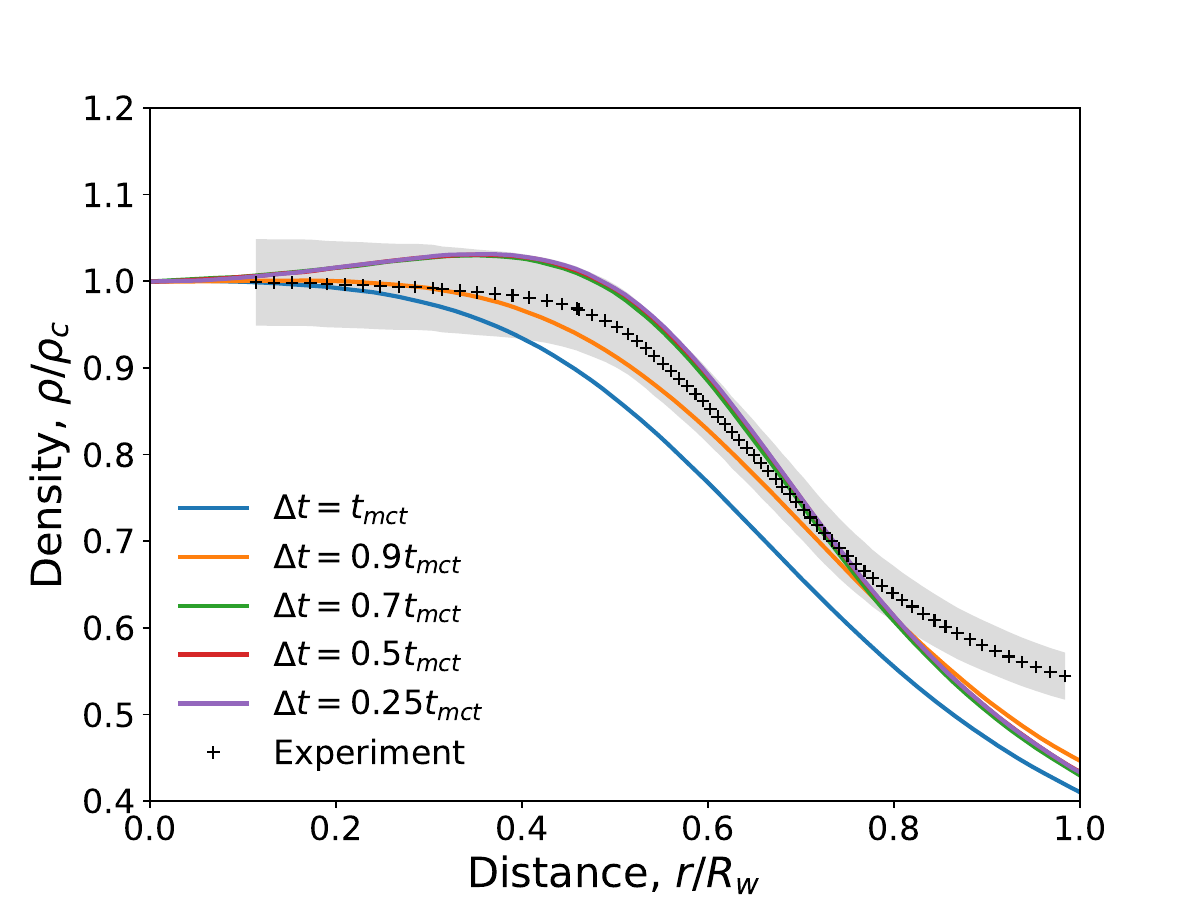}\\
	(a)&(b)\\
	\end{tabular}
    \includegraphics[width=0.5\textwidth]{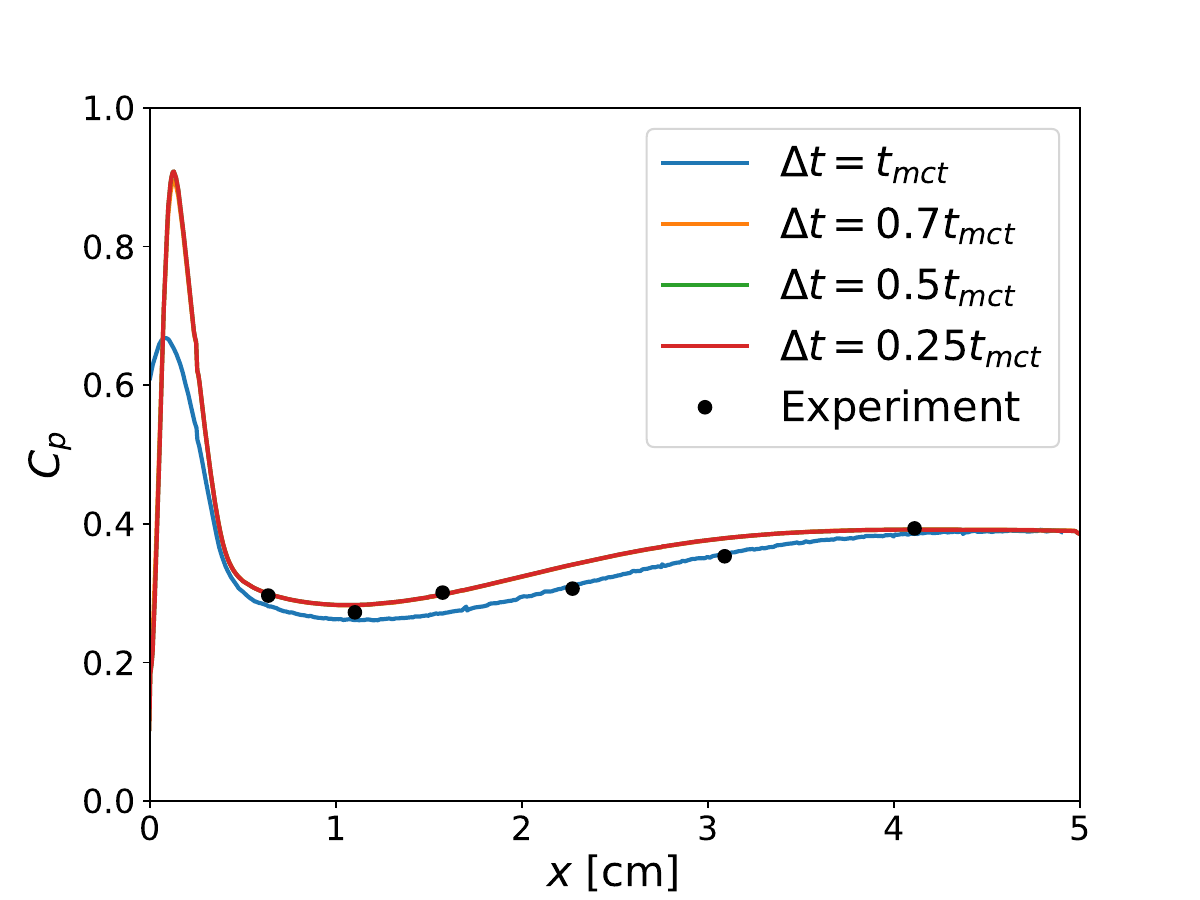}\\
    (c)\\
\caption{\textcolor{dgreen}{Effect of time step on (a) centerline rotational temperatures for case i.II, (b) radial density variation at cross-section $x/R_t = 3.7$ for case i.I and (c) pressure coefficient $C_p$ for case e.I.}}
	\label{fig:tstep_int_ext}
\end{figure}

\noindent Here, $\Bar{v}$ represents the average thermal speed
of the molecules which is determined from the kinetic theory
of gases, $\Bar{v} = \sqrt{{(8 \, K_B \, T)}/{(\pi \, m)}}$.
In DSMC, a smaller time step requires a larger
number of time steps needed to achieve a steady-state solution,
corresponding to increased CPU time.
Using much larger values of the time step may also
increase the CPU time~\cite{Eddi2002} since in DSMC the
probability of collisions between two particles increases with the size of the time step. Therefore, a trade-off
value of the time step $\Delta t$ has to be estimated
similar to the previous subsection. The convergence study
is performed by varying the time step $\Delta t$
using the trade-off values of \verb|fnum| 
and $\Delta x$.
For the calculation of the mean collision time
$t_\text{mct}$ in Eq.~(\ref{eq:mct}), the mean-free-path
value $\lambda = \lambda_{min}$ is chosen. The corresponding
values of $\lambda_{min}$ and $t_\text{mct}$ estimated for
different test cases are tabulated in~\autoref{tab:dsmc_inputs}
~in~\ref{app:dsmc_inputs}. \autoref{fig:tstep_int_ext}~(a)
shows the effect of the time step on the centerline
rotational temperature of case i.II. For this case, it can
be seen that in the limit of $\Delta x \leq t_\text{mct}$
there is no significant change observed in the results
which supports the assumption of~Eq.~(\ref{eq:mct}).
\autoref{fig:tstep_int_ext}~(b) depicts for case~i.I
the radial variation of the density near the throat region,
where the density is higher since it is near to the continuum
region and also due to presence of compression waves
near the throat. Here, it is obvious that a time-step
value of $\Delta t \leq 0.7~t_\text{mct}$ is required
to attain converged results. \autoref{fig:tstep_int_ext}~(c)
shows a similar trend in the calculation of $C_p$ for
case~e.I particularly near the vertex region,
where the density increases drastically due to the 
presence of a shock wave. Therefore, an optimal time-step
value of $\Delta t = 0.7~t_\text{mct}$ was chosen.
This value increases the simulation efficiency
by a factor of 1.1 for case~i.I and by a factor of 1.2
for case~e.I compared to the recommended value, i.e., 
$\Delta t = 0.25~t_\text{mct} $.

\color{black}

\subsubsection{Gas-surface interaction}
\label{sec:gas-surface interaction}

\begin{figure}[H] 
	\centering
	\begin{tabular}{cc}
	\includegraphics[width=0.5\textwidth]{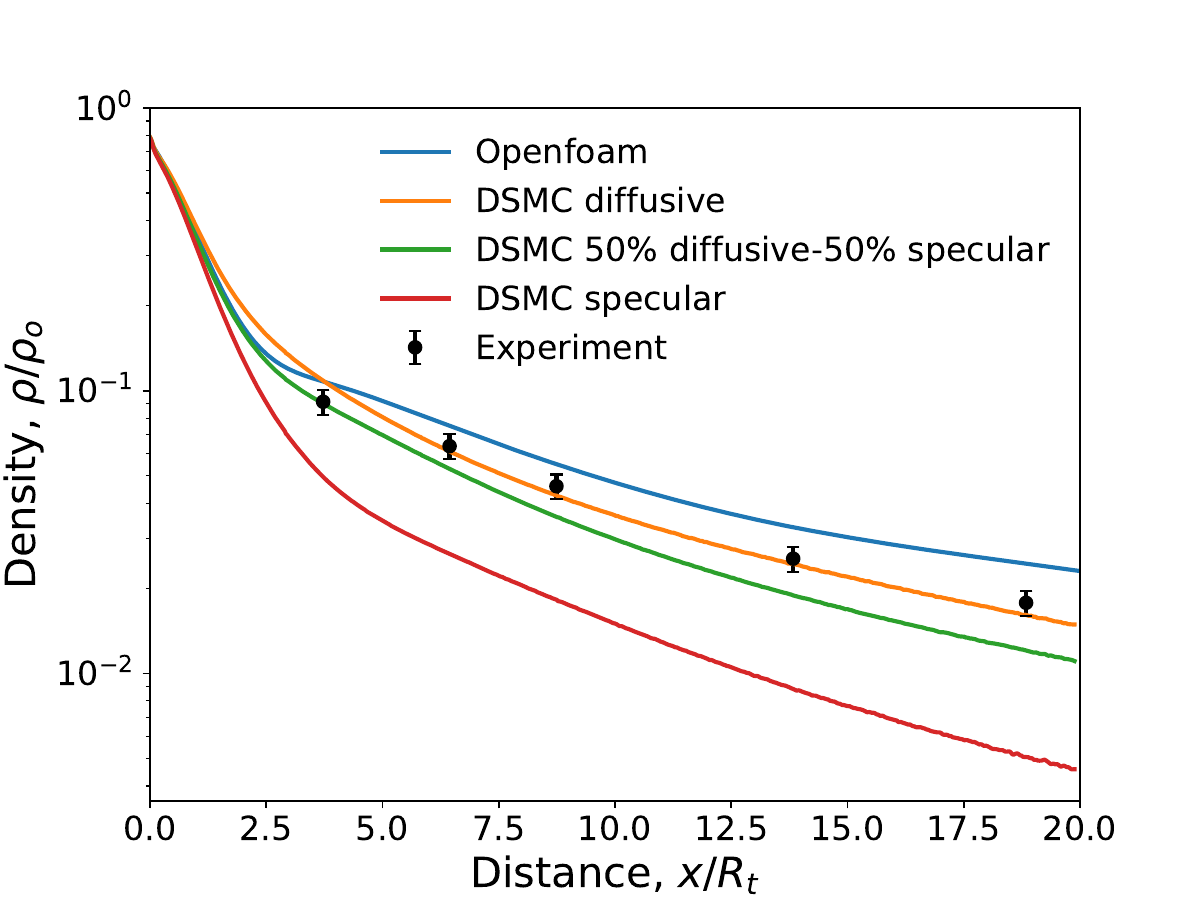}&\includegraphics[width=0.5\textwidth]{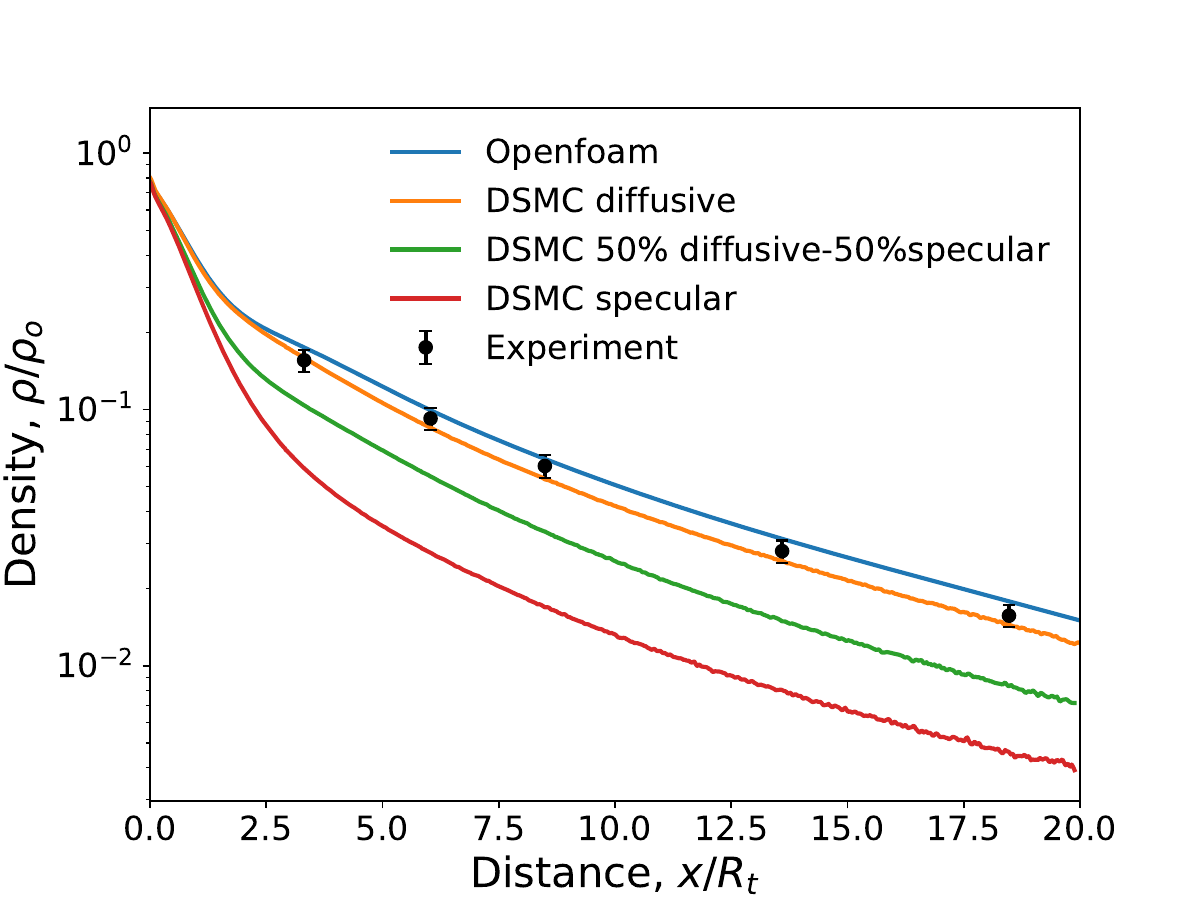}\\
	(a)&(b)\\
	
	\end{tabular}
	\caption{Effect of gas-surface interactions on densities along the nozzle 
             axis for the cases (a) i.I and (b) i.III.}
	\label{fig:clineDensity_BC}
\end{figure}

 For the internal flow cases, the simulations are performed using various gas-surface interaction models described in Section~\ref{sec:dsmc_bc}. The surface collisions are treated as either fully specular, fully diffusive or a combination of both. \autoref{fig:clineDensity_BC}~(a) and (b) show the effect of these surface collision models on the centerline densities for cases~i.I and i.III, respectively. In both cases the simulation with a completely diffusive gas-surface interaction matches well with the experimental results. The simulations performed with a specular gas-surface interaction model yield a faster expansion of the flows compared with the experiment and the simulation relying on the diffuse surface interaction. For simulations with the surface interactions modeled by the combination of $50~\%$ diffusive and $50~\%$ specular collisions, the centerline density curve lies in between the completely diffusive and completely specular simulations. As the proportion of specular collisions increases, the curve shifts toward the completely specular simulation curve and vice-versa.

\begin{figure}[H]
	\centering
	\begin{tabular}{cc}
	\includegraphics[width=0.5\textwidth]{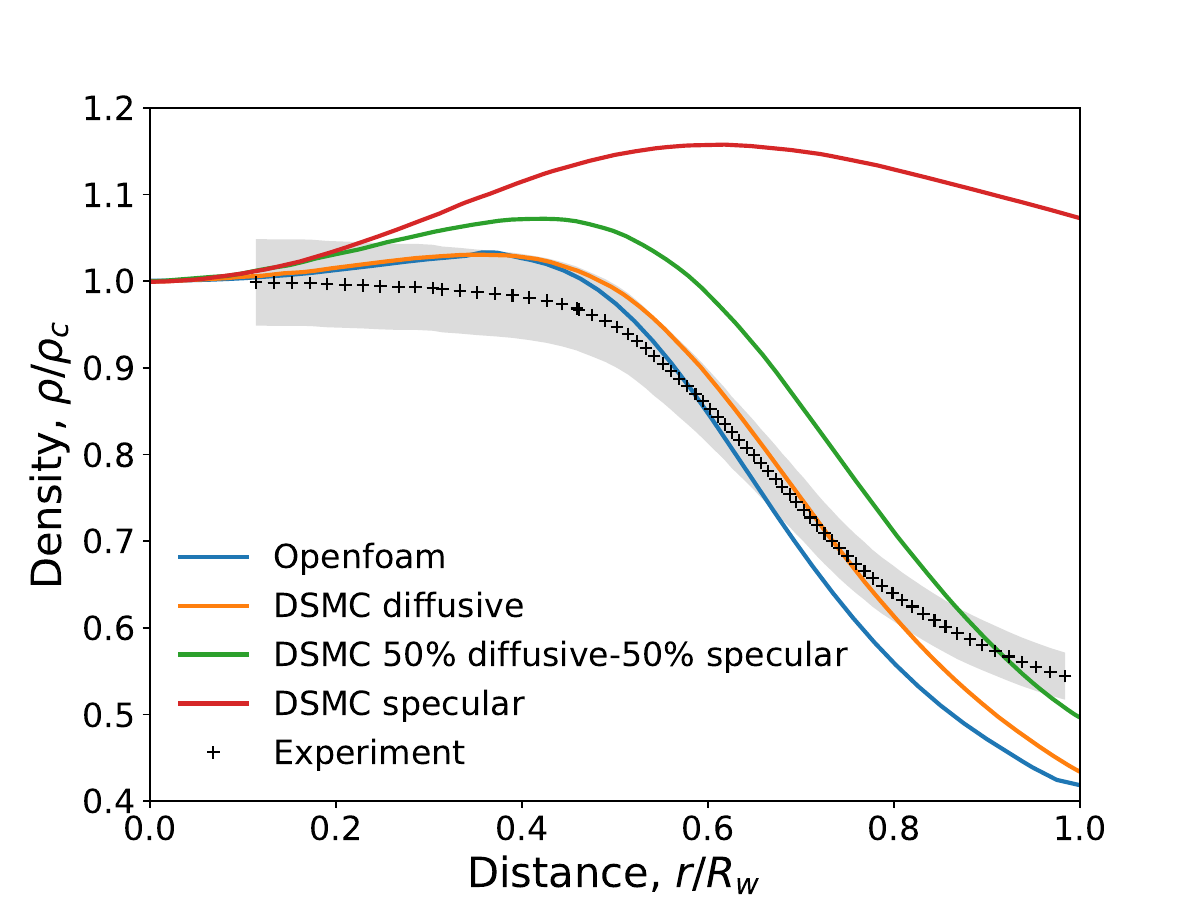}&\includegraphics[width=0.5\textwidth]{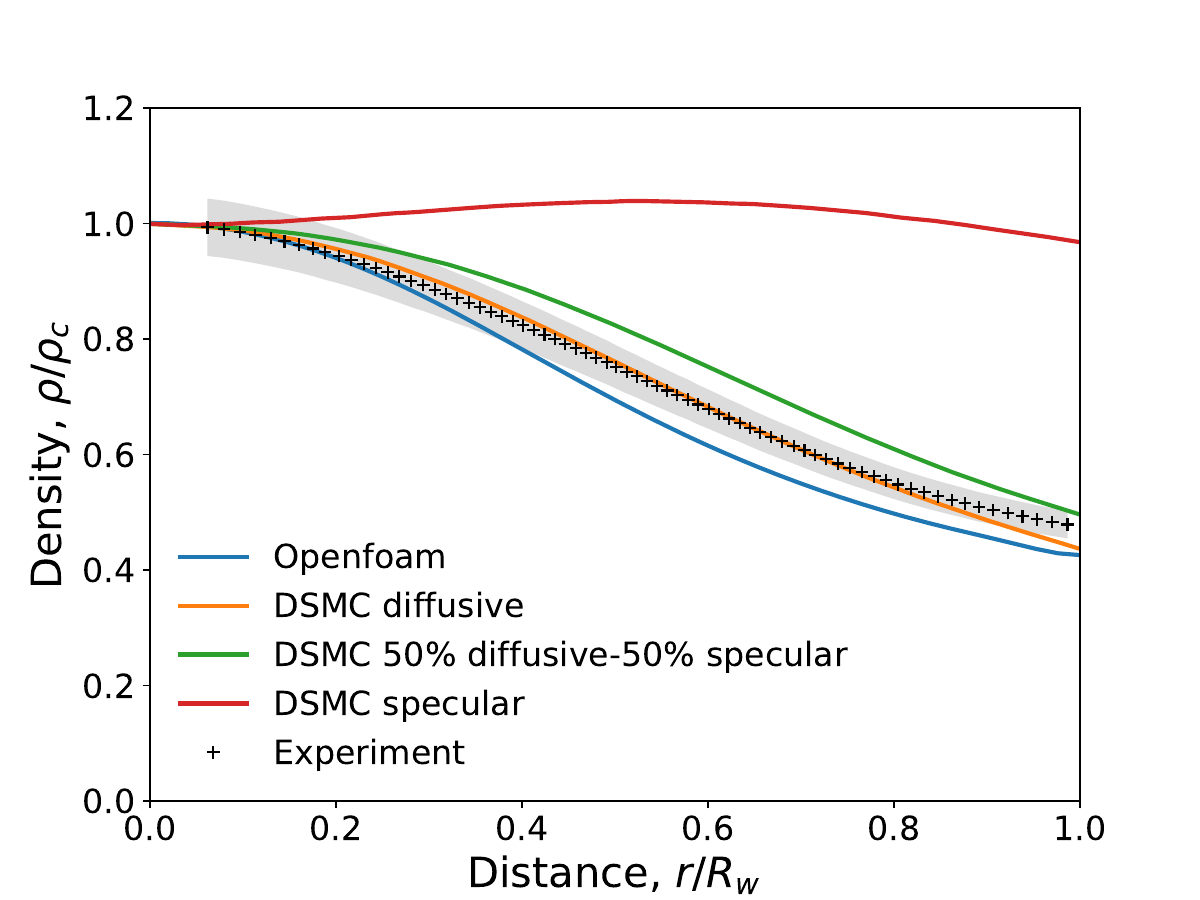}\\
	(a) Case i.I: $x/R_t = 3.7$&(b) Case i.III: $x/R_t = 3.7$\\\
	\includegraphics[width=0.5\textwidth]{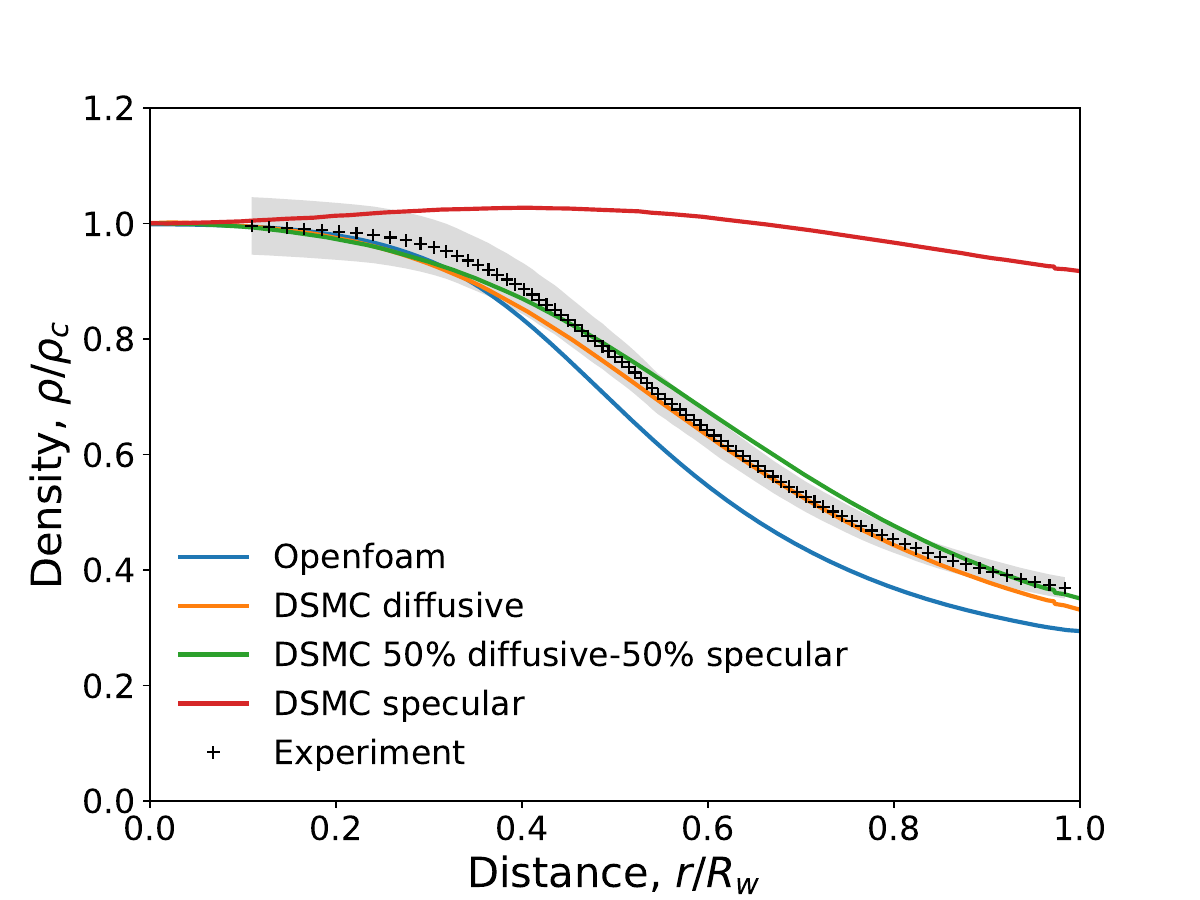}&\includegraphics[width=0.5\textwidth]{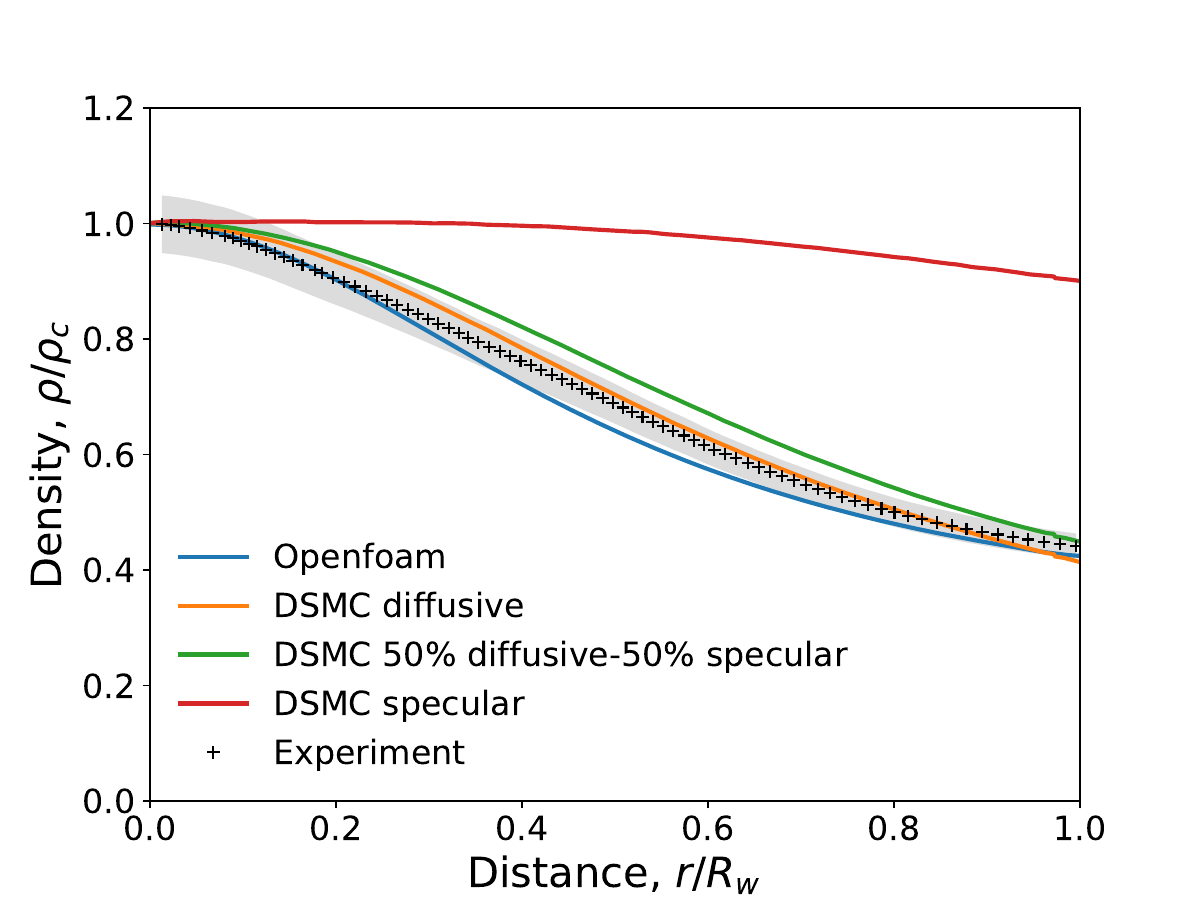}\\
	(c) Case i.I: $x/R_t = 6.2$&(d) Case i.III: $x/R_t = 6.2$
    
	\end{tabular}
	\caption{Effect of gas-surface interactions on densities at different cross-sections for case~i.I (left column) and case~i.III (right column).}
	\label{fig:radialDensity_bc}
\end{figure}

\begin{figure}[H]
	\centering
	\begin{tabular}{cc}
	\includegraphics[width=0.5\textwidth]{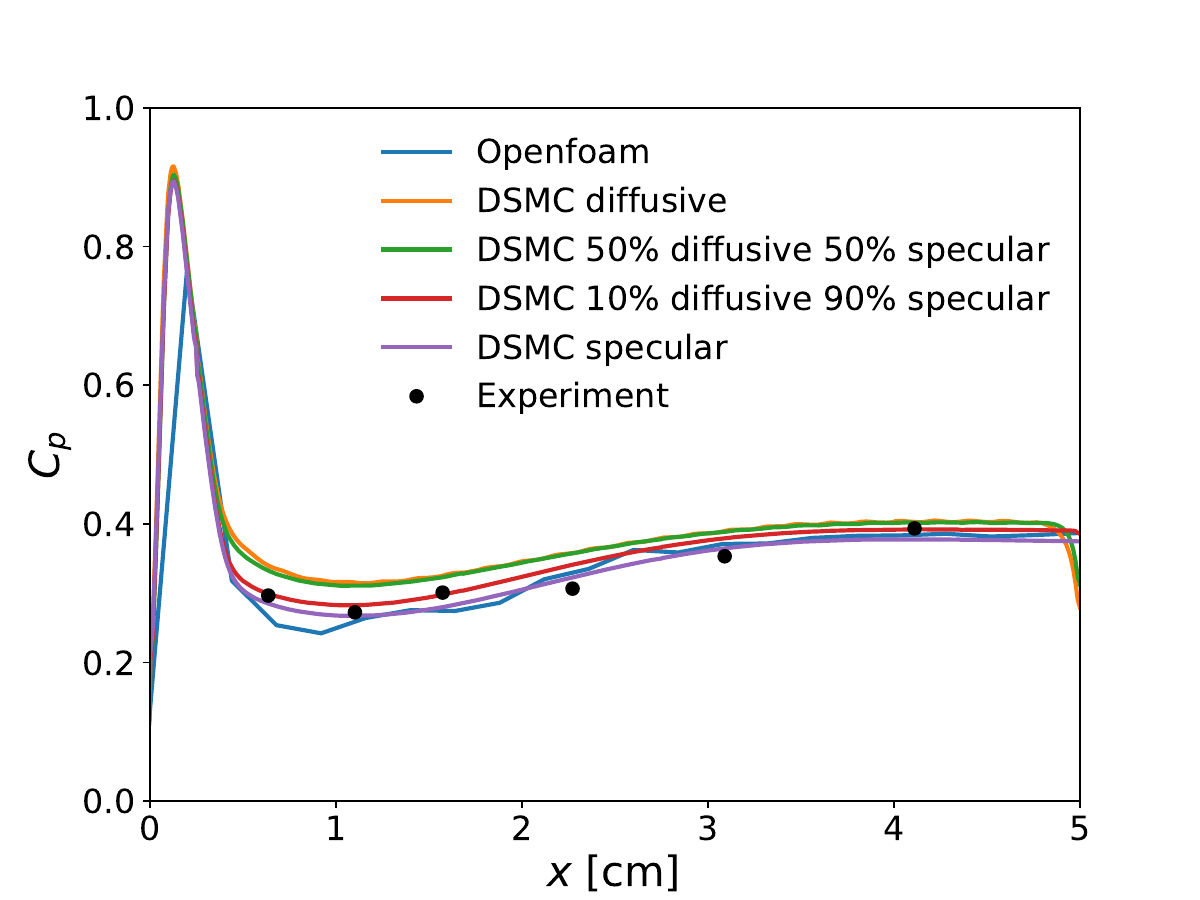}&\includegraphics[width=0.5\textwidth]{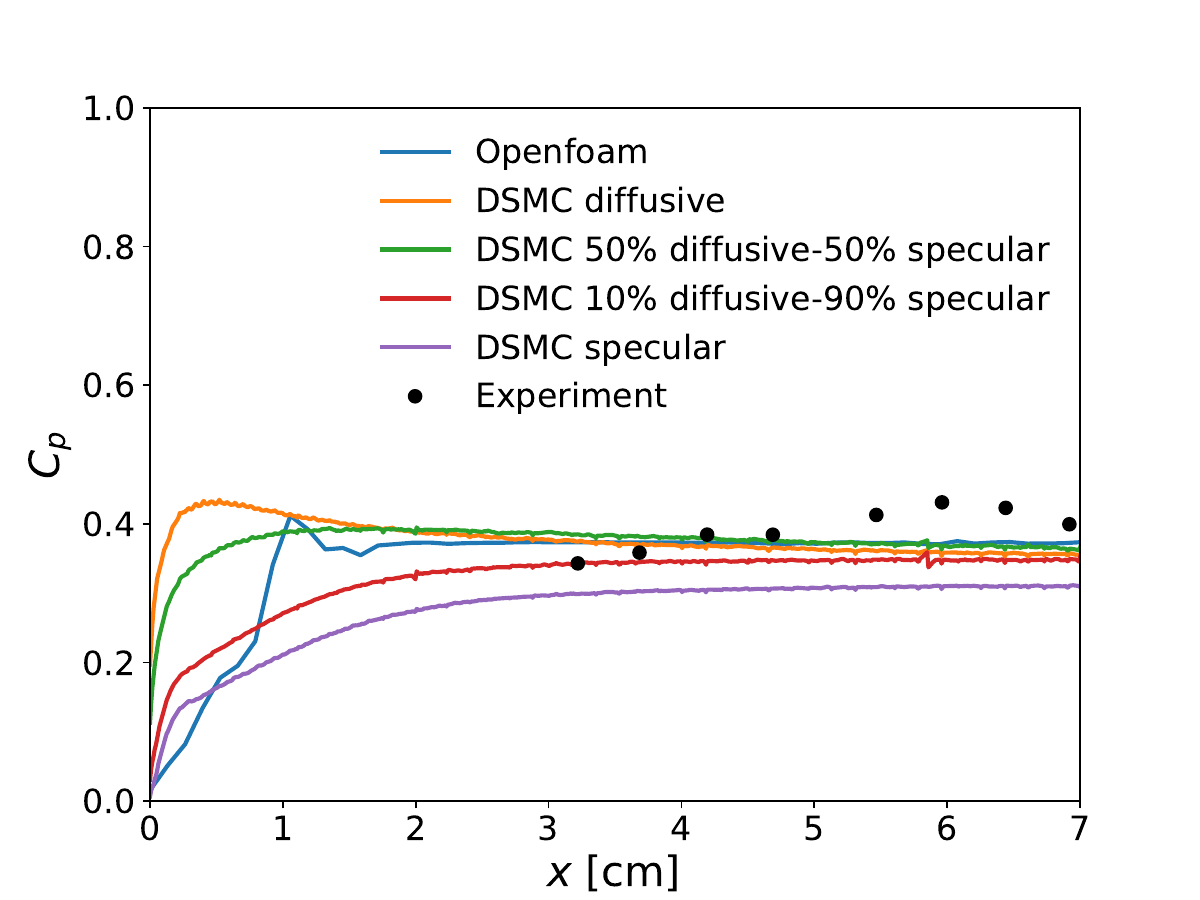}\\
	(a)&(b)\\
	
	\end{tabular}
	\caption{Effect of gas-surface interactions on $C_p$ for the cases
              (a) e.I. and (b) e.II.}
	\label{fig:accom}
\end{figure}

 \noindent The effect of the gas-surface interactions is also studied
 regarding radial variations of the density at three different
 cross-sections of the nozzle. \autoref{fig:radialDensity_bc} shows
 these distributions at the non-dimensional distances of $x/R_t =3.7$
 and $x/R_t=6.2$ from the throat for cases~i.I and i.III,
 respectively. Here, the densities are normalized by the corresponding
 density value at the axis $\rho_c$. The gray shaded regions in the
 plots represent the error margin in the
 experiments~\cite{rothe}. Similar trends as observed for the
 centerline data are visible in the results, i.e., the simulations
 with the diffusive-surface-interaction model matches well with the
 experiment. For case~i.I at the cross-section
   $x/R_t=3.7$ which is close to the nozzle throat, there is a slight
   deviation of the simulation results in comparison with the
   experimental density trend shown
   in~\autoref{fig:radialDensity_bc}~(a). The density first increases
   until a radial distance of $r/R_\text{w} = 0.4$ and then reduces
   with the distance toward the wall.  This density hump is due to the
   presence of a weaker compression wave near the throat~\cite{chung}
   which is also captured well by the continuum simulation.
   Furthermore, the density values do not coincide with the experiment
   near the nozzle wall. This deviation could be due to the
   collisional quenching effects of the electron beam technique used
   by~\citet{rothe} which reduces the quality of density measurements
   at higher pressure levels. As the flow of case~i.I progresses in
   the downstream direction, the simulations match well with the
   experiments due to low pressure levels. Furthermore, for case~i.III
   the quenching effects are reported to be negligible~\cite{rothe}
   which explains the good agreement of the simulation results with
   the experiments shown
   in~\autoref{fig:radialDensity_bc}~(b)~and~(d). For the external
   flow cases shown in~\autoref{fig:accom}, the gas-surface
   interaction models with fully diffusive and the interaction models
   consisting of fractions of specular collisions lead to a closer
   agreement with the experiments. Although the interaction models
   which are biased toward specular (e.g., $10~\%$ diffusive and
   $90~\%$ specular collisions) showed the best agreement with the
   experiments, the values of these specular to diffuse fractions can
   be case-specific and difficult to estimate. Therefore, it can be a
   safe option to assume the completely diffuse interaction
   model. Nevertheless, these two particular cases must the studied in
   more detail in the future.
 
\begin{figure}[H]
	\centering
	\begin{tabular}{cc}
	\includegraphics[width=0.5\textwidth]{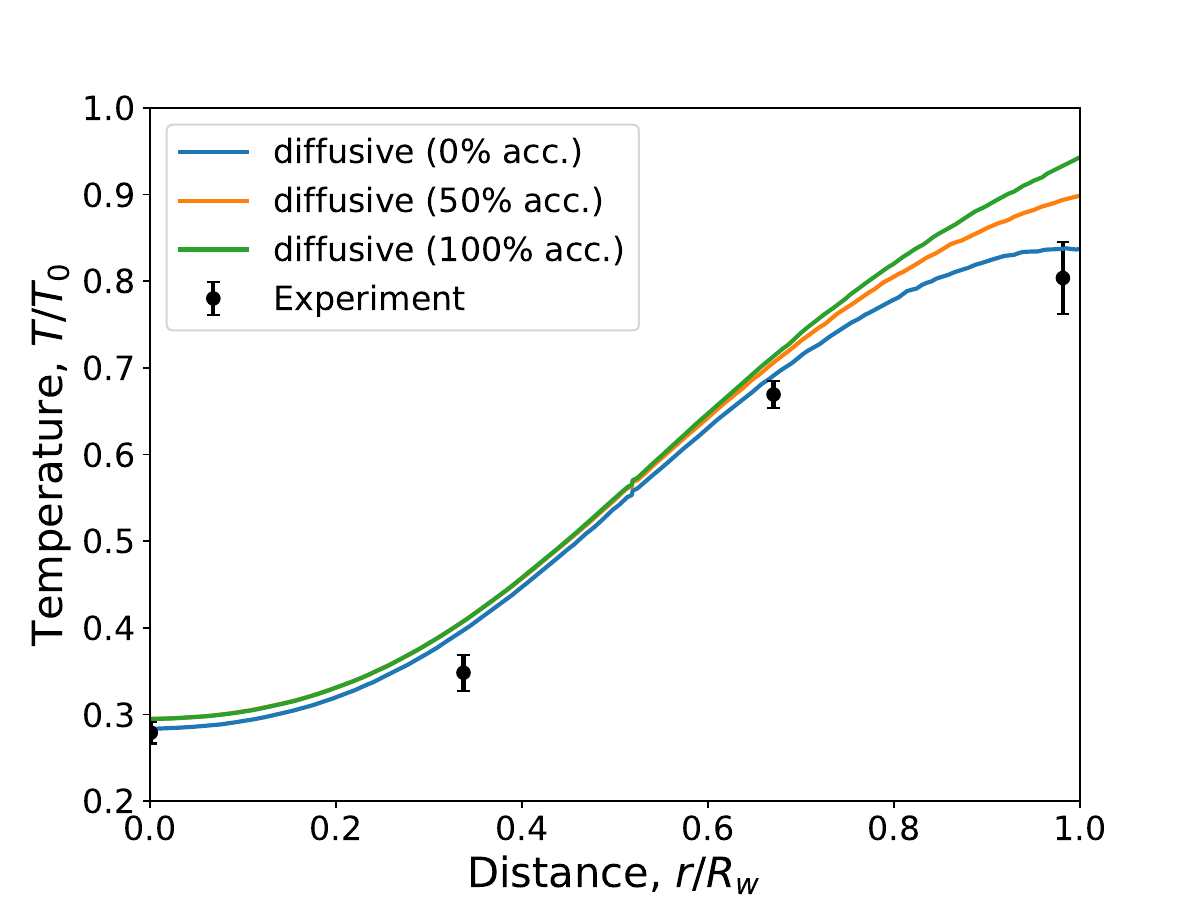}&\includegraphics[width=0.5\textwidth]{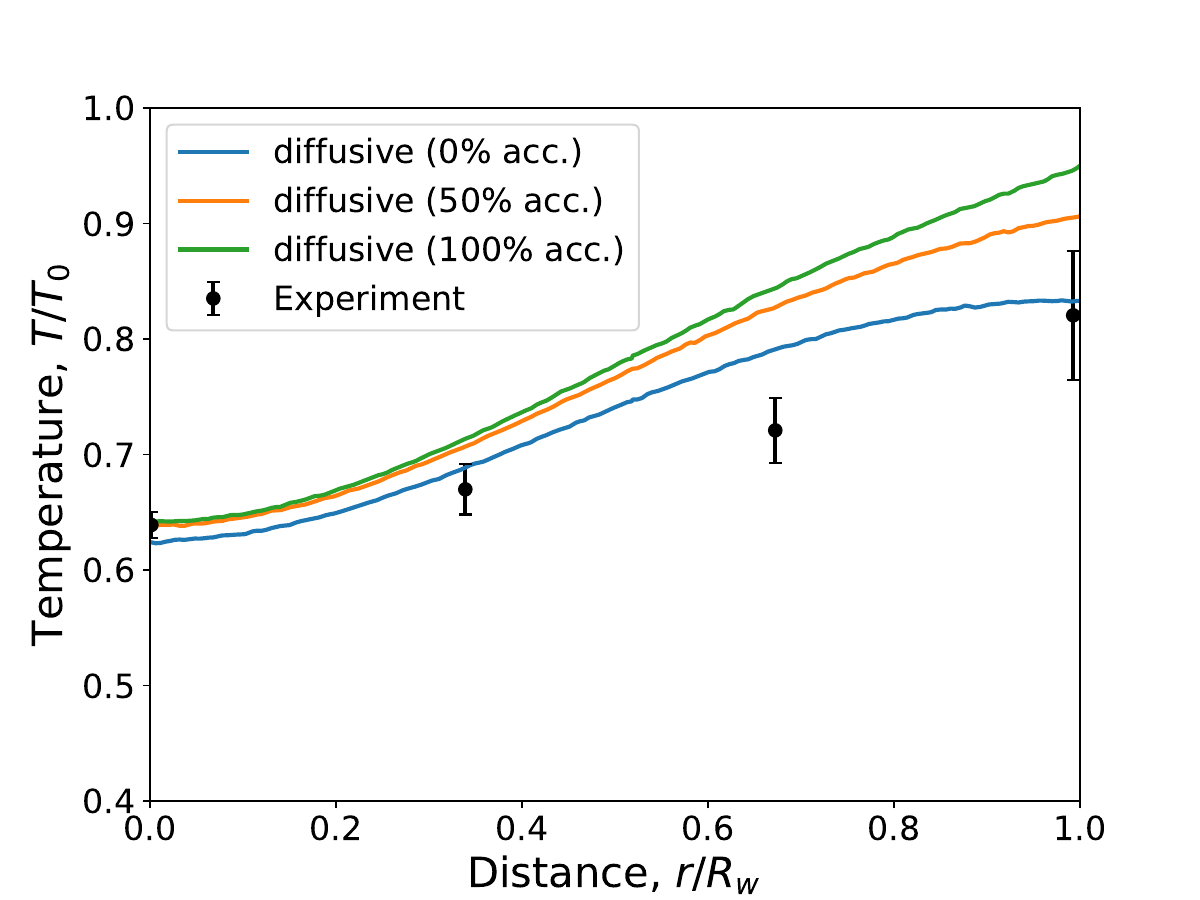}\\
	(a)&(b)\\
	
	\end{tabular}
	\caption{Effect of thermal accommodation coefficient on the rotational temperature at the cross-section $x/R_w = 13.7$: (a) case~i.I and (b) case~i.III.}
	\label{fig:474 and 141 radial temp validation}
\end{figure}

 Another important parameter in modeling the gas-surface interaction is the thermal accommodation coefficient. As described in Section~\ref{sec:dsmc_bc}, this parameter quantifies the energy exchange between the surface and the gas. For gas-surface interactions which are fully diffusive, this parameter has a negligible effect on the density. However, it has an influence on the temperature. \autoref{fig:474 and 141 radial temp validation}~(a) and (b) show the effect of the thermal accommodation coefficient on the rotational temperature at the cross-section $x/R_t=13.7$ for cases i.I and i.III, respectively. The thermal accommodation coefficient only mildly affects the centerline temperature. However, near the wall it has a significant influence. It can be seen that the fully diffusive gas-surface interaction model with an accommodation coefficient of $0~\%$ (adiabatic wall) is in close agreement with the experiment. As the value of the thermal accommodation coefficient increases, the temperatures near the wall increase and diverge from experiments. Although not shown, the effect of grid refinement near the surface on the results (\autoref{fig:clineDensity_BC} -- \ref{fig:474 and 141 radial temp validation}) was also studied. The grid near the wall is refined in the range of $\Delta x = \lambda_{min}~\text{to}~\frac{1}{3} \lambda_{min}$ and within this limit the grid refinement has negligible effect on the results. 
 
\subsubsection{Collision and energy exchange models}

\begin{figure}[H]
	\centering
	\begin{tabular}{cc}
	\includegraphics[width=0.5\textwidth]{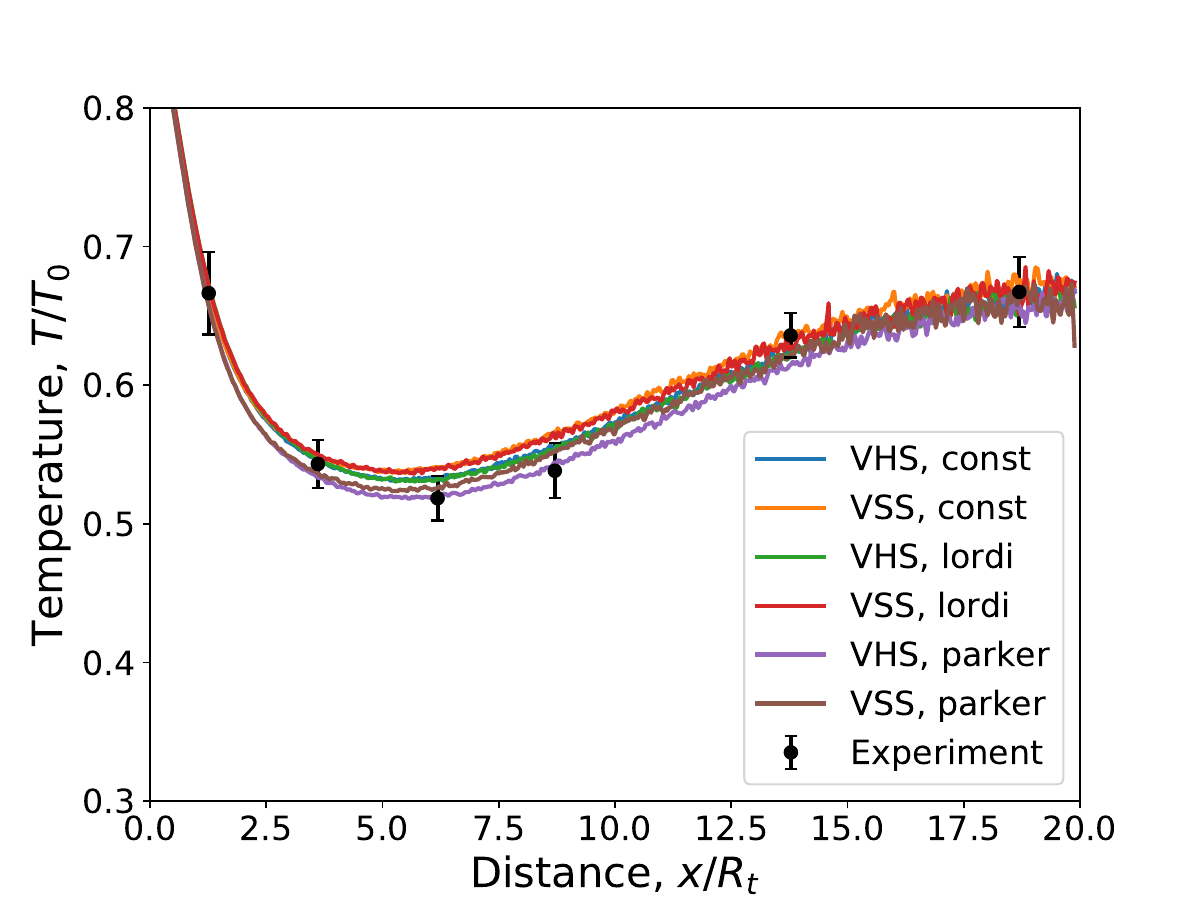}&\includegraphics[width=0.5\textwidth]{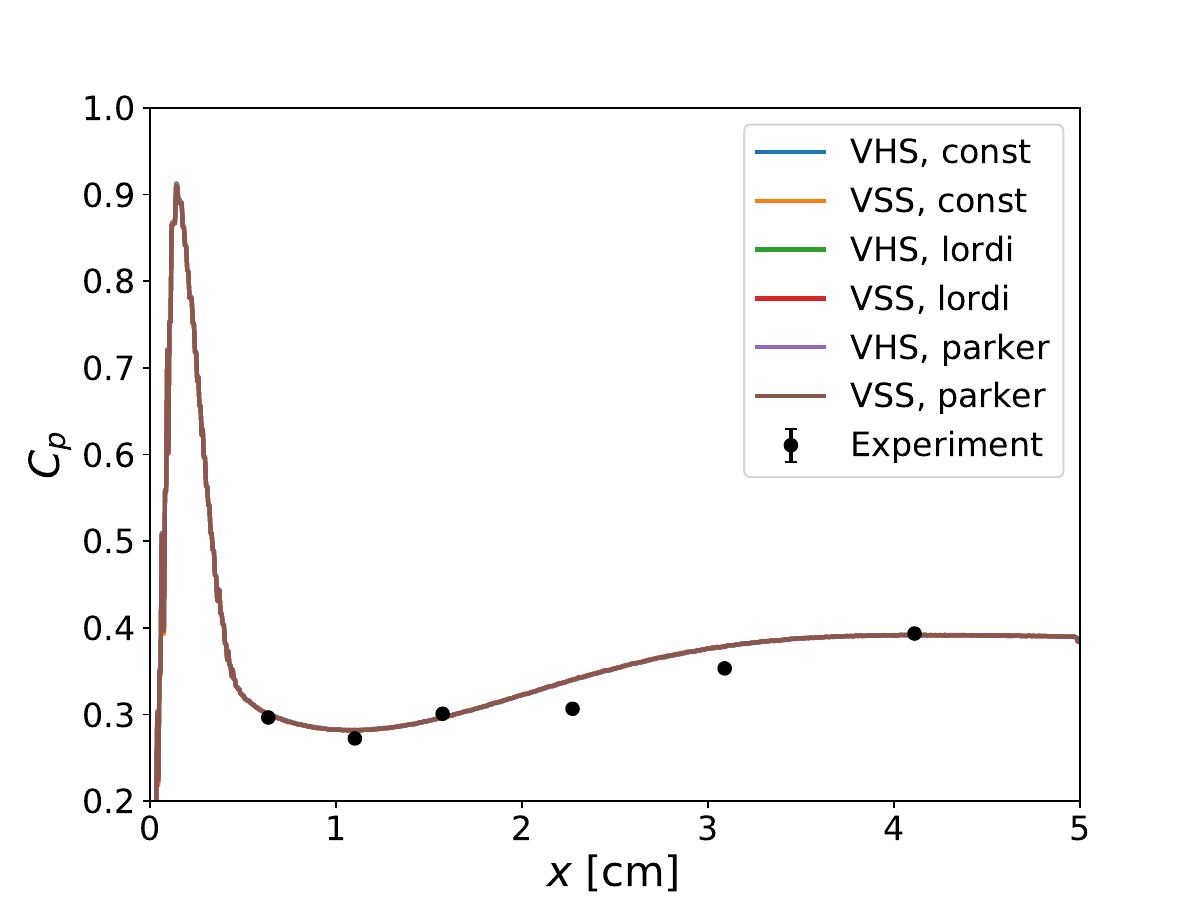}\\
	(a)&(b)\\
	
	\end{tabular}
	\caption{(a) Centerline rotational temperature profiles predicted with different collision models for case i.III. (b) Comparison of $ C_p $ distribution for various collision models for case e.I.}
	\label{fig:RotRelax1}
\end{figure}

More sensitivity studies have been conducted considering the choice of the molecular models (i.e., VHS and VSS) and the energy exchange models (constant and variable relaxation).
According to the evaluation in Section~\ref{sec:gas-surface interaction}, the fully diffusive gas-surface interaction model with $0~\%$ thermal accommodation is used in this study for the internal flow cases. The Maxwellian gas-surface interaction model with $10~\%$ diffusive and $90~\%$ specular is used for external flow cases. 
\autoref{fig:RotRelax1} shows the comparison of the various parameters obtained using different molecular and energy exchange models with the experimental data for cases i.III and e.I. It can be seen that the results predicted by these different models lie with in the experimental error estimates. However, in order to rank the best DSMC simulation configuration for internal flows, the accuracy is quantified on the basis of the error estimation given by:
\begin{equation}
    \text{error} = \frac{\psi_\text{simulation}-\psi_\text{experiment(mean)}}{\psi_\text{experiment(mean)}} \; ,
     \label{eq:error}
\end{equation}

where $\psi $ can be any flow quantity such as the rotational temperature. Again, the collision and energy exchange models have a negligible effect on the density. 

The simulation configuration with the variable hard-sphere model with a variable rotational relaxation (obtained using parameters from \cite{parker}) yield the best result for internal flow cases. For external flow cases the best result is achieved with the variable soft-sphere model with variable relaxation (obtained using parameters from \cite{parker}). However, in terms of performance the variable hard-sphere model with a constant rotational collision number (i.e., $\phi_{rot} =0.2$) is the best choice for both internal and external cases as the simulation speed is roughly $1.5$ times higher than the latter. More details on the simulation performance are discussed in Section~\ref{sec:performance}.

\subsubsection{DSMC sampling}

As mentioned in Section~\ref{sec:dsmc} 
the macroscopic flow properties shown above are
obtained in the DSMC method by averaging the microscopic
properties per grid cell. However, DSMC being a probabilistic
method is very much prone to statistical fluctuations. The
statistical noise in the flow field is typically filtered
out by time averaging of the cell properties to obtain mean
macroscopic properties. This procedure is performed after
a steady state of the flow is established.
In~\autoref{fig:fnum_int_ext}~and~\autoref{fig:fnum_int_ext2}, it was
evaluated that a sufficient number of particles (defined in terms of
{\tt{fnum}}) is required to obtain reliable macroscopic values.
Furthermore, it can be observed from these figures that using
less particles than the trade-off values also induces statistical
noise. In this section, the statistical error associated with
DSMC according to the effect of the number of particles per cell $N_c$
and the number of time steps $N_T$ used for averaging is studied.
Here, the root-mean-square (rms) error is used as an indicator
to quantify the statistical fluctuations. The  relative
root-mean-square error $\epsilon_\psi$ in
dimensionless form is given by:
\begin{equation}
    \epsilon_\psi =  \sqrt{\frac{1}{N_{cell}} \sum_{i=1}^{N_{cell}} \left ( \frac{\psi_i -  \psi_{ref,i}}{\psi_{max}} \right )} \; .
\end{equation}

\begin{figure}[H]
	\centering
	\begin{tabular}{cc}
	\includegraphics[width=0.5\textwidth]{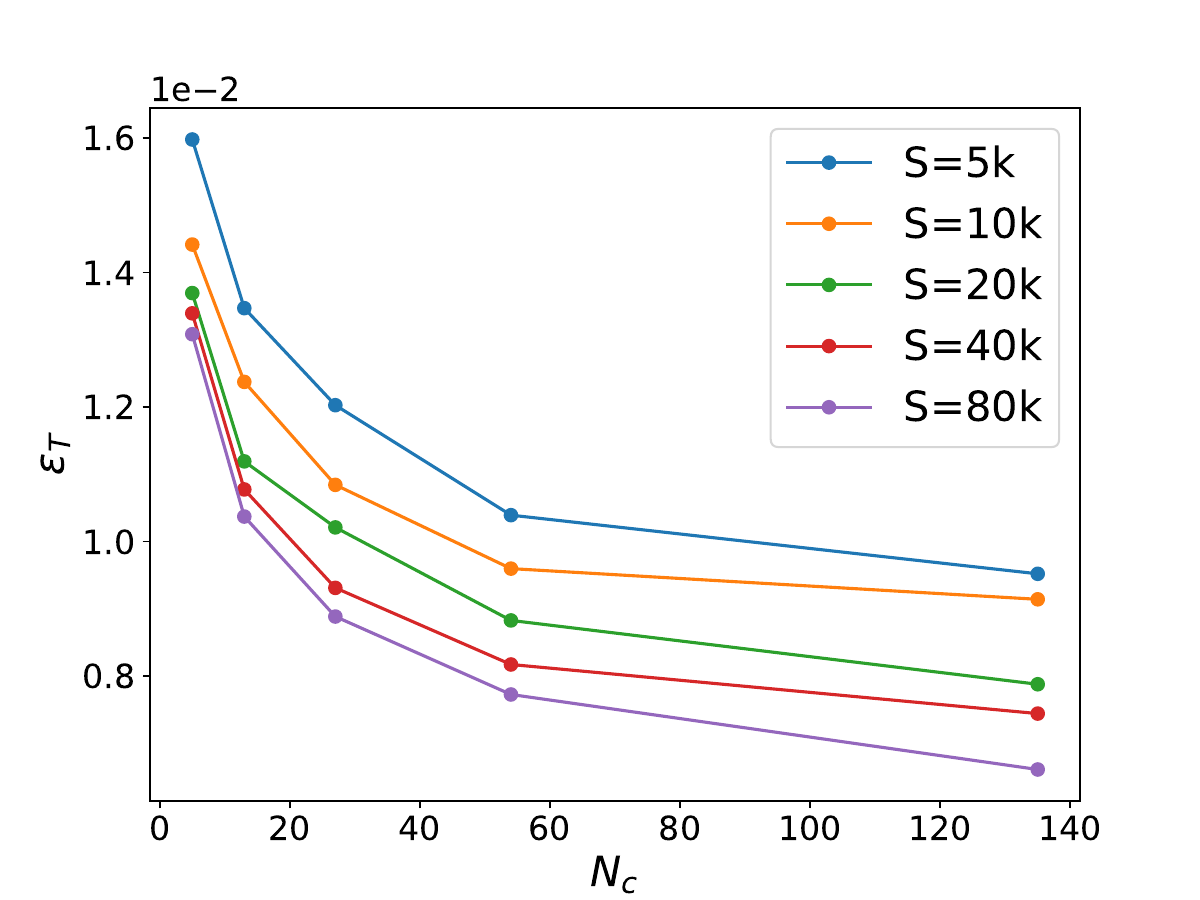}&\includegraphics[width=0.5\textwidth]{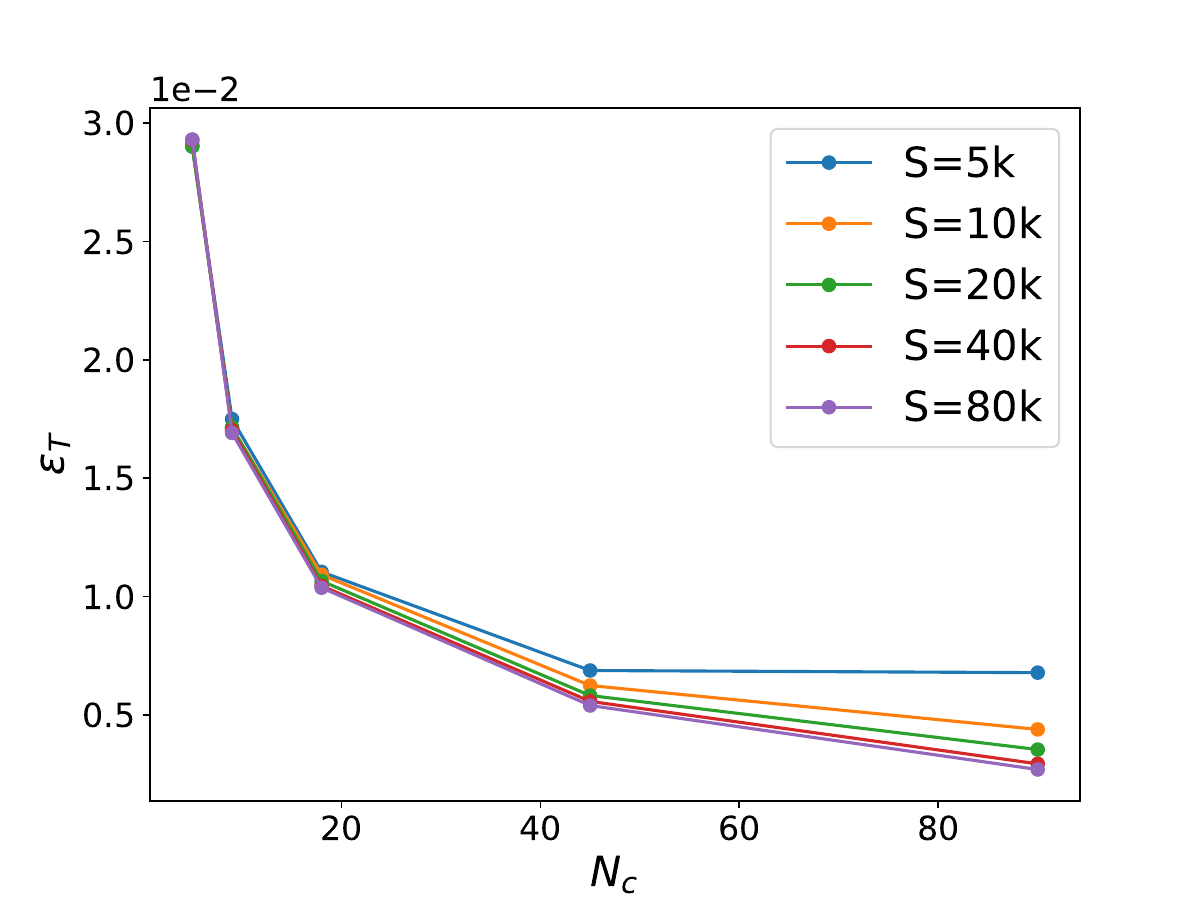}\\
	(a)&(b)\\
	
	\end{tabular}
	\caption{\textcolor{dblue}{RMS error $\epsilon_T$ based on the 
            translational temperature for (a) case~i.III
            and (b) case e.I.}}
            
	\label{fig:rms}
\end{figure}

Again, $\psi$ can be any macroscopic property
where $\psi_{max}$ is its maximum value in the flow field and
$N_{cell}$ is the total number of grid cells in the simulation
domain. $\psi_i$ and $\psi_{ref,i}$ represent the computed and
the reference solutions in the grid cell $i$. A sample size $S$
is defined as  the product of $N_c$ and $N_T$ through which a
macroscopic property is sampled. In the study, it was observed
that the translational temperature had a higher sensitivity to
the statistical fluctuations and hence it is chosen to describe
the results. The rms error $\epsilon_T$ based on the translational
temperature is determined for different values of $S$ obtained
by varying $N_c$ and $N_T$, see~\autoref{fig:rms}. The
reference solution used to estimate the rms error has a sample
size of $S= 2700\,k$ ($N_c=270$, $N_T =10,000$) for the internal
flow and $S= 1800 \, k$ ($N_c=180$, $N_T =10,000$) for the
external flow case.
Peak temperatures of $300~K$ and $4100~K$ were used for
normalizing the rms errors for internal and external cases,
respectively.

\autoref{fig:rms} shows that increasing 
$N_c$ has a major effect in reducing the statistical errors. For a fixed value of $N_c$ the rms error reduces until it reaches an asymptotic value. 
This trend was also  observed in the study  by~\citet{CHEN_boyd}. 
The computational cost for a particular sample size $S$ is
observed to be constant for small to large values of $N_c$.
Therefore, it is advantageous to use greater values of $N_c$
for a specific $S$ leading to the smallest rms error at given 
CPU-time resources.

\subsection{Performance study}
\label{sec:performance}

The last part deals with the performance and computational costs of the DSMC simulations running on the high-performance computer HSUper at the Helmut-Schmidt University. HSUper consists of 581 nodes in total, 571 compute nodes with 256~GB RAM per node, 5 nodes with 1~TB RAM per node, 5 nodes with 2 NVIDIA A100 GPUs. Each node contains two sockets equipped with Intel Xeon Platinum 8360Y (36 cores, 2.4~GHz) processors, hence 72 physical cores or 144 virtual CPUs via hyperthreading. The memory is provided by $ 16\times 16 $ GB DDR4 RDIMM 3200 MHz ECC-registered modules. HSUper utilizes InfiniBand HDR, 100~Gb/s, non-blocking fat tree networking.
The test cases with the optimized simulation setup (i.III and e.I) were
used to study the performance.

The  energy consumption of the CPU and DRAM reported in the following are based on the running average power limit \cite{Georgiou2014} as well as the CPU time of the job reported by the SLURM workload manager.
In order to keep the result comparable for internal and external flow cases with different configurations, the run-time is normalized in all cases to 1 core/node. SPARTA uses a hierarchical Cartesian grid, which is not a body-fitted grid. Since a big portion of the grid points is therefore
outside the computational domain and without simulation particles, the usage of
load balancing methods is necessary to optimize computational efforts. 

\begin{figure}[H]
	\centering
	\begin{tabular}{cc}
		   \includegraphics[width=0.5\textwidth]{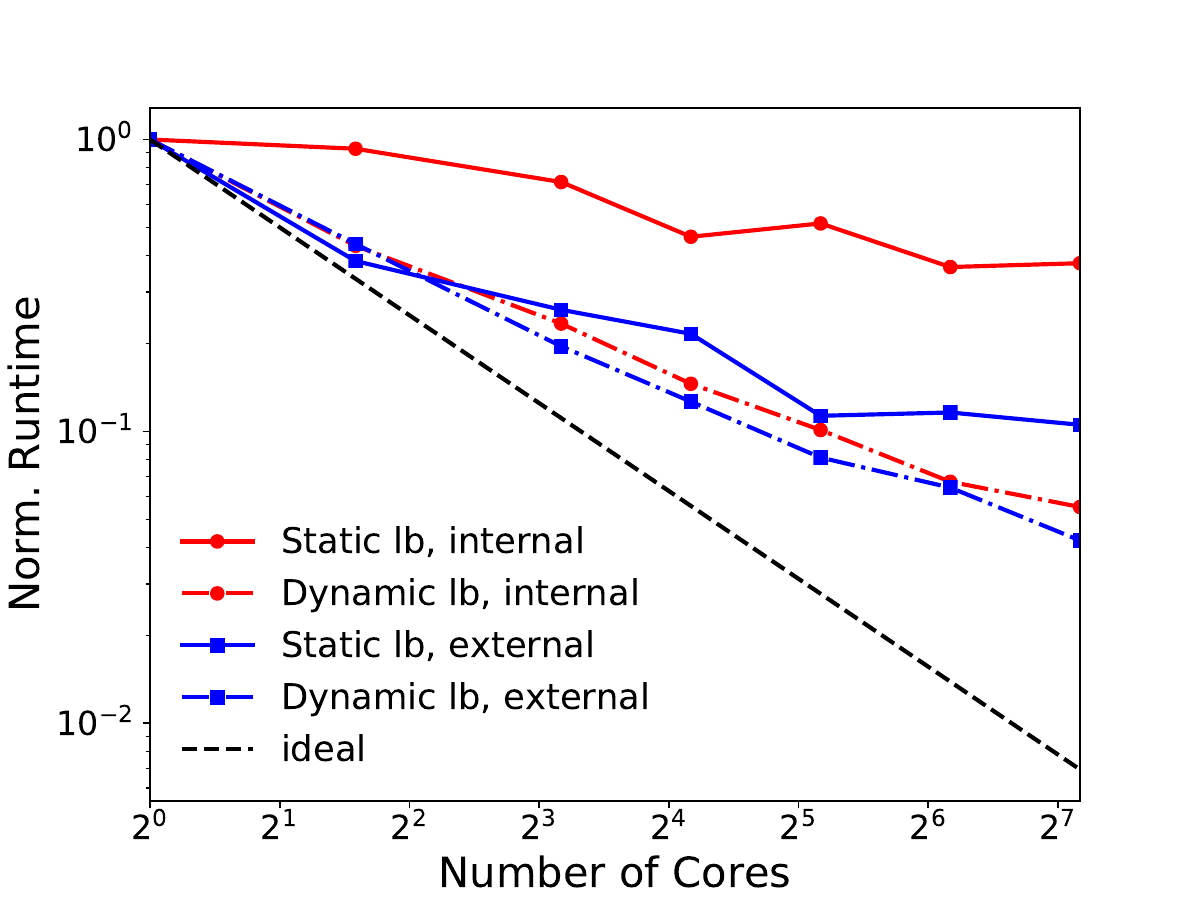}
         &\includegraphics[width=0.5\textwidth]{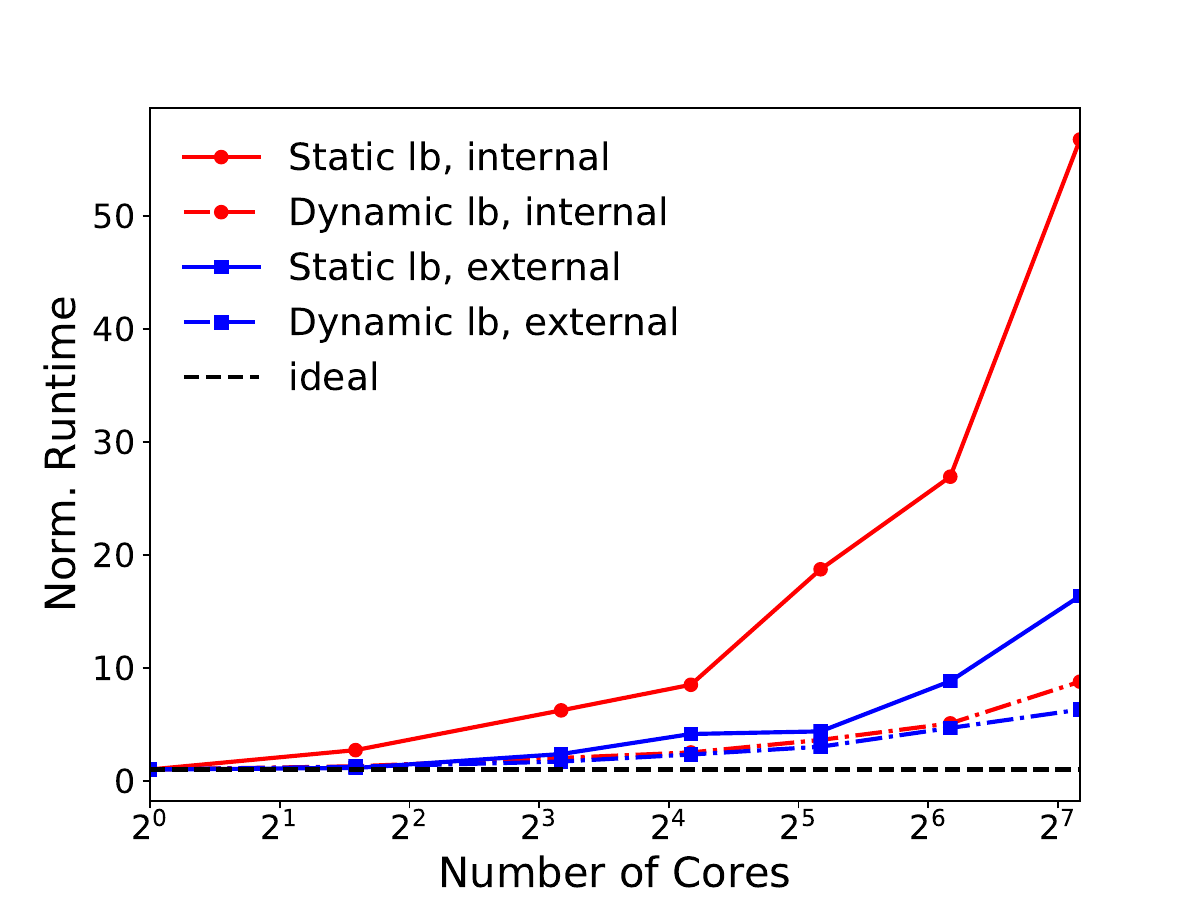}\\
		(a)&(b)\\
		\includegraphics[width=0.5\textwidth]{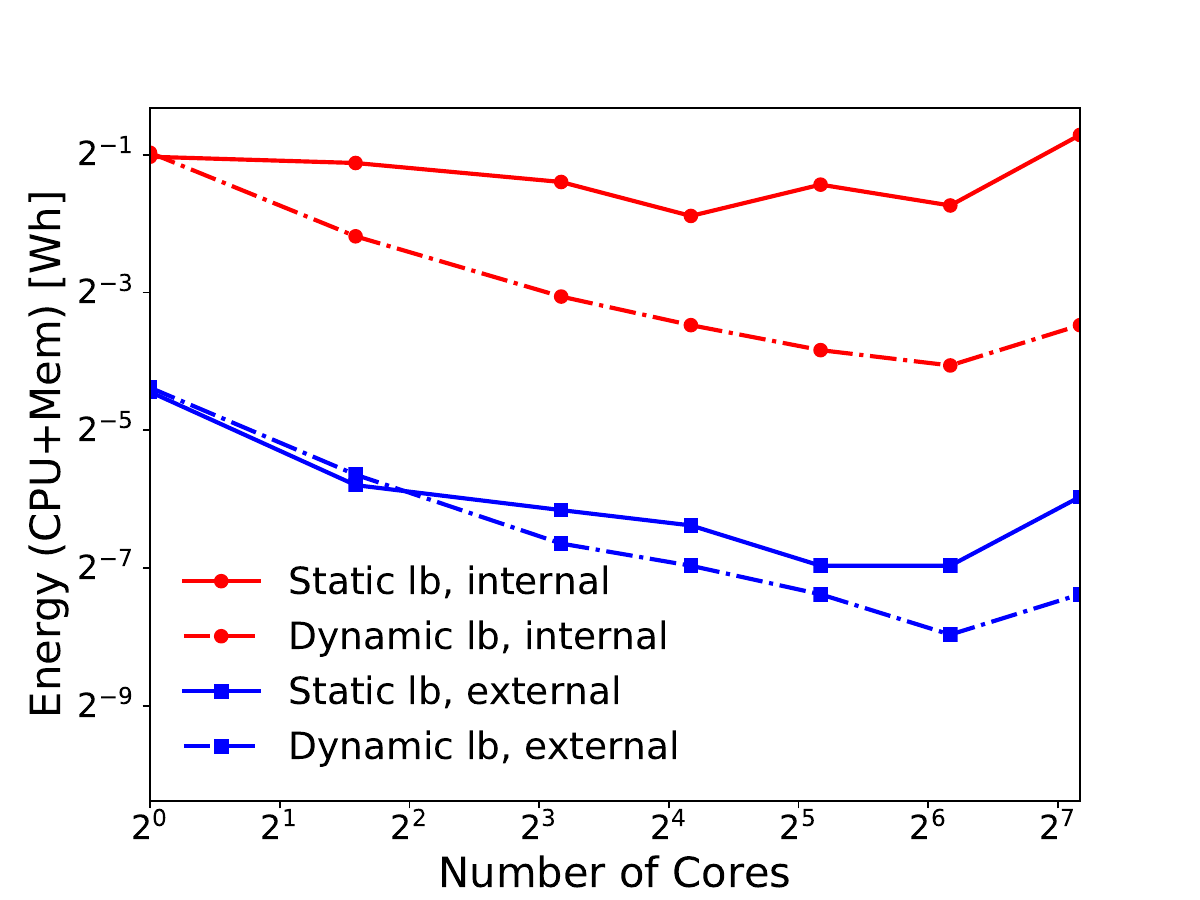}
       &\includegraphics[width=0.5\textwidth]{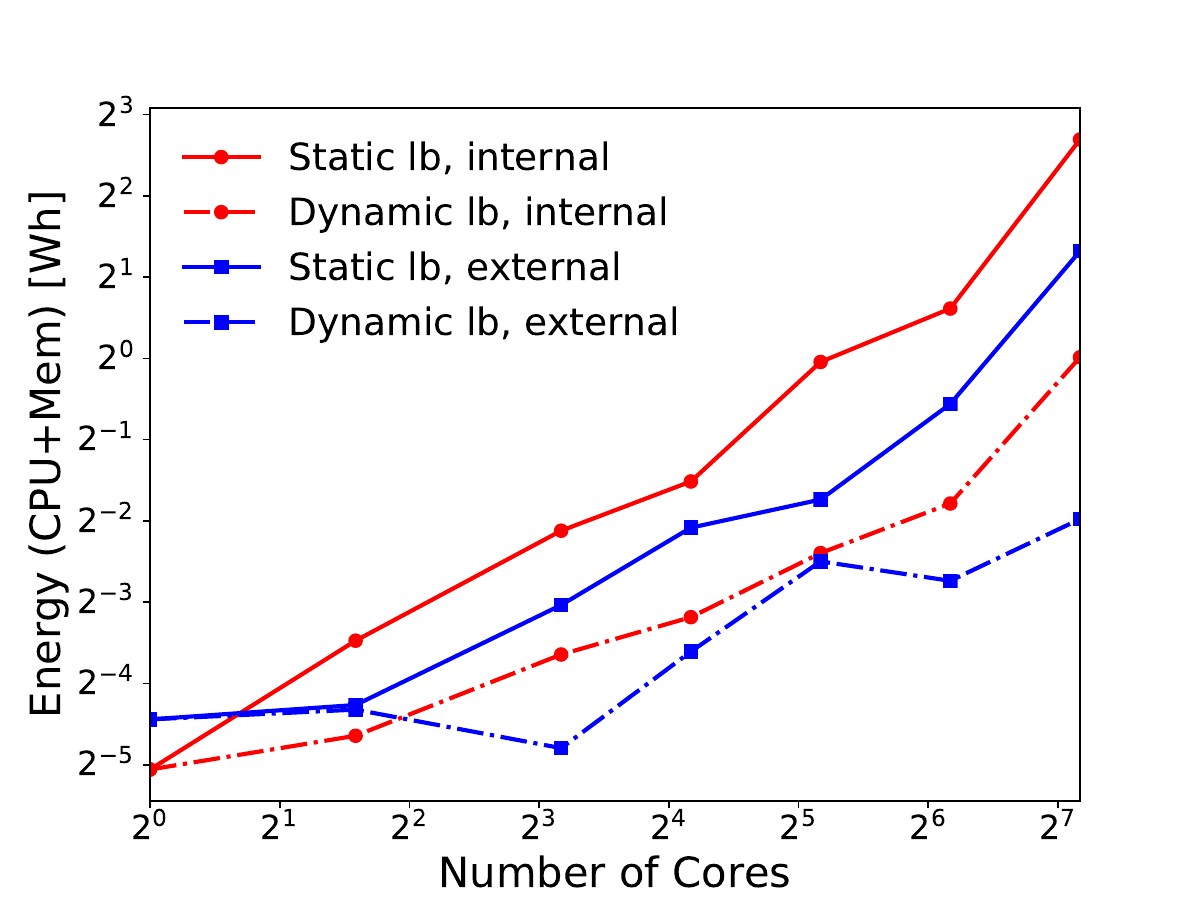}\\
		(c)&(d)\\
	\end{tabular}
	\caption{Scaling results on up to 144 cores. Left column: strong scaling. 
    Right column: weak scaling. 
    (a) \& (b) normalized run-time. (c) \& (d) energy consumption for one time step.}
	\label{fig:perf1}
\end{figure}

\autoref{fig:perf1} compares the strong scaling results on 1 core up to 144 cores equivalent to 2 nodes on HSUper using static and dynamic load balancing for the internal and external cases. The dynamic load balancing had a huge impact on the speedup of the simulation. The effect of the dynamic load balancing is stronger for the internal case than for the external flow because the external case is a hypersonic flow and due to the very high flow velocity the computational domain gets saturated very quickly with sufficient simulation particles. Thus, the static load balancing at the beginning of the simulation already leads to good results contrary to the internal case where the flow velocity is much smaller than in the external case and the simulation needs therefore a much longer time to reach the steady-state solution. 
In \autoref{fig:perf1}~(c) and (d), a similar behavior is also observed in the energy consumption results, which is strongly related to the run-time.
 
  Weak scaling is achieved by varying the parameter \texttt{fnum}, e.g., the number of particles is increased by a factor of three when running on three cores, compared to the single-core case. The results also shown in \autoref{fig:perf1} are in full agreement with the argumentation given for the strong scaling results. Note that the \texttt{fnum} parameter decreases correspondingly as the number of cores increases, e.g.,  the \verb|fnum| value for 1 core is decreased by a factor of 3 for 3 cores. So the number of simulation particles increases for 3 cores and the number of simulation particles per core remains the same as for 1 core. The \texttt{fnum} parameter is set for 9, 18, 36 and 72 cores, correspondingly. Note that in \autoref{fig:perf1}~(d) for the internal test case, the weak scaling configuration starts from a different number of particles on 1 core  than the strong scaling configuration ($\texttt{fnum}=5e15$) in \autoref{fig:perf1}~(c).
 
\begin{figure}[H]
	\centering
	\begin{tabular}{cc}
		\includegraphics[width=0.5\textwidth]{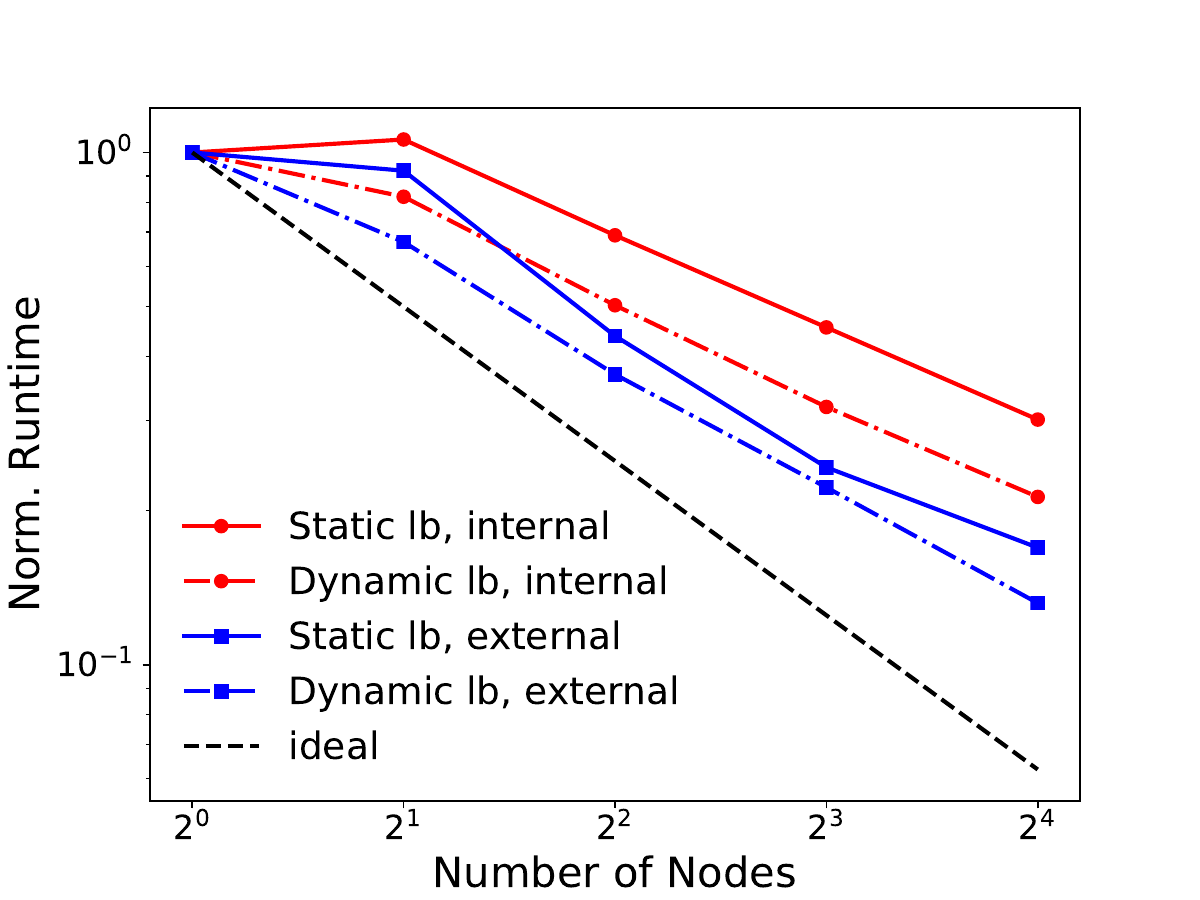}&\includegraphics[width=0.5\textwidth]{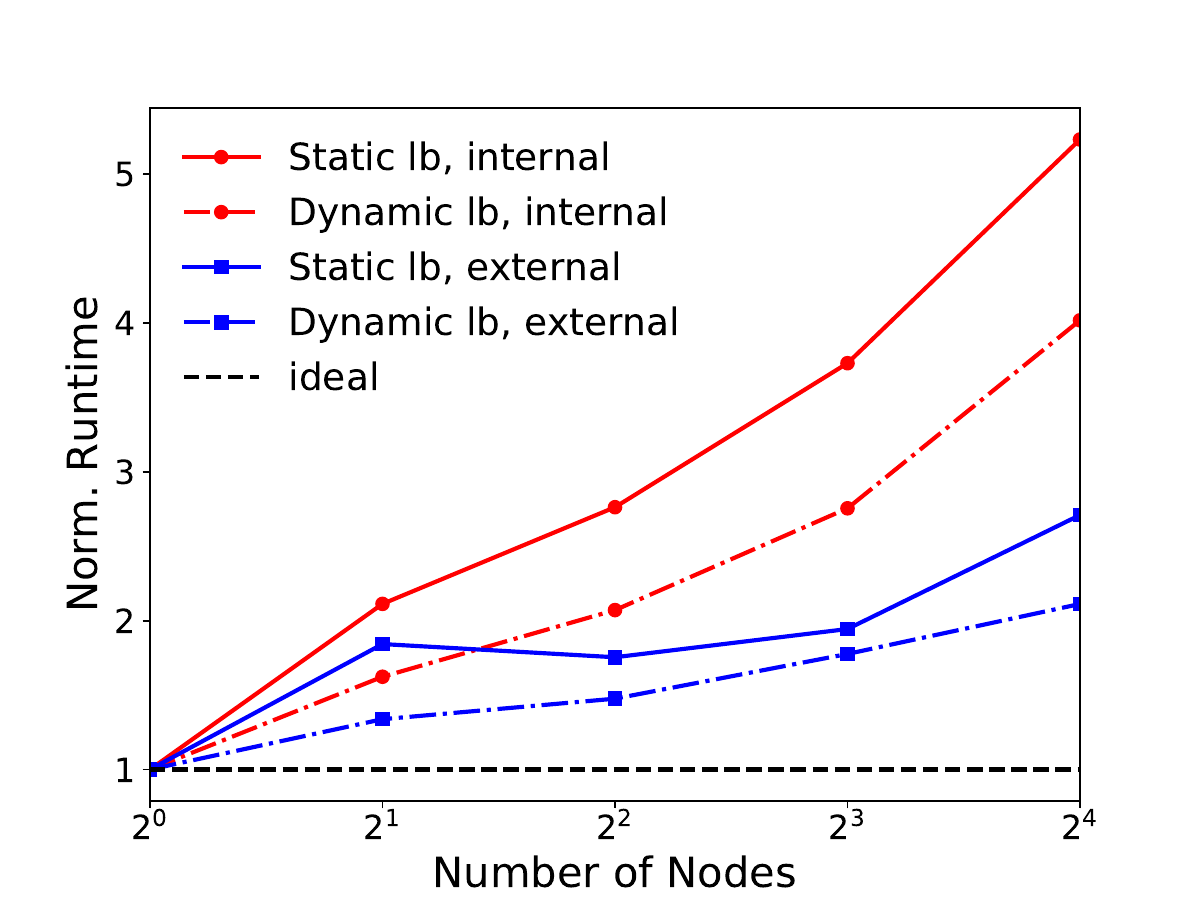}\\
		(a)&(b)\\
		\includegraphics[width=0.5\textwidth]{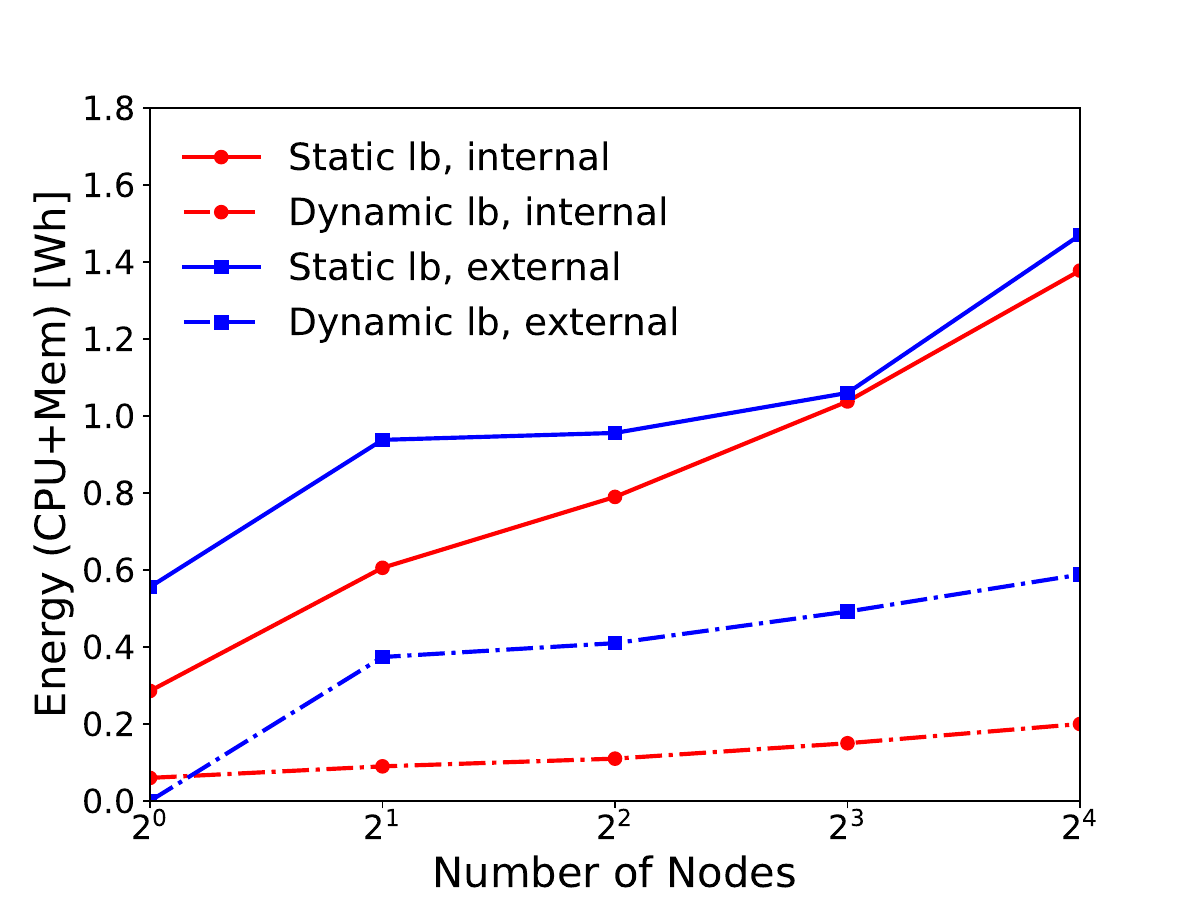}&\includegraphics[width=0.5\textwidth]{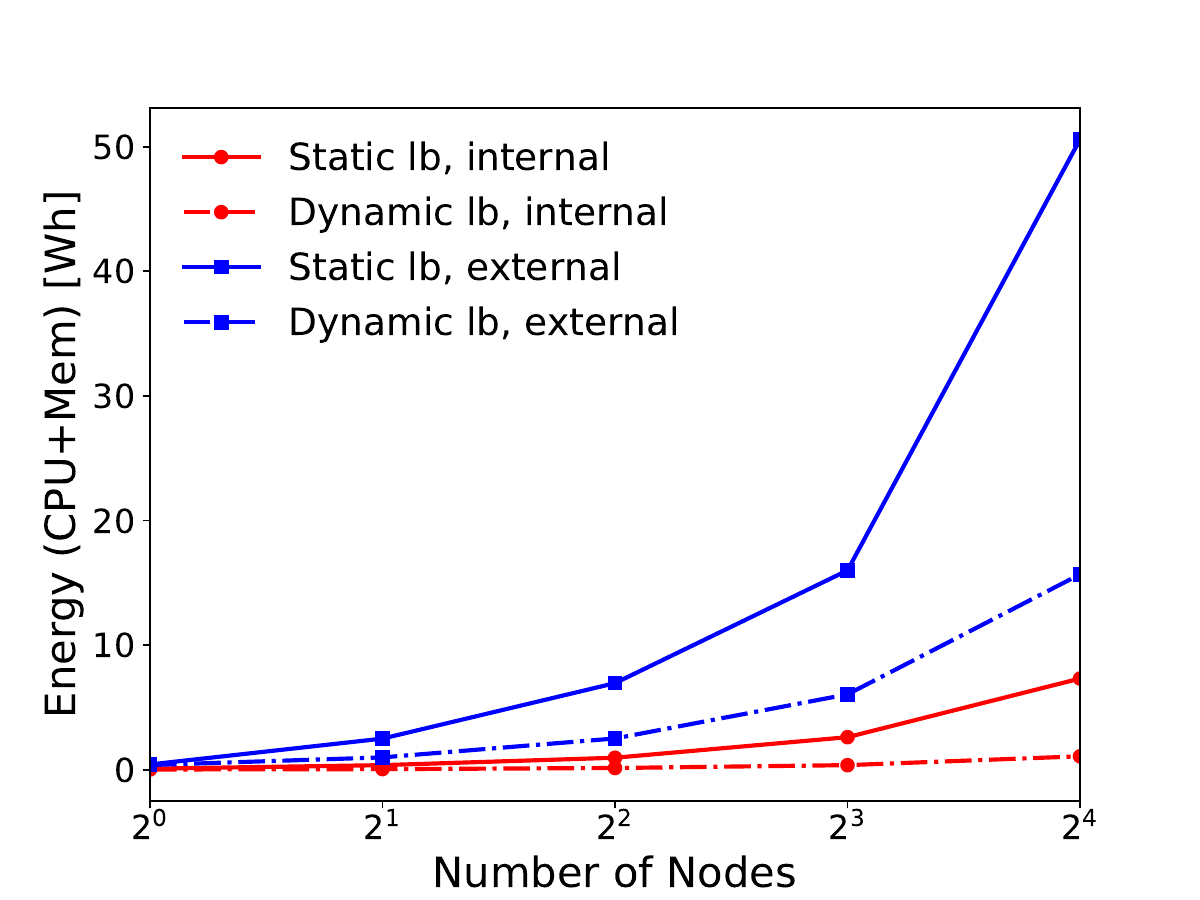}\\
		(c)&(d)\\
	\end{tabular}
	\caption{Scaling results on 1 to 16 nodes. Left column: strong scaling. 
    Right column: weak scaling. 
    (a) \& (b) normalized run-time.\cmmnt{ (c) \& (d) memory usage.} (c) \& (d) energy consumption for one time step.}
	\label{fig:perf3}
\end{figure}

\autoref{fig:perf3} demonstrates the strong scaling results on 1 to 16 nodes (i.e. up to 1152 cores). It has to be mentioned that all 72 cores of each HSUper node are occupied for the simulation. Unlike in \autoref{fig:perf1}~(c) and (d), the energy consumption on more than one node increases constantly in \autoref{fig:perf3}~(c) and (d). By using a few processors of a node, the other processors of the node are in the idle mode and still consume energy. Therefore, using more cores of one node which leads to a smaller run-time may require less energy, because the reduction of the computation time may compensate the effect of using more cores. \autoref{fig:perf3}~(c) shows that this is not the case when using more than one node. Considering simulations using 1 or 2 nodes, if the speedup of the simulation from 1 to 2 nodes would be exactly 2, then the energy consumption would be the same, since the speedup of the simulation would fully compensate the additional energy consumption of more nodes. But as can be seen in  \autoref{fig:perf3}~(a), the speedup of the simulation from 1 to 2 nodes is less than two. So the computational time reduction can not compensate the energy consumption of the additional node and therefore the total energy consumption increases (by a factor of 2). It can be summarized that since an entire node is always allocated exclusively, the faster the simulation is finalized  on it, the less energy is used and this 
rationalizes the decrease in energy consumption up to 72 cores. For more cores the
energy consumption increases, because more nodes are allocated. Based on this explanation, in the SPARTA cases it is necessary in terms of energy consumption to always fully use all cores of a node as long as runtime decreases are observed. The weak scaling results in \autoref{fig:perf3} are in full agreement with the argumentation given for the strong scaling.

\section{Conclusions}\label{sec:conclusion}

The rarefied nitrogen gas flow through a convergent-divergent nozzle (internal flow) and over a conical body (external flow) were numerically investigated using a continuum-based Navier-Stokes solver and a stochastic Boltzmann solver (DSMC) for different Knudsen- and Reynolds-number regimes. The numerical results were validated against the available experimental data. In the higher Knudsen number range ($\Knud > 0.1$), the DSMC method demonstrated superior accuracy in predicting state properties such as density, pressure, and rotational temperature. However, in the continuum regime ($\Knud < 0.1$) the Navier-Stokes solver exhibited higher accuracy and computational efficiency than the DSMC method. Consequently, for internal-flow cases encompassing variable Knudsen number regimes, a one-way-coupled hybrid approach combining the Navier-Stokes and the DSMC solver was preferred to accurately
resolve the flow in the entire domain while mitigating computational costs. 

The sensitivity study of the DSMC simulations unveiled several
essential insights, particularly regarding parameters such as
the ratio of real molecules to simulation particles \verb|fnum|,
the grid size, the time step, the sampling and the surface
and molecular collision models:
\begin{itemize}
 \item Threshold values for \verb|fnum| or the number of particles to approach optimal simulation accuracy were investigated and are reported for both internal and external flow cases in 2D and 3D configurations. 
 
 \item Our study revealed that varying grid sizes $\Delta x$ of the DSMC domain in the range of the minimum mean free path $\lambda_{min}$ have negligible effects on the results.
 
 \item We found that the value of the time step $\Delta t$ must 
 definitely be a fraction of the mean collision time $t_\text{mct}$. For resolving
 near-continuum flow regions in DSMC domains, 
$\Delta t$ values of $0.7~t_\text{mct}$ or smaller 
are required. However, for highly rarefied regions the time-step
size can be relaxed to $\Delta t = t_\text{mct}$. 
 
 \item Gas-surface interactions play a major role in obtaining accurate results. For all cases the results obtained with the 
 fully diffusive interaction model (isotropic scattering) showed good agreement with the experiments. Furthermore, the adiabatic nature of the wall in the internal-flow scenarios necessitates the utilization of a diffusive model with incomplete thermal accommodation, accomplished through the CLL model. 
 
  \item The sensitivity study on molecular and energy exchange models did not yield  major significance on the results and all predictions lie within the experimental margins. However, from the performance point of view, the VHS collision model with constant relaxation is found to have a higher computational speed for all cases.

  \item The statistical error analysis of the 
  DSMC method showed that the translational temperature had the highest
  sensitivity to statistical fluctuations. The study favored the use
  of more particles per cell $N_c$ compared to the number of sampling
  time steps $N_T$ for a specific sample size $S = N_c \times N_T$.
  This approach significantly reduces statistical errors while
  maintaining the same computational costs for the sample size.
\end{itemize}

In addition, performance studies concerning both strong and weak scaling
were conducted to analyze the computational speedup 
and the energy consumption of DSMC simulations.
This investigation revealed that the dynamic load
balancing feature of SPARTA provided the most efficient
solution. The iterative study presented here
exploits an optimal parametric setup for hybrid DSMC/CFD
simulations without compromising the computational efficiency.

This benchmarked hybrid methodology holds huge
potential for simulating molecular beam experiments,
particularly those involving gas flows spanning a wide 
Knudsen number range. At Deutsches Elektronen-Synchrotron
(DESY), molecular beams are generated using an aerosol
injector or a cryogenic buffer gas cell for
nanoparticle imaging experiments~\cite{ROTH201817, samantaBGC}.
The integration of the hybrid CFD/DSMC methodology into the
simulation framework currently underway at DESY aims to
accurately simulate such molecular beam experiments. Furthermore,
this simulation framework can facilitate the study of
gas-particle interactions within the experiments and 
contribute to optimizing experimental designs. 

\section{Acknowledgements}

\noindent This work was supported by the 
HELMHOLTZ Data Science Graduate School for the
Structure of Matter (DASHH, HIDSS-0002), by the Helmut-Schmidt
University, University of the Armed Forces Hamburg, and by
Deutsches Elektronen-Synchrotron DESY, a member of the 
Helmholtz Association (HGF). 

Computational resources (HPC-cluster HSUper) have been
provided by the project hpc.bw, funded by dtec.bw -- 
Digitalization and Technology Research Center of the Bundeswehr.
dtec.bw is funded by the European Union -- NextGenerationEU.
Furthermore, we acknowledge 
the computational resource Maxwell operated at DESY. 

\appendix
\section{Mach number profiles of the low density nozzle}\label{app:machno}

\begin{figure}[H]
	\centering
	\begin{tabular}{cc}
	\includegraphics[width=0.5\textwidth]{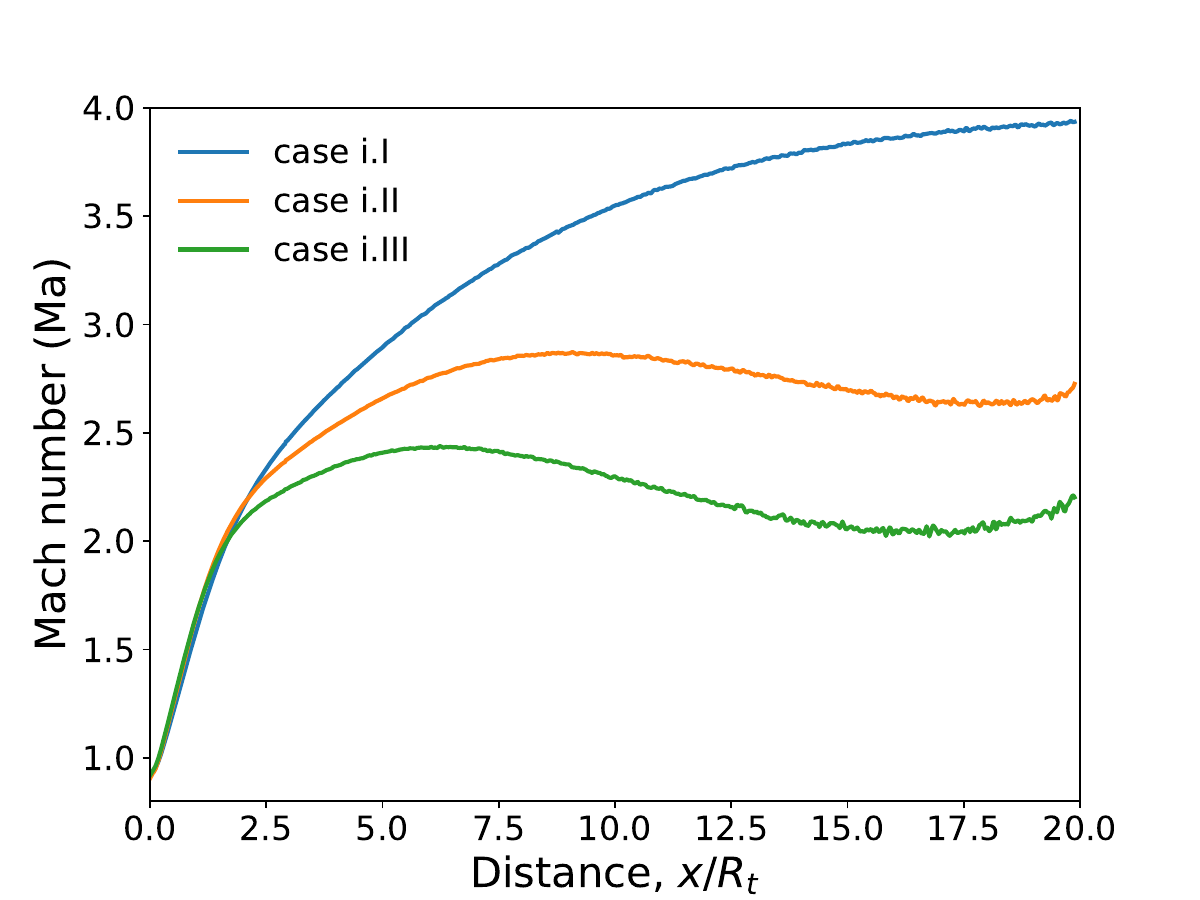}&\includegraphics[width=0.5\textwidth]{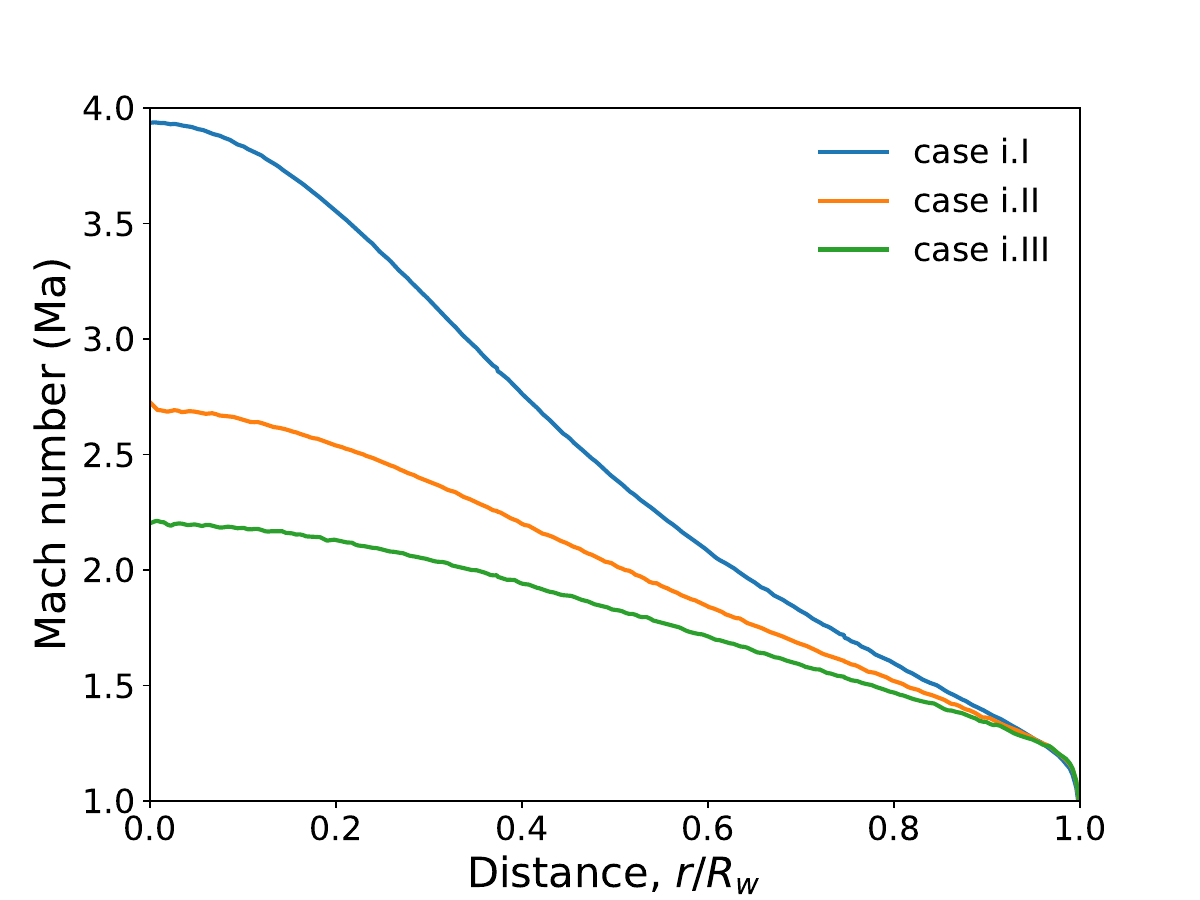}\\
	(a)&(b)\\
	
	\end{tabular}
	\caption{\textcolor{dgreen}{(a) Centerline Mach number profiles from throat to exit.
 (b) Radial Mach number profiles at the exit plane of the nozzle.}}
	\label{fig:mach_nozzle}
\end{figure}

\autoref{fig:mach_nozzle}~(a) shows the variation of the Mach number
along he centerline and \autoref{fig:mach_nozzle}~(b) shows the radial
variation of the Mach number at the exit plane for the internal flow
cases (i.I, i.II and i.III). It can be seen that the flow at the outlet
is supersonic and hence the properties at the outflow are determined
according to a supersonic flow. This proves the validity of the outlet
condition described in Section~\ref{sec:dsmc}.

\section{DSMC inputs}\label{app:dsmc_inputs}

Inputs for the DSMC computations are summarized in \autoref{tab:dsmc_inputs}.

\begin{table}[!h!]
	\centering
	\color{black}
	\begin{tabular}{l c c }
		\hline
		\textbf{Case}&{$\lambda_{min}$ [m]}&{$t_\text{mct}$} [s]\\
		\hline
        
		i.I& 1.95e-5& 4.3e-8\\
		
		i.II& 4.44e-5& 1e-7\\
		
		i.III& 6.6e-5 &1.5e-7\\
		
		e.I& 1.7e-6 & 3.5e-9 \\
		
		e.II& 1.5e-5 & 3.16e-8 \\
		
		\hline
	\end{tabular}
	\caption{\label{tab:dsmc_inputs} \textcolor{dgreen}{Values of
     the minimum mean free path $\lambda_{min}$ and the mean
     collision time $t_\text{mct}$ estimated based on CFD calculations.}}
\end{table}

\section{OpenFoam computational grids}\label{app:grids}

\begin{figure}[H]
    \centering
    \includegraphics[width=1\textwidth]{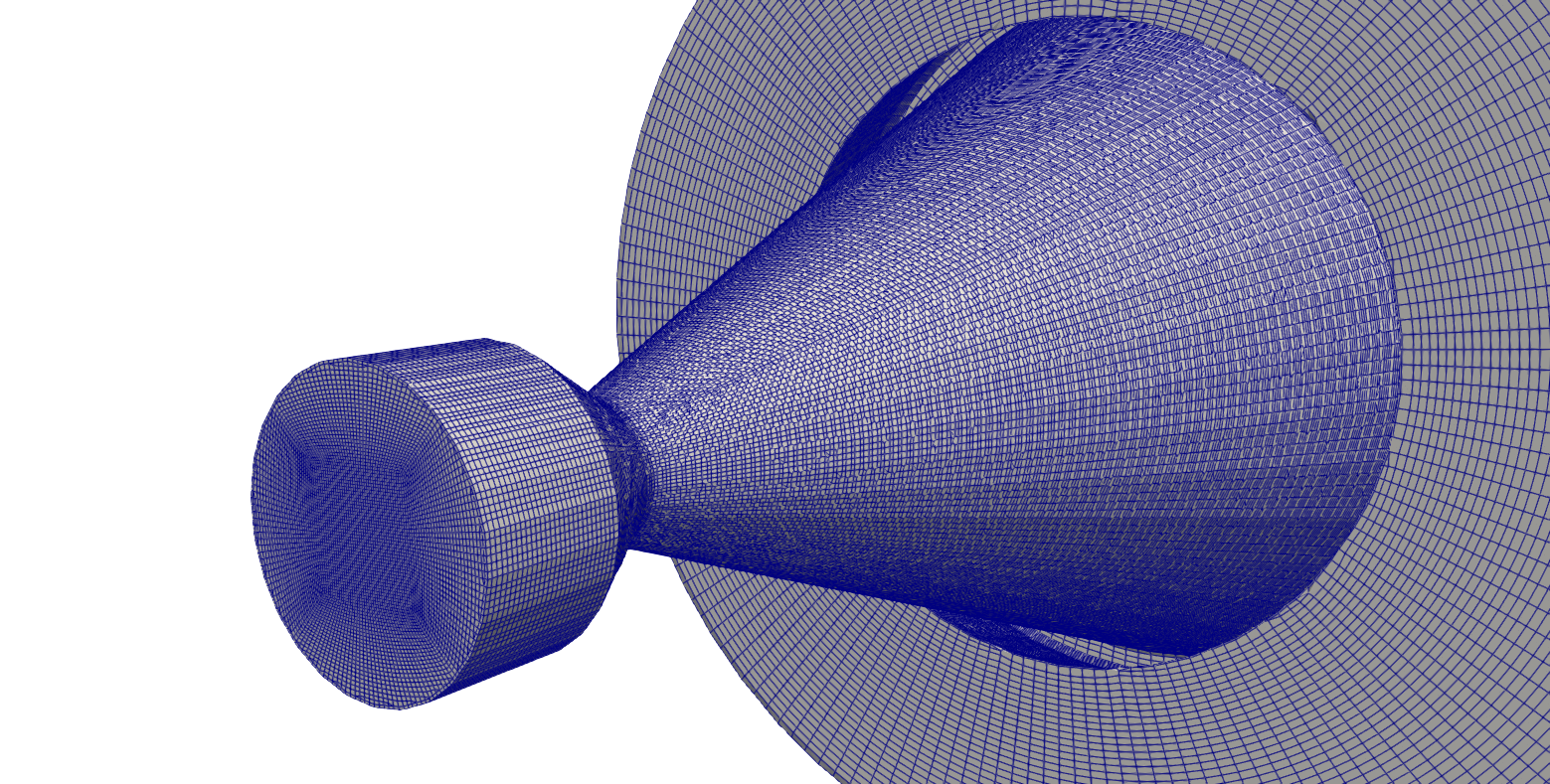}
    \caption{\textcolor{dblue}{Structured O-grid of the nozzle geometry.}}
    \label{fig:nozzle_mesh}
\end{figure}

\begin{figure}[H]
    \centering
    \begin{tabular}{cc}
    \includegraphics[width=0.6\textwidth]{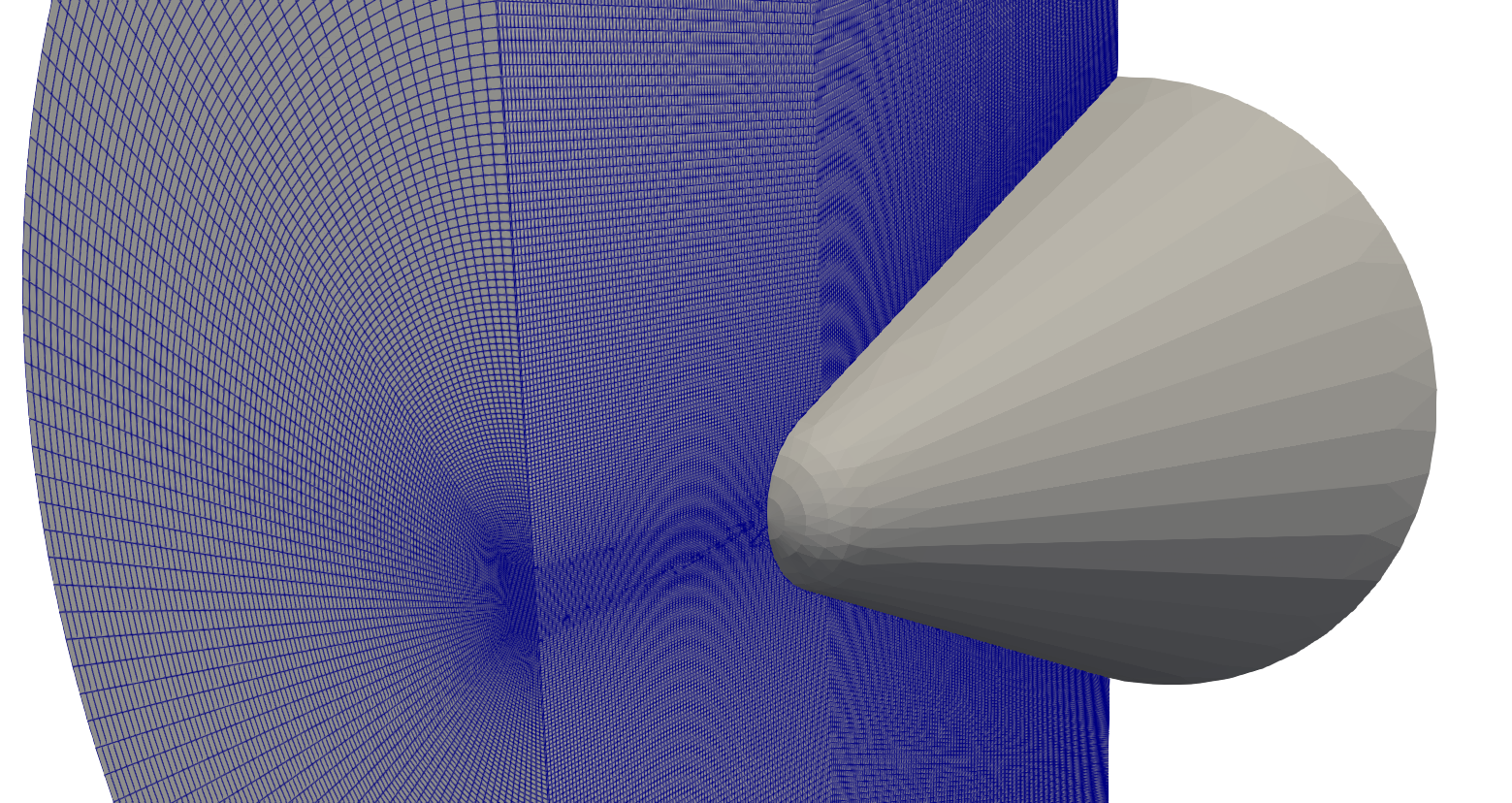}&
    \includegraphics[width=0.6\textwidth]{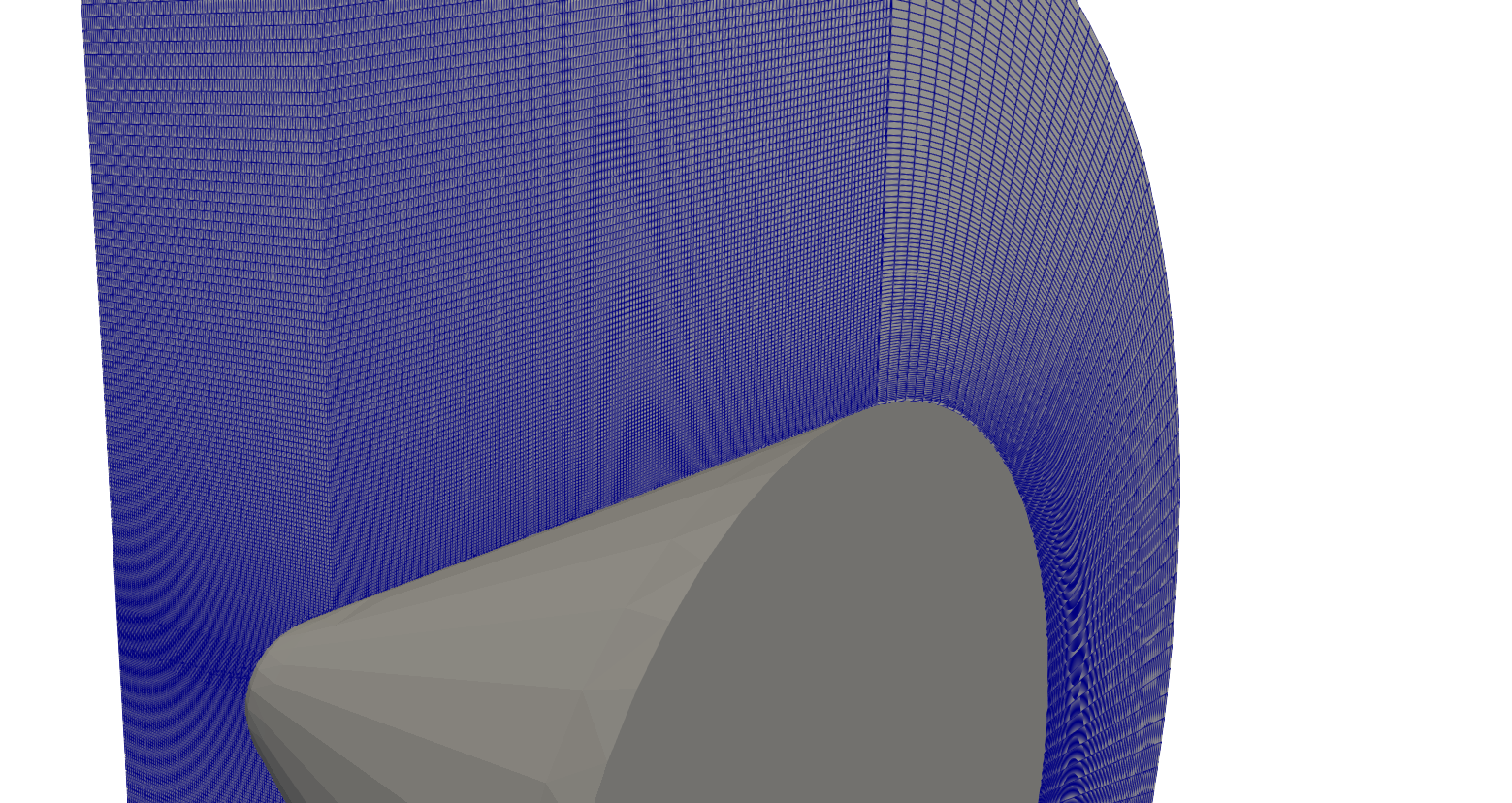}\\
	(a)&(b)\\
    \end{tabular}
    \caption{\textcolor{dblue}{Structured grid (clipped) of a typical conical body viewed from different directions.}}
    \label{fig:blunt_cone_mesh}
\end{figure}

The computational grids for the CFD simulations are generated using
the \texttt{blockMesh} and \texttt{snappyHexMesh} utilities in OpenFoam.
Applying these tools proper structured body-fitted grids can be
obtained as shown in~\autoref{fig:nozzle_mesh}
and \autoref{fig:blunt_cone_mesh}. The final grid for the internal
flow case (nozzle) consists of 513,035 cells. For the blunt cone
case~(e.I) and the sharp cone case~(e.II) grids consisting of
5,143,944 and 8,240,616 cells are employed, respectively.
These numbers of control volumes are obtained based on grid-independence
studies not detailed here since the CFD simulations are not the main 
topic of the present study.

\bibliographystyle{elsarticle-num-names}
\bibliography{flows-cfd-dsmc.bib}

\begin{thebibliography}{61}
\providecommand{\natexlab}[1]{#1}
\providecommand{\url}[1]{\texttt{#1}}
\providecommand{\urlprefix}{URL }
\expandafter\ifx\csname urlstyle\endcsname\relax
  \providecommand{\doi}[1]{doi:\discretionary{}{}{}#1}\else
  \providecommand{\doi}[1]{doi:\discretionary{}{}{}\begingroup
  \urlstyle{rm}\url{#1}\endgroup}\fi
\providecommand{\bibinfo}[2]{#2}

\bibitem[{Bird(1994)}]{Bird1994}
\bibinfo{author}{G.~A. Bird}, \bibinfo{title}{Molecular Gas Dynamics and Direct
  Simulation of Gas Flows}, \bibinfo{publisher}{Oxford}, \bibinfo{year}{1994}.

\bibitem[{Bird(2013)}]{Bird2013}
\bibinfo{author}{G.~A. Bird}, \bibinfo{title}{The DSMC Method: Version 1.2},
  \bibinfo{publisher}{CreateSpace}, \bibinfo{year}{2013}.

\bibitem[{Alexander and Garcia(1997)}]{francis}
\bibinfo{author}{F.~Alexander}, \bibinfo{author}{A.~Garcia},
  \bibinfo{title}{The direct simulation {M}onte {C}arlo method},
  \bibinfo{journal}{Computers in Physics} \bibinfo{volume}{11}
  (\bibinfo{year}{1997}) \bibinfo{pages}{588--593},
  \doi{\bibinfo{doi}{10.1063/1.168619}}.

\bibitem[{Shen(2005)}]{Shen2005}
\bibinfo{author}{C.~Shen}, \bibinfo{title}{Rarefied Gas Dynamics: Fundamentals,
  Simulations and Micro Flows}, \bibinfo{publisher}{Springer},
  \bibinfo{year}{2005}.

\bibitem[{La~Torre et~al.(2009)La~Torre, Kenjeres, Kleijn, and Moerel}]{torre}
\bibinfo{author}{F.~La~Torre}, \bibinfo{author}{S.~Kenjeres},
  \bibinfo{author}{C.~R. Kleijn}, \bibinfo{author}{J.-L. P.~A. Moerel},
  \bibinfo{title}{Evaluation of Micronozzle Performance through {DSMC},
  {N}avier-{S}tokes and Coupled {DSMC/N}avier-{S}tokes Approaches}, in:
  \bibinfo{booktitle}{Computational Science--ICCS 2009: 9th Int. Conf. Baton
  Rouge, LA, USA, May 25-27, 2009 Proc., Part I 9},
  \bibinfo{organization}{Springer}, ISBN \bibinfo{isbn}{978-3-642-01969-2},
  \bibinfo{pages}{675--684}, \doi{\bibinfo{doi}{10.1007/978-3-642-01970-8_67}},
  \bibinfo{year}{2009}.

\bibitem[{Wang and Boyd(2003)}]{wang}
\bibinfo{author}{W.-L. Wang}, \bibinfo{author}{I.~Boyd}, \bibinfo{title}{Hybrid
  DSMC-CFD Simulations of Hypersonic Flow over Sharp and Blunted Bodies}, in:
  \bibinfo{booktitle}{36th AIAA Thermophysics Conference, Orlando, Florida,
  USA}, \bibinfo{pages}{1--13}, \doi{\bibinfo{doi}{10.2514/6.2003-3644}},
  \bibinfo{year}{2003}.

\bibitem[{Glass and Horvath(2003)}]{glass}
\bibinfo{author}{C.~E. Glass}, \bibinfo{author}{T.~J. Horvath},
  \bibinfo{title}{Comparison of a {3-D} {CFD-DSMC} solution methodology with a
  wind tunnel experiment}, \bibinfo{journal}{AIP Conference Proceedings}
  \bibinfo{volume}{663}~(\bibinfo{number}{1}) (\bibinfo{year}{2003})
  \bibinfo{pages}{441--448}, \doi{\bibinfo{doi}{10.1063/1.1581580}},
  \urlprefix\url{https://aip.scitation.org/doi/abs/10.1063/1.1581580}.

\bibitem[{Ghazanfari et~al.(2022)Ghazanfari, Shademan, and
  Mansourzadeh}]{ghazan}
\bibinfo{author}{V.~Ghazanfari}, \bibinfo{author}{M.~Shademan},
  \bibinfo{author}{F.~Mansourzadeh}, \bibinfo{title}{Investigation of feed flow
  effect using {CFD-DSMC} method in a gas centrifuge},
  \bibinfo{journal}{Journal of Nuclear Research and Applications}
  \bibinfo{volume}{2} (\bibinfo{year}{2022}) \bibinfo{pages}{7--14},
  \doi{\bibinfo{doi}{10.24200/jon.2022.1027}}.

\bibitem[{Virgile et~al.(2022)Virgile, Albert, and Julien}]{virgile}
\bibinfo{author}{C.~Virgile}, \bibinfo{author}{A.~Albert},
  \bibinfo{author}{L.~Julien}, \bibinfo{title}{Optimisation of a hybrid
  {NS–DSMC} methodology for continuous–rarefied jet flows},
  \bibinfo{journal}{Acta Astronautica} \bibinfo{volume}{195}
  (\bibinfo{year}{2022}) \bibinfo{pages}{295--308}, ISSN
  \bibinfo{issn}{0094-5765},
  \doi{\bibinfo{doi}{10.1016/j.actaastro.2022.03.012}},
  \urlprefix\url{https://www.sciencedirect.com/science/article/pii/S0094576522001102}.

\bibitem[{Malaikannan et~al.(2018)Malaikannan, Kumar, and
  Chinnappan}]{malaikannan}
\bibinfo{author}{G.~Malaikannan}, \bibinfo{author}{R.~Kumar},
  \bibinfo{author}{A.~K. Chinnappan}, \bibinfo{title}{A novel efficient hybrid
  DSMC–dynamic collision limiter algorithm for multiscale transitional
  flows}, \bibinfo{journal}{International Journal for Numerical Methods in
  Fluids} \bibinfo{volume}{86}~(\bibinfo{number}{9}) (\bibinfo{year}{2018})
  \bibinfo{pages}{565--581}, \doi{\bibinfo{doi}{10.1002/fld.4466}},
  \urlprefix\url{https://onlinelibrary.wiley.com/doi/abs/10.1002/fld.4466}.

\bibitem[{Espinoza et~al.(2016)Espinoza, Casseau, Scanlon, and
  Brown}]{espinoza}
\bibinfo{author}{D.~E.~R. Espinoza}, \bibinfo{author}{V.~Casseau},
  \bibinfo{author}{T.~J. Scanlon}, \bibinfo{author}{R.~E. Brown},
  \bibinfo{title}{An open-source hybrid {CFD-DSMC} solver for high speed
  flows}, \bibinfo{journal}{AIP Conference Proceedings}
  \bibinfo{volume}{1786}~(\bibinfo{number}{1}) (\bibinfo{year}{2016})
  \bibinfo{pages}{050007}, \doi{\bibinfo{doi}{10.1063/1.4967557}},
  \urlprefix\url{https://aip.scitation.org/doi/abs/10.1063/1.4967557}.

\bibitem[{Farber et~al.(2016)Farber, Farber, Gr\"{a}bel, Krick, Reitz, and
  Ueberholz}]{farber}
\bibinfo{author}{K.~Farber}, \bibinfo{author}{P.~Farber},
  \bibinfo{author}{J.~Gr\"{a}bel}, \bibinfo{author}{S.~Krick},
  \bibinfo{author}{J.~Reitz}, \bibinfo{author}{P.~Ueberholz},
  \bibinfo{title}{Development and validation of a coupled
  {N}avier-{S}tokes/{DSMC} simulation for rarefied gas flow in the production
  process for OLEDs}, \bibinfo{journal}{Appl. Math. Comput.}
  \bibinfo{volume}{272}~(\bibinfo{number}{P3}) (\bibinfo{year}{2016})
  \bibinfo{pages}{648--656}, ISSN \bibinfo{issn}{0096-3003},
  \doi{\bibinfo{doi}{10.1016/j.amc.2015.05.040}}.

\bibitem[{Abbate et~al.(2008)Abbate, Kleijn, and Thijsse}]{giannandrea}
\bibinfo{author}{G.~Abbate}, \bibinfo{author}{C.~R. Kleijn},
  \bibinfo{author}{B.~J. Thijsse}, \bibinfo{title}{Validation of a hybrid
  {N}avier-{S}tokes/{DSMC} method for multiscale transient and steady-state gas
  flows}, \bibinfo{journal}{International Journal for Multiscale Computational
  Engineering} \bibinfo{volume}{6} (\bibinfo{year}{2008})
  \bibinfo{pages}{1--12}.

\bibitem[{Auweter-Kurtz et~al.(2005)Auweter-Kurtz, Fertig, Petkow, Stindl,
  Quandt, Munz, Adamidis, Resch, Roller, D'Andrea, and Schneider}]{kurtz}
\bibinfo{author}{M.~Auweter-Kurtz}, \bibinfo{author}{M.~Fertig},
  \bibinfo{author}{D.~Petkow}, \bibinfo{author}{T.~Stindl},
  \bibinfo{author}{M.~Quandt}, \bibinfo{author}{C.-D. Munz},
  \bibinfo{author}{P.~Adamidis}, \bibinfo{author}{M.~Resch},
  \bibinfo{author}{S.~Roller}, \bibinfo{author}{D.~D'Andrea},
  \bibinfo{author}{R.~Schneider}, \bibinfo{title}{Development of a hybrid
  {PIC/DSMC} Code}, in: \bibinfo{booktitle}{29th Int. Electric Propulsion
  Conference}, \bibinfo{pages}{1--15}, \bibinfo{year}{2005}.

\bibitem[{Gallis et~al.(2014)Gallis, Torczynski, Plimpton, Rader, and
  Koehler}]{gallis}
\bibinfo{author}{M.~A. Gallis}, \bibinfo{author}{J.~R. Torczynski},
  \bibinfo{author}{S.~J. Plimpton}, \bibinfo{author}{D.~J. Rader},
  \bibinfo{author}{T.~Koehler}, \bibinfo{title}{Direct simulation {M}onte
  {C}arlo: {T}he quest for speed}, \bibinfo{journal}{AIP Conference
  Proceedings} \bibinfo{volume}{1628}~(\bibinfo{number}{1})
  (\bibinfo{year}{2014}) \bibinfo{pages}{27--36},
  \doi{\bibinfo{doi}{10.1063/1.4902571}}.

\bibitem[{Plimpton et~al.(2019)Plimpton, Moore, Borner, Stagg, Koehler,
  Torczynski, and Gallis}]{plimp}
\bibinfo{author}{S.~J. Plimpton}, \bibinfo{author}{S.~G. Moore},
  \bibinfo{author}{A.~Borner}, \bibinfo{author}{A.~K. Stagg},
  \bibinfo{author}{T.~P. Koehler}, \bibinfo{author}{J.~R. Torczynski},
  \bibinfo{author}{M.~A. Gallis}, \bibinfo{title}{Direct simulation {M}onte
  {C}arlo on petaflop supercomputers and beyond}, \bibinfo{journal}{Physics of
  Fluids} \bibinfo{volume}{31}~(\bibinfo{number}{8}) (\bibinfo{year}{2019})
  \bibinfo{pages}{086101}, \doi{\bibinfo{doi}{10.1063/1.5108534}}.

\bibitem[{Li et~al.(2022)Li, Zhang, Deng, Ding, Jiang, and Wang}]{li}
\bibinfo{author}{J.~Li}, \bibinfo{author}{Z.~Zhang}, \bibinfo{author}{L.~Deng},
  \bibinfo{author}{X.~Ding}, \bibinfo{author}{D.~Jiang},
  \bibinfo{author}{P.~Wang}, \bibinfo{title}{Massively Parallel Acceleration of
  Unstructured DSMC Computing}, in: \bibinfo{booktitle}{Proceedings of the 2022
  6th High Performance Computing and Cluster Technologies Conference}, HPCCT
  '22, \bibinfo{publisher}{Association for Computing Machinery},
  \bibinfo{address}{New York, NY, USA}, ISBN \bibinfo{isbn}{9781450396646},
  \bibinfo{pages}{1--6}, \doi{\bibinfo{doi}{10.1145/3560442.3560443}},
  \bibinfo{year}{2022}.

\bibitem[{Ivanov et~al.(1997)Ivanov, Markelov, Taylor, and Watts}]{ivanov}
\bibinfo{author}{M.~Ivanov}, \bibinfo{author}{G.~Markelov},
  \bibinfo{author}{S.~Taylor}, \bibinfo{author}{J.~Watts},
  \bibinfo{title}{Parallel {DSMC} strategies for {3D} computations}, in:
  \bibinfo{editor}{P.~Schiano}, \bibinfo{editor}{A.~Ecer},
  \bibinfo{editor}{J.~Periaux}, \bibinfo{editor}{N.~Satofuka} (Eds.),
  \bibinfo{booktitle}{Parallel Computational Fluid Dynamics 1996},
  \bibinfo{publisher}{North-Holland}, \bibinfo{address}{Amsterdam}, ISBN
  \bibinfo{isbn}{978-0-444-82327-4}, \bibinfo{pages}{485--492},
  \doi{\bibinfo{doi}{10.1016/B978-044482327-4/50128-5}},
  \urlprefix\url{https://www.sciencedirect.com/science/article/pii/B9780444823274501285},
  \bibinfo{year}{1997}.

\bibitem[{Roohi and Darbandi(2012)}]{roohi}
\bibinfo{author}{E.~Roohi}, \bibinfo{author}{M.~Darbandi},
  \bibinfo{title}{Recommendations on performance of parallel {DSMC} algorithm
  in solving subsonic nanoflows}, \bibinfo{journal}{Applied Mathematical
  Modelling} \bibinfo{volume}{36}~(\bibinfo{number}{5}) (\bibinfo{year}{2012})
  \bibinfo{pages}{2314--2321}, ISSN \bibinfo{issn}{0307-904X},
  \doi{\bibinfo{doi}{10.1016/j.apm.2011.08.036}},
  \urlprefix\url{https://www.sciencedirect.com/science/article/pii/S0307904X11005361}.

\bibitem[{Fedosov et~al.(2005)Fedosov, Rogasinsky, Zeifman, Ivanov, ALexeenko,
  and Levin}]{Fedo}
\bibinfo{author}{D.~A. Fedosov}, \bibinfo{author}{S.~V. Rogasinsky},
  \bibinfo{author}{M.~I. Zeifman}, \bibinfo{author}{M.~S. Ivanov},
  \bibinfo{author}{A.~A. ALexeenko}, \bibinfo{author}{D.~A. Levin},
  \bibinfo{title}{Analysis of Numerical Errors in the {DSMC} Method},
  \bibinfo{journal}{AIP Conference Proceedings}
  \bibinfo{volume}{762}~(\bibinfo{number}{1}) (\bibinfo{year}{2005})
  \bibinfo{pages}{589--594}, ISSN \bibinfo{issn}{0094-243X},
  \doi{\bibinfo{doi}{10.1063/1.1941600}}.

\bibitem[{Chung et~al.(1995)Chung, Kim, Stubbs, and De~Witt}]{chung}
\bibinfo{author}{C.-H. Chung}, \bibinfo{author}{S.~C. Kim},
  \bibinfo{author}{R.~M. Stubbs}, \bibinfo{author}{K.~J. De~Witt},
  \bibinfo{title}{Low-density nozzle flow by the direct simulation {M}onte
  {C}arlo and continuum methods}, \bibinfo{journal}{Journal of Propulsion and
  Power} \bibinfo{volume}{11}~(\bibinfo{number}{1}) (\bibinfo{year}{1995})
  \bibinfo{pages}{64--70}, \doi{\bibinfo{doi}{10.2514/3.23841}}.

\bibitem[{Koura and Matsumoto(1991)}]{koura}
\bibinfo{author}{K.~Koura}, \bibinfo{author}{H.~Matsumoto},
  \bibinfo{title}{Variable soft sphere molecular model for inverse power law or
  {L}ennard-{J}ones potential}, \bibinfo{journal}{Physics of Fluids A: Fluid
  Dynamics} \bibinfo{volume}{3} (\bibinfo{year}{1991}) \bibinfo{pages}{2459},
  \doi{\bibinfo{doi}{10.1063/1.858184}}.

\bibitem[{Liu et~al.(2006)Liu, Zhang, Zhang, and Chen}]{liu}
\bibinfo{author}{M.~Liu}, \bibinfo{author}{X.~Zhang},
  \bibinfo{author}{G.~Zhang}, \bibinfo{author}{Y.~Chen}, \bibinfo{title}{Study
  on micronozzle flow and propulsion performance using {DSMC} and continuum
  methods}, \bibinfo{journal}{Acta Mechanica Sinica/Lixue Xuebao}
  \bibinfo{volume}{22} (\bibinfo{year}{2006}) \bibinfo{pages}{409--416},
  \doi{\bibinfo{doi}{10.1007/s10409-006-0020-y}}.

\bibitem[{Moss et~al.(1997)Moss, Wilmoth, and Price}]{moss}
\bibinfo{author}{J.~Moss}, \bibinfo{author}{R.~Wilmoth},
  \bibinfo{author}{J.~Price}, \bibinfo{title}{{DSMC} Simulations of blunt body
  flows for {M}ars entries: {M}ars pathfinder and mars microprobe capsules},
  in: \bibinfo{booktitle}{32nd Thermophysics Conference},
  \bibinfo{pages}{2508}, \doi{\bibinfo{doi}{10.2514/6.1997-2508}},
  \bibinfo{year}{1997}.

\bibitem[{Braunstein and Cline(2003)}]{braunstein}
\bibinfo{author}{M.~Braunstein}, \bibinfo{author}{J.~Cline},
  \bibinfo{title}{Progress on Parallelizing a General Purpose Direct Simulation
  Monte Carlo ({DSMC}) Code for High Performance Computing Applications}, in:
  \bibinfo{booktitle}{NASA Publications 20030107284, AMOS 2003 Technical
  Conference, Maui, HI, United States, 10 Sep. 2003}, \bibinfo{year}{2003}.

\bibitem[{Khanlarov and Lukianov(1999)}]{khanlarov}
\bibinfo{author}{G.~O. Khanlarov}, \bibinfo{author}{G.~A. Lukianov},
  \bibinfo{title}{{DSMC} of the inner atmosphere of a comet on shared memory
  multiprocessors}, in: \bibinfo{editor}{P.~Sloot}, \bibinfo{editor}{M.~Bubak},
  \bibinfo{editor}{A.~Hoekstra}, \bibinfo{editor}{B.~Hertzberger} (Eds.),
  \bibinfo{booktitle}{High-Performance Computing and Networking},
  \bibinfo{publisher}{Springer Berlin Heidelberg}, \bibinfo{address}{Berlin,
  Heidelberg}, ISBN \bibinfo{isbn}{978-3-540-48933-7},
  \bibinfo{pages}{1187--1189}, \bibinfo{year}{1999}.

\bibitem[{Celoria(2022)}]{celoria}
\bibinfo{author}{M.~Celoria}, \bibinfo{title}{Porting of {DSMC} to multi-GPUs
  using OpenACC}, in: \bibinfo{booktitle}{Master Thesis},
  \bibinfo{publisher}{SISSA}, \bibinfo{pages}{5--63}, \bibinfo{year}{2022}.

\bibitem[{{ESI-OpenCFD}(sent)}]{openfoam}
\bibinfo{author}{{ESI-OpenCFD}}, \bibinfo{title}{{OpenFOAM - The Open Source
  CFD Toolbox}}, \bibinfo{howpublished}{\url{https://www.openfoam.com}},
  \bibinfo{note}{version v2112}, \bibinfo{year}{2004--present}.

\bibitem[{{Sandia National Laboratories}(2022)}]{sparta2022}
\bibinfo{author}{{Sandia National Laboratories}}, \bibinfo{title}{{SPARTA -
  Stochastic PArallel Rarefied-gas Time-accurate Analyzer}},
  \bibinfo{howpublished}{\url{https://www.sparta.github.io}},
  \bibinfo{note}{release 18Jul2022}, \bibinfo{year}{2022}.

\bibitem[{Swaminathan-Gopalan and Stephani(2016)}]{gopal}
\bibinfo{author}{K.~Swaminathan-Gopalan}, \bibinfo{author}{K.~A. Stephani},
  \bibinfo{title}{Recommended direct simulation {M}onte {C}arlo collision model
  parameters for modeling ionized air transport processes},
  \bibinfo{journal}{Physics of Fluids}
  \bibinfo{volume}{28}~(\bibinfo{number}{2}) (\bibinfo{year}{2016})
  \bibinfo{pages}{027101}, \doi{\bibinfo{doi}{10.1063/1.4939719}}.

\bibitem[{Ze(1705)}]{ze}
\bibinfo{author}{X.~Ze}, \bibinfo{title}{{DSMC} Simulation on Couette Flow in
  Micro-channels}, in: \bibinfo{booktitle}{Proceedings of the 2017 2nd Int.\
  Conference on Materials Science, Machinery and Energy Engineering (MSMEE
  2017)}, \bibinfo{publisher}{Atlantis Press}, ISBN
  \bibinfo{isbn}{978-94-6252-346-3}, ISSN \bibinfo{issn}{2352-5401},
  \bibinfo{pages}{991--994}, \doi{\bibinfo{doi}{10.2991/msmee-17.2017.192}},
  \bibinfo{year}{2017/05}.

\bibitem[{Weaver and Alexeenko(2015)}]{VSSparam2}
\bibinfo{author}{A.~B. Weaver}, \bibinfo{author}{A.~A. Alexeenko},
  \bibinfo{title}{Revised variable soft sphere and {L}ennard-{J}ones model
  parameters for eight common gases up to 2200 {K}}, \bibinfo{journal}{Journal
  of Physical and Chemical Reference Data} \bibinfo{volume}{44}
  (\bibinfo{year}{2015}) \bibinfo{pages}{023103},
  \doi{\bibinfo{doi}{10.1063/1.4921245}}.

\bibitem[{{Stefanov} et~al.(2005){Stefanov}, {Barber}, {Ota}, and
  {Emerson}}]{stefanov}
\bibinfo{author}{S.~K. {Stefanov}}, \bibinfo{author}{R.~W. {Barber}},
  \bibinfo{author}{M.~{Ota}}, \bibinfo{author}{D.~R. {Emerson}},
  \bibinfo{title}{{Comparison between {N}avier-{S}tokes and {DSMC} Calculations
  for Low {R}eynolds Number Slip Flow Past a Confined Microsphere}}, in:
  \bibinfo{editor}{M.~{Capitelli}} (Ed.), \bibinfo{booktitle}{Rarefied Gas
  Dynamics: 24th International Symposium on Rarefied Gas Dynamics}, vol.
  \bibinfo{volume}{762} of \emph{\bibinfo{series}{American Institute of Physics
  Conference Series}}, \bibinfo{pages}{701--706},
  \doi{\bibinfo{doi}{10.1063/1.1941617}}, \bibinfo{year}{2005}.

\bibitem[{Larsen and Borgnakke(1974)}]{L-B1}
\bibinfo{author}{P.~S. Larsen}, \bibinfo{author}{C.~Borgnakke},
  \bibinfo{title}{Rarefied Gas Dynamics 1: {S}tatistical collision model for
  simulating polyatomic gas with restricted energy exchange},
  vol.~\bibinfo{volume}{1}, \bibinfo{publisher}{DFVLR-Press},
  \bibinfo{year}{1974}.

\bibitem[{Larsen and Borgnakke(1975)}]{L-B2}
\bibinfo{author}{P.~S. Larsen}, \bibinfo{author}{C.~Borgnakke},
  \bibinfo{title}{Statistical collision model for {M}onte {C}arlo simulation of
  polyatomic gas mixture}, \bibinfo{journal}{J. Comput. Phys}
  \bibinfo{volume}{18} (\bibinfo{year}{1975}) \bibinfo{pages}{405}.

\bibitem[{Xiao et~al.(2014)Xiao, Shang, and Wu}]{xiao}
\bibinfo{author}{H.~Xiao}, \bibinfo{author}{Y.~Shang}, \bibinfo{author}{D.~Wu},
  \bibinfo{title}{{DSMC} Simulation and Experimental Validation of Shock
  Interaction in Hypersonic Low Density Flow}, \bibinfo{journal}{The Scientific
  World Journal} \bibinfo{volume}{2014} (\bibinfo{year}{2014})
  \bibinfo{pages}{732765}, \doi{\bibinfo{doi}{10.1155/2014/732765}}.

\bibitem[{Klothakis and Nikolos(2015)}]{kloth}
\bibinfo{author}{A.~Klothakis}, \bibinfo{author}{I.~Nikolos},
  \bibinfo{title}{Modeling of rarefied hypersonic flows using the massively
  parallel {DSMC} kernel "{SPARTA}"}, in: \bibinfo{booktitle}{8th Int.\
  Congress on Computational Mechanics}, vol.~\bibinfo{volume}{12},
  \bibinfo{pages}{1--10}, \bibinfo{year}{2015}.

\bibitem[{Wu et~al.(2007)Wu, Lian, Cheng, and Chen}]{wu}
\bibinfo{author}{J.~S. Wu}, \bibinfo{author}{Y.~Y. Lian},
  \bibinfo{author}{G.~Cheng}, \bibinfo{author}{Y.-S. Chen},
  \bibinfo{title}{Parallel hybrid particle-continuum ({DSMC-NS}) flow
  simulations using {3-D} unstructured mesh}, \bibinfo{journal}{Parallel
  Computational Fluid Dynamics 2006}  (\bibinfo{year}{2007})
  \bibinfo{pages}{1--10}\doi{\bibinfo{doi}{10.1016/B978-044453035-6/50003-1}}.

\bibitem[{Chen and Boyd(1996)}]{CHEN_boyd}
\bibinfo{author}{G.~Chen}, \bibinfo{author}{I.~D. Boyd},
  \bibinfo{title}{Statistical Error Analysis for the Direct Simulation Monte
  Carlo Technique}, \bibinfo{journal}{Journal of Computational Physics}
  \bibinfo{volume}{126}~(\bibinfo{number}{2}) (\bibinfo{year}{1996})
  \bibinfo{pages}{434--448}, ISSN \bibinfo{issn}{0021-9991},
  \doi{\bibinfo{doi}{https://doi.org/10.1006/jcph.1996.0148}},
  \urlprefix\url{https://www.sciencedirect.com/science/article/pii/S0021999196901485}.

\bibitem[{Plotnikov and Shkarupa(2012)}]{plotnikov}
\bibinfo{author}{M.~Y. Plotnikov}, \bibinfo{author}{E.~V. Shkarupa},
  \bibinfo{title}{Selection of sampling numerical parameters for the {DSMC}
  method}, \bibinfo{journal}{Computers \& Fluids}
  \bibinfo{volume}{58}~(\bibinfo{number}{12}) (\bibinfo{year}{2012})
  \bibinfo{pages}{102--111},
  \doi{\bibinfo{doi}{10.1016/j.compfluid.2012.01.007}}.

\bibitem[{Boyd(2003)}]{Boyd1}
\bibinfo{author}{I.~Boyd}, \bibinfo{title}{Predicting breakdown of the
  continuum equations under rarefied flow conditions}, \bibinfo{journal}{AIP
  Conference Proceedings} \bibinfo{volume}{663} (\bibinfo{year}{2003})
  \bibinfo{pages}{899--906}, \doi{\bibinfo{doi}{10.1063/1.1581636}}.

\bibitem[{Hedahl and Wilmoth(1996)}]{MaxCLL_nasa}
\bibinfo{author}{M.~Hedahl}, \bibinfo{author}{R.~Wilmoth},
  \bibinfo{title}{Comparisons of the Maxwell and CLL gas/surface interaction
  models using DSMC}, \bibinfo{type}{Tech. Rep.}, \bibinfo{institution}{NASA
  Langley Research Center Hampton, VA, United States},
  \urlprefix\url{https://ntrs.nasa.gov/citations/19960009098},
  \bibinfo{year}{1996}.

\bibitem[{Pfeiffer et~al.(2019)Pfeiffer, Mirza, and Nizenkov}]{Pfeiffer}
\bibinfo{author}{M.~Pfeiffer}, \bibinfo{author}{A.~Mirza},
  \bibinfo{author}{P.~Nizenkov}, \bibinfo{title}{Evaluation of particle-based
  continuum methods for a coupling with the direct simulation {M}onte {C}arlo
  method based on a nozzle expansion}, \bibinfo{journal}{Physics of Fluids}
  \bibinfo{volume}{31} (\bibinfo{year}{2019}) \bibinfo{pages}{073601},
  \doi{\bibinfo{doi}{10.1063/1.5098085}}.

\bibitem[{Falchi et~al.(2017)Falchi, Minisci, Vasile, and Kubicek}]{falchi}
\bibinfo{author}{A.~Falchi}, \bibinfo{author}{E.~Minisci},
  \bibinfo{author}{M.~Vasile}, \bibinfo{author}{M.~Kubicek},
  \bibinfo{title}{Aero-thermal re-entry sensitivity analysis using {DSMC} and a
  high dimensional model representation-based approach}, in:
  \bibinfo{booktitle}{7th European Conference on Space Debris},
  \bibinfo{pages}{1--11}, \bibinfo{year}{2017}.

\bibitem[{Rothe(1970)}]{rothe}
\bibinfo{author}{D.~E. Rothe}, \bibinfo{title}{Experimental Study of Viscous
  Low-Density Nozzle Flows}, \bibinfo{type}{Tech. Rep.},
  \bibinfo{institution}{CAL-AI-2590-A-2, Cornell Aeronautical Laboratory, Inc.,
  Cornell University, Buffalo, New York}, \bibinfo{year}{1970}.

\bibitem[{Greenshields et~al.(2010)Greenshields, Weller, Gasparini, and
  J.}]{Greenshields}
\bibinfo{author}{C.~Greenshields}, \bibinfo{author}{H.~Weller},
  \bibinfo{author}{L.~Gasparini}, \bibinfo{author}{R.~J.},
  \bibinfo{title}{Implementation of semi-discrete, non-staggered central
  schemes in a colocated, polyhedral, finite volume framework, for high-speed
  viscous flows}, \bibinfo{journal}{International Journal for Numerical Methods
  in Fluids} \bibinfo{volume}{62}~(\bibinfo{number}{1}) (\bibinfo{year}{2010})
  \bibinfo{pages}{1--21}.

\bibitem[{Kurganov and Tadmor(2000)}]{Kurganov}
\bibinfo{author}{A.~Kurganov}, \bibinfo{author}{E.~Tadmor}, \bibinfo{title}{New
  High-Resolution Central Schemes for Nonlinear Conservation Laws and
  Convection–Diffusion Equations}, \bibinfo{journal}{Journal of Computational
  Physics} \bibinfo{volume}{160} (\bibinfo{year}{2000})
  \bibinfo{pages}{241--282}, \doi{\bibinfo{doi}{10.1006/jcph.2000.6459}}.

\bibitem[{Kraposhin et~al.(2015)Kraposhin, Bovtrikova, and
  Strijhak}]{Kraposhin}
\bibinfo{author}{M.~Kraposhin}, \bibinfo{author}{A.~Bovtrikova},
  \bibinfo{author}{S.~Strijhak}, \bibinfo{title}{Adaptation of Kurganov- Tadmor
  numerical scheme for applying in combination with the PISO method in
  numerical simulation of flows in a wide range of mach numbers},
  \bibinfo{journal}{Procedia Computer Science} \bibinfo{volume}{56}
  (\bibinfo{year}{2015}) \bibinfo{pages}{43--52}.

\bibitem[{Corporation(2022)}]{SPARTA}
\bibinfo{author}{S.~Corporation}, \bibinfo{title}{Sparta Users Manual},
  \bibinfo{organization}{Sandia National Laboratories}, \bibinfo{year}{2022}.

\bibitem[{spa(2022)}]{spartaBench}
\bibinfo{title}{Sparta Benchmarks}, \bibinfo{note}{accessed: 2010-09-30},
  \bibinfo{year}{2022}.

\bibitem[{Parker(1959)}]{parker}
\bibinfo{author}{J.~G. Parker}, \bibinfo{title}{Rotational and vibrational
  relaxation in diatomic gases}, \bibinfo{journal}{Physics of Fluids}
  \bibinfo{volume}{2}~(\bibinfo{number}{4}) (\bibinfo{year}{1959})
  \bibinfo{pages}{449--462}, ISSN \bibinfo{issn}{0031-9171},
  \doi{\bibinfo{doi}{10.1063/1.1724417}}.

\bibitem[{Lordi and Mates(1970)}]{lordi}
\bibinfo{author}{J.~A. Lordi}, \bibinfo{author}{R.~E. Mates},
  \bibinfo{title}{Rotational relaxation in nonpolar diatomic gases},
  \bibinfo{journal}{Physics of Fluids}
  \bibinfo{volume}{13}~(\bibinfo{number}{2}) (\bibinfo{year}{1970})
  \bibinfo{pages}{291--308}, ISSN \bibinfo{issn}{0031-9171},
  \doi{\bibinfo{doi}{10.1063/1.1692920}}.

\bibitem[{Hinchen and Foley(1966)}]{lobularExp}
\bibinfo{author}{J.~J. Hinchen}, \bibinfo{author}{W.~M. Foley},
  \bibinfo{title}{Scattering Molecular beams by metallic surfaces}, in:
  \bibinfo{booktitle}{5th International Symposium on Rarefied Gas Dynamics},
  \bibinfo{pages}{505}, \bibinfo{year}{1966}.

\bibitem[{Lord(1991)}]{Lord_ext1}
\bibinfo{author}{R.~G. Lord}, \bibinfo{title}{Some extensions to the
  {C}ercignani-{L}ampis gas-surface scattering kernel},
  \bibinfo{journal}{Physics of Fluids} \bibinfo{volume}{A3}
  (\bibinfo{year}{1991}) \bibinfo{pages}{706--710}.

\bibitem[{Lord(1995)}]{Lord_ext2}
\bibinfo{author}{R.~G. Lord}, \bibinfo{title}{Some further extensions to the
  {C}ercignani-{L}ampis gas-surface interaction model},
  \bibinfo{journal}{Physics of Fluids} \bibinfo{volume}{7}
  (\bibinfo{year}{1995}) \bibinfo{pages}{1159--1161}.

\bibitem[{Wang et~al.(2002)Wang, Sun, and Boyd}]{wang_old}
\bibinfo{author}{W.-L. Wang}, \bibinfo{author}{Q.~Sun},
  \bibinfo{author}{I.~Boyd}, \bibinfo{title}{Towards development of a hybrid
  {DSMC-CFD} method for simulating hypersonic interacting flows},
  \bibinfo{journal}{AIAA Paper}  (\bibinfo{year}{2002})
  \bibinfo{pages}{2002--3099,~}\doi{\bibinfo{doi}{10.2514/6.2002-3099}}.

\bibitem[{Moss and Bird(2004)}]{moss2}
\bibinfo{author}{J.~Moss}, \bibinfo{author}{G.~Bird}, \bibinfo{title}{DSMC
  Simulations of Hypersonic Flows With Shock Interactions and Validation With
  Experiments}, in: \bibinfo{booktitle}{37th AIAA Thermophysics Conference},
  \bibinfo{pages}{2585}, \doi{\bibinfo{doi}{10.2514/6.2004-2585}},
  \bibinfo{year}{2004}.

\bibitem[{Ng and Liu(2002)}]{Eddi2002}
\bibinfo{author}{E.~Y.-K. Ng}, \bibinfo{author}{N.~Liu}, \bibinfo{title}{The
  impacts of time-step size in the application of the direct simulation Monte
  Carlo method to ultra-thin gas film lubrication}, \bibinfo{journal}{Journal
  of Micromechanics and Microengineering}
  \bibinfo{volume}{12}~(\bibinfo{number}{5}) (\bibinfo{year}{2002})
  \bibinfo{pages}{567}, \doi{\bibinfo{doi}{10.1088/0960-1317/12/5/309}},
  \urlprefix\url{https://dx.doi.org/10.1088/0960-1317/12/5/309}.

\bibitem[{Georgiou et~al.(2014)Georgiou, Cadeau, Glesser, Auble, Jette, and
  Hautreux}]{Georgiou2014}
\bibinfo{author}{Y.~Georgiou}, \bibinfo{author}{T.~Cadeau},
  \bibinfo{author}{D.~Glesser}, \bibinfo{author}{D.~Auble},
  \bibinfo{author}{M.~Jette}, \bibinfo{author}{M.~Hautreux},
  \bibinfo{title}{Energy Accounting and Control with {SLURM} Resource and Job
  Management System}, in: \bibinfo{booktitle}{Distributed Computing and
  Networking}, \bibinfo{publisher}{Springer Berlin Heidelberg},
  \bibinfo{address}{Berlin, Heidelberg}, \bibinfo{pages}{96--118},
  \doi{\bibinfo{doi}{10.1007/978-3-642-45249-9_7}}, \bibinfo{year}{2014}.

\bibitem[{Roth et~al.(2018)Roth, Awel, Horke, and Küpper}]{ROTH201817}
\bibinfo{author}{N.~Roth}, \bibinfo{author}{S.~Awel}, \bibinfo{author}{D.~A.
  Horke}, \bibinfo{author}{J.~Küpper}, \bibinfo{title}{Optimizing aerodynamic
  lenses for single-particle imaging}, \bibinfo{journal}{Journal of Aerosol
  Science} \bibinfo{volume}{124} (\bibinfo{year}{2018})
  \bibinfo{pages}{17--29}, ISSN \bibinfo{issn}{0021-8502},
  \doi{\bibinfo{doi}{10.1016/j.jaerosci.2018.06.010}}.

\bibitem[{Samanta et~al.(2020)Samanta, Amin, Estillore, Roth, Worbs, Horke, and
  Küpper}]{samantaBGC}
\bibinfo{author}{A.~K. Samanta}, \bibinfo{author}{M.~Amin},
  \bibinfo{author}{A.~D. Estillore}, \bibinfo{author}{N.~Roth},
  \bibinfo{author}{L.~Worbs}, \bibinfo{author}{D.~A. Horke},
  \bibinfo{author}{J.~Küpper}, \bibinfo{title}{{Controlled beams of
  shock-frozen, isolated, biological and artificial nanoparticles}},
  \bibinfo{journal}{Structural Dynamics}
  \bibinfo{volume}{7}~(\bibinfo{number}{2}) (\bibinfo{year}{2020})
  \bibinfo{pages}{024304}, ISSN \bibinfo{issn}{2329-7778},
  \doi{\bibinfo{doi}{10.1063/4.0000004}}.

\end{thebibliography}

\end{document}